\documentclass{elsart}

\usepackage{natbib}

\usepackage{graphicx}
\newcommand{\dg}{$^\circ$\,}
\usepackage{amssymb}
\usepackage{morefloats}
\usepackage{pdflscape}
\usepackage{longtable}

\begin{document}

\begin{frontmatter}

\title{Asteroid pairs: a complex picture}

% use optional labels to link authors explicitly to addresses:
%\author[label1,label2]{}
% \address[label1]{}
% \address[label2]{}

\author[a]{P. Pravec},
\author[a,b]{P. Fatka},
\author[b]{D. Vokrouhlick\'y},
\author[a]{P. Scheirich},
\author[b]{J. \v{D}urech},
\author[S]{D. J. Scheeres},
\author[a]{P. Ku\v{s}nir\'ak},
\author[a]{K. Hornoch},
\author[d]{A. Gal\'ad},           % 2110, 3749,7343,10123,25884,88604 (Modra)
\author[j]{D. P. Pray},           % 1741, 2110, 2897, 3749, 4765, 4905, 7343, 7813, 5026,10123,11677,16815,19289,21028,42946,52852,88259
\author[p]{Yu. N. Krugly},        % 3749, 4905, 5026, 7343, 9068,11677,13284,17198,19289,42946,44612,54041,60677,70511,A5247
\author[U]{O. Burkhonov},         % 13284,17198,19289,44612,54041,60677,70511,88259,A5247
\author[U]{Sh. A. Ehgamberdiev},
\author[P]{J. Pollock},           % 2110, 3749, 4765, 4905, 5026,10123,19289,25884,51866
\author[F]{N. Moskovitz},         %43008
\author[O]{J. L. Ortiz},          % 2110, 3749, 5026, 7343, 9068,13284,17288,21436
\author[O]{N. Morales},           % 2110, 3749, 5026, 7343, 9068,13284,17288,21436
\author[n]{M. Hus\'arik},         % 2897, 3749, 7343, 9068,13284,21028,88604
\author[W,w]{R. Ya. Inasaridze},  % 3749, 7343, 9068,11677,13284,19289
\author[J]{J. Oey},               % 1741,13284,16126,16815,25884
\author[K]{D. Polishook},         % 4765,25884,42946,63440
\author[b]{J. Hanu\v{s}},
\author[a,b]{H. Ku\v{c}\'akov\'a},
\author[b]{J. Vra\v{s}til},
\author[d]{J. Vil\'agi},          % 2110, 3749, 7343,10123,25884
\author[d]{\v{S}. Gajdo\v{s}},    % 2110, 3749, 7343,88604
\author[d]{L. Korno\v{s}},        % 3749, 7343,25884,88604
\author[d,H]{P. Vere\v{s}},       %10123,25884
\author[v]{N. M. Gaftonyuk},      % 5026, 7343,38184
\author[p]{T. Hromakina},         % 13284,17198,19289,54041,60677,A5247
\author[p]{A. V. Sergeyev},       % 42946,A5247
\author[p]{I. G. Slyusarev},      % 70511,88259
\author[W,w]{V. R. Ayvazian},     % 13284
\author[l]{W. R. Cooney},         % 2110, 7343
\author[l]{J. Gross},             % 2110, 7343
\author[l,L]{D. Terrell},         % 2110, 7343
\author[V]{F. Colas},             % 3749, 7343
\author[V]{F. Vachier},           % 1741
%\author[N]{J. Brinsfield},       % 3749 (available online)
\author[s]{S. Slivan},            % 1741
\author[F]{B. Skiff},             % 4905
\author[M,m]{F. Marchis},         % 3749
\author[U]{K. E. Ergashev},       %19289
\author[k,x]{D.-H. Kim}           % 1741
\author[A]{A. Aznar},             % 3749
\author[r,R]{M. Serra-Ricart},    % 3749
\author[B]{R. Behrend},           % 3749
\author[C]{R. Roy},               % 3749
\author[D]{F. Manzini},           % 3749
\author[M]{I. E. Molotov}
%\author[D]{S. Mottola},
%\author[W]{V. R. Ayvazian},       % 14627, 16598
%\author[W]{O. I. Kvaratskhelia},  % 14627, 16598
%\author[W]{V. T. Zhuzhunadze},    % 14627
%\author[R]{Z. Donchev},           % 81337
%\author[R]{G. Borisov},           % 81337
%\author[R]{T. Bonev},             % 81337
%\author[w]{V. V. Rumyantsev},     % 57738

\address[a]{Astronomical Institute, Academy of Sciences of the Czech Republic, Fri\v{c}ova 1, CZ-25165 Ond\v{r}ejov, Czech Republic}
\address[b]{Institute of Astronomy, Charles University, Prague, V Hole\v{s}ovi\v{c}k\'ach 2, CZ-18000 Prague 8, Czech Republic}
\address[S]{Department of Aerospace Engineering Sciences, The University of Colorado at Boulder, Boulder, CO, USA}
\address[d]{Modra Observatory, Department of Astronomy, Physics of the Earth, and Meteorology, FMPI UK, Bratislava SK-84248, Slovakia}
\address[j]{Sugarloaf Mountain Observatory, South Deerfield, MA, USA}
\address[p]{Institute of Astronomy of Kharkiv National University, Sumska Str. 35, Kharkiv 61022, Ukraine}
\address[U]{Ulugh Beg Astronomical Institute, Astronomicheskaya Street\,33, 100052 Tashkent, Uzbekistan}
\address[P]{Physics and Astronomy Department, Appalachian State University, Boone, NC~28608, USA}
\address[F]{Lowell Observatory, 1400 W Mars Hill Road, Flagstaff, AZ 86001, USA}
\address[O]{Instituto de Astrof\'{\i}sica de Andaluc\'{\i}a (CSIC), Glorieta de la Astronomía S/N, 18008-Granada, Spain}
\address[n]{Astronomical Institute of the Slovak Academy of Sciences, SK-05960 Tatransk\'a Lomnica, Slovakia}
\address[W]{Kharadze Abastumani Astrophysical Observatory, Ilya State University, K.~Cholokashvili Avenue~3/5, Tbilisi~0162, Georgia}
\address[w]{Samtskhe-Javakheti State University, Rustaveli Street~113,  Akhaltsikhe~0080, Georgia}
\address[J]{Blue Mountains Observatory, Leura, N.S.W., Australia}
\address[K]{Department of Earth and Planetary Sciences, Weizmann Institute of Science, Rehovot 0076100, Israel}
\address[H]{Harvard-Smithsonian Center for Astrophysics, 60 Garden St., MS 51, Cambridge, MA 02138, USA}
\address[v]{Crimean Astrophysical Observatory of Russian Academy of Sciences, 298409 Nauchny, Ukraine}
\address[l]{Sonoita Research Observatory, 77 Paint Trail, Sonoita, AZ~85637, USA}
\address[L]{Deptartment of Space Studies, Southwest Research Institute, Boulder, CO~80302, USA}
%\address[N]{XXXX}
\address[V]{IMCCE-CNRS-Observatoire de Paris, 77~avenue Denfert Rochereau, 75014 Paris, France}
\address[s]{Department of Earth, Atmospheric and Planetary Sciences, Massachusetts Institute of Technology, Cambridge, MA~02139, USA}
\address[M]{SETI Institute, Carl Sagan Center, 189 Bernardo Avenue, Suite 200, Mountain View, CA~94127, USA}
\address[m]{Observatoire de Paris, LESIA, 5 Place Jules Janssen, 92190 Meudon, France}
\address[k]{Korea Astronomy and Space Science Institute, 776, Daedeok-daero, Yuseong-gu, Daejeon, Korea}
\address[x]{Chungbuk National University, 1, Chungdae-ro, Seowon-gu, Cheongju-si, Chungcheongbuk-do, Korea}
\address[A]{Observatorio Isaac Aznar, Grupo de Observatorios APT C/La Plana, 44, 13, E-46530 Pucol, Valencia, Spain}
\address[r]{Instituto de Astrof\'isica de Canarias, C/V\'ia L\'actea, s/n, 38205 La Laguna, Tenerife, Spain}
\address[R]{Departamento de Astrof\'isica, Universidad de La Laguna, 38205 La Laguna, Tenerife, Spain}
\address[B]{Observatoire de Gen\`eve, CH-1290 Sauverny, Switzerland}
\address[C]{293 Chemin de St Guillaume, 84570 Blauvac, France}
\address[D]{Stazione Astronomica di Sozzago, I-28060 Sozzago, Italy}
\address[M]{Keldysh Institute of Applied Mathematics, RAS, Miusskaya sq.\,4, Moscow 125047, Russia}
%\address[D]{German Aerospace Center (DLR), Institute of Planetary Research, Rutherfordstr.\,2, 12489 Berlin, Germany}
%\address[R]{Institute of Astronomy and NAO, Bulgarian Academy of Sciences, 72, Tsarigradsko Chaussee Blvd., 1784 Sofia, Bulgaria}
%\address[w]{Crimean Astrophysical Observatory, RAS, 298409 Nauchny, Crimea}

%\newpage
\vspace{2cm}
%\small{Submitted to Icarus}

%Pages: 44 of text, 73 including tables and figures \\
%Tables: 6    \\
%Figures: 26  \\

\newpage
%\bigskip
Proposed running head: Asteroid pairs complex

\vspace{1cm}
Editorial correspondence to: \\
Dr. Petr Pravec              \\
Astronomical Institute AS CR \\
Fri\v{c}ova 1                \\
Ond\v{r}ejov                 \\
CZ-25165                     \\
Czech Republic               \\
Phone: 00420-323-620352      \\
Fax: 00420-323-620263        \\
E-mail address: petr.pravec@asu.cas.cz \\

\newpage
\begin{abstract}
We studied a sample of 93 asteroid pairs, i.e., pairs of genetically related asteroids that are on highly similar heliocentric orbits.
We estimated times elapsed since separation of pair members (i.e., pair age) that are between $7 \times 10^3$~yr and a few $10^6$~yr.
With photometric observations, we derived the rotation periods $P_1$ for all the primaries (i.e., the larger members of asteroid pairs)
and a sample of secondaries (the smaller pair members).  We derived the absolute magnitude differences of the studied asteroid pairs that provide
their mass ratios $q$.  For a part of the studied pairs, we refined their WISE geometric albedos and collected or estimated their taxonomic
classifications.  For 17 asteroid pairs, we also determined their pole positions.  In two pairs where we obtained the spin poles for both pair components,
we saw the same sense of rotation for both components and constrained the angles between their original spin vectors at the time of their separation.
We found that the primaries of 13 asteroid pairs in our sample are actually binary or triple systems, i.e., they have one or two bound, orbiting secondaries (satellites).
As a by-product, we found also 3 new young asteroid clusters (each of them consisting of three known asteroids on highly similar heliocentric orbits).
We compared the obtained asteroid pair data with theoretical predictions and discussed their implications.
We found that 86 of the 93 studied asteroid pairs follow the trend of primary rotation period vs mass ratio that was found by
Pravec et~al.~(2010).  Of the 7 outliers, 3 appear insignificant (may be due to our uncertain or incomplete knowledge of the three pairs),
but 4 are high mass ratio pairs that were unpredicted by the theory of asteroid pair formation by rotational fission.  We discuss a (remotely) possible way that they could
be created by rotational fission of flattened parent bodies followed by re-shaping of the formed components.  The 13 asteroid pairs with binary primaries are particularly
interesting systems that place important constraints on formation and evolution of asteroid pairs.   We present two hypotheses for their formation:
The asteroid pairs having both bound and unbound secondaries could be ``failed asteroid clusters'', or they could be formed by a cascade primary spin
fission process.  Further studies are needed to reveal which of these two hypotheses for formation of the paired binary systems is real.
\end{abstract}


\end{frontmatter}

\newpage
\section{Introduction}
\label{IntrSect}

In the main belt of asteroids, there exist pairs of asteroids that are on highly similar heliocentric orbits.  They were discovered by
Vokrouhlick\'y and Nesvorn\'y (2008) who showed that the asteroid pairs cannot be random coincidences of unrelated asteroids from the local asteroid population, but most of them must be genetically related pairs.
They proposed 60 such asteroid pairs.  Pravec and Vokrouhlick\'y (2009) developed a statistical asteroid pair identification procedure and they
found 73 statistically significant pairs (most of them have been confirmed by backward orbit integrations; see Section~\ref{PairIdentsect}).
Pravec et~al.~(2010) studied a sample of 32 asteroid pairs and found a correlation between the rotation frequencies of asteroid pair primaries\footnote{We call `primary' and `secondary', respectively,
the larger and the smaller member of an asteroid pair.} and the asteroid pair mass ratios.  Following the theory by Scheeres (2007), they found the correlation to be an evidence
for formation of asteroid pairs by rotational fission.

A number of dynamical and physical studies of asteroid pairs were published since then.  Vokrouhlick\'y and Nesvorn\'y~(2009) and Vokrouhlick\'y et~al.~(2011, 2017) studied the young asteroid pair of (6070) Rheinland
and (54827) Kurpfalz.\footnote{Hereafter we for short designate asteroid pairs with the primary and secondary asteroid numbers (or principal designations), e.g., 6070--54827 for the pair
of (6070) Rheinland and (54827) Kurpfalz.}  They determined its age of $16.34 \pm 0.04$~kyr and found that the spin vectors of the two asteroids are both retrograde, but they were not colinear but tilted by $38^\circ \pm 12^\circ$
at the time of separation. \v{Z}i\v{z}ka et~al.~(2016) studied asteroid pair 87887--415992 and found it to have a probable age of $7.4 \pm 0.3$~kyr, that is probably the youngest one of known asteroid pairs.
Vokrouhlick\'y~(2009) found that the triple asteroid (3749) Balam is paired with asteroid 2009~BR60, which was the first known case of such complex system with both bound and unbound secondaries.
Polishook~(2014a) found that the members of pair 2110--44612 have the same sense of rotation (retrograde), as expected for an asteroid pair formed by rotational fission.
Pravec et~al.~(2018) studied 13 young asteroid clusters (i.e., groups of three or more genetically related asteroids on highly similar heliocentric orbits) and found that
the properties of 11 of them are consistent with the rotational fission formation process, linking them to asteroid pairs.

Spectral or color observations of paired asteroids\footnote{We use the term `paired asteroid' as a synonym for `member of an asteroid pair'.} were done by
Moskovitz~(2012),
Duddy et~al.~(2012, 2013),
Wolters et~al.~(2014) and
Polishook et~al.~(2014a).
They found that the pairs belong to a variety of taxonomic classes, indicating that the asteroid structure and not their composition, is the main property that enabled their fission.
They also found that in most asteroid pairs, the two components have the same or similar spectra and colors, consistent with same composition of both components as expected for the secondary formed by fission from the primary.
Some silicate pairs, belonging to the S-complex, present subtle spectral/color differences between the primary and secondary, which they suggested could be due to different degrees of ``space weathering'' of the surfaces of the pair members.
As they found no large-scale spectral non-uniformity on the surfaces of young asteroid pairs, Polishook et~al.~(2014b) suggested that the rotational fission was followed by a spread of dust that covered the primary body uniformly.

Being motivated by the progress in our knowledge and understanding of asteroid pairs as briefly outlined above, we underwent a thorough photometric study of a sample of nearly 100 asteroid pairs.  This study has not only enlarged the sample of studied asteroid pairs
by nearly a factor of 3, but it also went to smaller asteroid sizes than before, extending our knowledge of asteroid pair properties to sizes about 1~km where we start seeing new features in the asteroid pair population.
And we also performed observations within this survey thoroughly so that to be able to resolve also potential binary nature of studied asteroids.  We outline our results in this paper.

\bigskip
\section{Pair identification and age estimation}
\label{PairIdentsect}

We identified candidate asteroid pairs by analyzing the distribution of asteroid distances in the five-dimensional space
of mean orbital elements $(a, e, i, \varpi, \Omega)$ using the method of Pravec and Vokrouhlick\'y~(2009).\footnote{Pravec and Vokrouhlick\'y~(2009) originally
used osculating orbital elements, but later we amended the method with the use of mean elements, following suggestion by D.~Nesvorn\'y (2010, personal communication;
see also Ro{\. z}ek et~al.~2011).  See also an application of the method for asteroid clusters in Pravec et~al.~(2018).
We took the mean elements from the AstDyS catalog webpage (Kne\v{z}evi\'c et~al. 2002, Kne\v{z}evi\'c and Milani 2003).}
The distance ($d_{\rm mean}$) between two asteroid orbits was computed with a positive-definite quadratic form
\begin{equation}
 \left ( \frac{d_{\rm mean}}{n a} \right )^2 = k_a \left(\frac{\delta a}{a}\right)^2
  + k_e (\delta e)^2 + k_i (\delta \sin i)^2 + k_\Omega (\delta \Omega)^2
  + k_\varpi (\delta \varpi)^2 \; ,
 \label{dEq}
\end{equation}
where $n$ and $a$ are the mean motion and semimajor axis of either of the
two asteroids and $(\delta a, \delta e, \delta \sin i, \delta \varpi,
\delta \Omega)$ is the separation vector of their mean orbital elements.
Following Zappal\`a et~al.~(1990) and Pravec and Vokrouhlick\'y~(2009), we used $k_a = 5/4$, $k_e = k_i = 2$ and
$k_\varpi = k_\Omega = 10^{-4}$.  The distance $d_{\rm mean}$ between two asteroid orbits is an approximate gauge for the relative velocity of the asteroids at close encounter (see Ro\.zek et~al. 2011 for explicit tests and a comparison with other metric
functions used in meteoritics).
For most asteroid pairs, it is in the range from a few $10^{-1}$ to a~few~10~m/s.

To confirm the pair membership suggested by the asteroid distances in the space of mean orbital elements,
we integrated a set of geometric clones (1000 clones for each asteroid) with the Yarkovsky effect acting on each clone differently.
The Yarkovsky effect was represented using a fake transverse acceleration acting on the clone with a magnitude
providing secular change in semimajor axis $\dot{a}_{\rm{Yark}}$ (see Farnocchia et al., 2013).
It was chosen from the range $\langle -\dot{a}_{\rm{max}},\dot{a}_{\rm{max}} \rangle$, where $\dot{a}_{\rm{max}}$ was estimated from the asteroid size (see Vokrouhlick\'{y}, 1999).
These limit values for the semimajor axis drift rate correspond to bodies with zero obliquity, for which the diurnal variant of the effect is optimized,
and the diurnal thermal parameter equal to the square root of two, for which the magnitude of the Yarkovsky effect is maximal (see, e.g., Vokrouhlick\'{y}, 1999). 
The goal of our backward orbital integrations was to find low relative-velocity close encounters between the clones of the members of a tested pair.
We chose following limits on the physical distance and relative velocity between the clones $r_{\rm{rel}} \leq 5-10 R_{\rm{Hill}}$ and $v_{\rm{rel}} \leq 2-4 v_{\rm{esc}}$, where $R_{\rm{Hill}}$ and $v_{\rm{esc}}$ are the radius of the Hill sphere and the surface escape velocity, respectively, of the primary body.
The narrower limits were used for better converging clones (e.g., younger ones, or those in non-chaotic zones of the main asteroid belt), while the loosened limits were typically used for pairs with the orbits affected by
some orbital chaoticity.
The radius of the Hill sphere was estimated as $R_{\rm{Hill}} \sim a D_1 \frac{1}{2} \left( \frac{4\pi}{9} \frac{G\rho_1}{\mu} \right)^{1/3} $,
where $a$ is the semi-major axis of the primary's heliocentric orbit, $D_1$ is its diameter, $\rho_1$ is its bulk density (assumed 2 g/cm$^3$),
$G$ is the gravitational constant and $\mu$ is the gravitational parameter of the Sun.
The escape velocity was estimated as $v_{\rm{esc}} \sim D_1 \frac{1}{2} \left( \frac{8\pi}{3} G \rho_1 \right)^{1/2}$ (both formulas from Pravec et~al.,~2010, Supplementary Information).

For numerical integration we used the fast and accurate implementation of a Wisdom-Holman symplectic integrator WHFast (Rein and Tamayo, 2015)
%\footnote{Rein and Tamayo (2015), ``WHFAST: a fast and unbiased implementation of a symplectic Wisdom-Holman integrator for long-term gravitational simulations''.}
from the REBOUND package (Rein and Liu, 2012).
%\footnote{Rein and Liu (2012), ``REBOUND: An open-source multi-purpose N-body code for collisional dynamics'', http://github.com/hannorein/rebound.}
We implemented the Yarkovsky effect into the code following Nesvorn\'y and Vokrouhlick\'y~(2006). We included gravitational attraction of the Sun, the 8 major planets, two dwarf planets Pluto and Ceres and two large asteroids Vesta and Pallas. We chose the time-step to be six hours, this allows us to detect close and fast encounters with massive bodies in our simulation. The geometric clones were created in the six-dimensional space of equinoctial elements $\textbf{E}$ using the probability distribution $p\left(\textbf{E}\right) \propto \rm{exp} \left(-\frac{1}{2}\Delta\textbf{E}\cdot\Sigma\cdot\Delta\textbf{E}\right)$, where $\Delta \textbf{E} = \textbf{E} - \textbf{E}^*$ is the difference with respect to the best-fit orbital values $\textbf{E}^*$ and $\Sigma$ is the normal matrix of the orbital solution downloaded from AstDyS website at the initial epoch MJD 58000 (Milani and Groncchi, 2010).
Each geometric clone was given a random value of $\dot{a}_{\rm{Yark}}$ from the range $\langle -\dot{a}_{\rm{max}}, \dot{a}_{\rm{max}} \rangle$.

For each pair, we estimated a time since separation of the secondary from the primary (designated $T_{\rm{sep}}$) from the distribution of the calculated past times of close and slow encounters between their clones.
With the output frequency of 10 days we checked all the clone combinations ($1000 \times 1000$) between the primary and the secondary and
for each we found their minimum distance $r_{\rm{rel}}$ and relative velocity $v_{\rm{rel}}$ at their encounter.
%If an encounter of two clones lasted longer than is our output period (consecutive detection of encounter of the same two clones), we chose only the event with the smallest $v_{\rm{rel}}$.
Encounters satisfying the chosen distance and velocity limits (see above) were counted and their time histogram was used for estimating $T_{\rm{sep}}$ for given tested pair.  The histograms for individual asteroid pairs
are shown in Section~\ref{IndivPairsSect} or in the Electronic Supplementary Information.
The bin widths in the histograms are 10 or 20~kyr for the past time axis spanning to $< 1500$ or $\ge 1500$~kyr, respectively.
As the distributions of $T_{\rm{sep}}$ are non-Gaussian and often strongly asymmetric, we used the median (i.e., the 50th~percentile) value of the distribution as a nominal estimate
for the time of separation of the members of given pair (i.e., the pair age).
For an uncertainty (error bar) of the separation time, we adopted the 5th and the 95th~percentile of the distribution for the lower
and upper limit on the separation time, respectively.

\bigskip
\section{Asteroid pair properties}
\label{IndivPairsSect}

We collected available lightcurve photometric data and run new photometric observations for all the primaries and some secondaries of our sample of 93 asteroid pairs.
We used our standard photometric observational and reduction techniques that we describe in Electronic Supplementary Information.
The obtained data were analysed using the methods described in Pravec et~al.~(2006) that provided rotation periods of the studied paired asteroids and revealed the binary nature of several of them.
For paired asteroids with sufficient data, we derived their spin vectors and constructed their convex shape models
using the technique of Kaasalainen et~al.~(2001), with confidence ranges estimated as in Vokrouhlick\'y et~al.~(2011).
%In most cases, we obtained two possible solutions for the pole direction with similar values of ecliptic latitude $B$ and the values of ecliptic longitude $L$ about 180$^\circ$ apart.
For most primaries and some secondaries, we also derived their accurate absolute magnitudes, from which we calculated their $\Delta H \equiv (H_2 - H_1)$ values\footnote{We designate quantities belonging to the primary and secondary
with the indices `1' and `2', respectively.}
and propagated their uncertainties for pairs where we had the accurate absolute magnitudes for both members of a given pair.
However, the absolute magnitudes for some primaries and many secondaries for which we did not obtain accurate $H$ values
were taken from the MPC catalog\footnote{http://www.minorplanetcenter.org/iau/MPCORB.html.}.
The uncertainties of $\Delta H$ in such cases were assumed to be $\pm 0.3$ (see Pravec et~al., 2012a, for analysis of the uncertainties of absolute magnitudes reported in asteroid orbit catalogs).
The asteroid pair mass ratio $q$ is estimated from its $\Delta H$ value with
\begin{equation}
q = 10^{-0.6 \Delta H}.
\label{q_DeltaH}
\end{equation}
Where available, we took the diameters and geometric albedos of studied asteroids from their WISE observations (Masiero et~al.~2011) and
refined them using our accurate $H$ values using the method described in Pravec et~al.~(2012b).  For three asteroids, we derived their diameters and geometric albedos from thermophysical modeling (Appendix~\ref{AppendTPM}).
%For primaries where no thermal WISE observations were obtained, we estimated their diameters assuming the mean geometric albedos for their probable spectral types.
For a sample of paired asteroids, we also measured their colors with Sloan Digital Sky Survey (SDSS) filters or took their SDSS color measurements
from the SDSS Moving Object Catalog (Ivezi\'c et~al.~2001), and used them to estimate their taxonomic classifications.
SDSS photometric data for individual objects were obtained in sequences employing the g'r'i'z' filters.
The mean color indices and the mean $r$ for individual measured asteroids are given in Suppl.~Table~2.
Taxonomic classification was achieved by down-sampling the resolution of the spectral envelopes in the Bus taxonomic system (Bus and Binzel, 2002) to the SDSS filter set
and then minimizing the RMS residual between the photometric data and the taxonomic envelopes.
In some cases the data were equivalently well fit with more than one taxonomic type, in which case we present two possible assignments.
For several paired asteroids, we also collected published taxonomic classification from spectral data.

The data for the studied asteroid pairs are presented in Tables~\ref{AstPairsDatatable} to \ref{AstPairSatstable}.
In Table~\ref{AstPairsDatatable}, for each studied asteroid pair, we give the distance of its members in the space of mean orbital elements ($d_{\rm mean}$), its age estimated from the backward orbital integrations of the pair members,
the primary and secondary absolute magnitudes ($H_1, H_2$), its $\Delta H$, the primary diameter ($D_1$) derived from the WISE observations (if given to a tenth of km) or estimated assuming the mean geometric albedo of its derived or
estimated taxonomic type (if rounded to 1~km), the primary and secondary rotation periods and mean observed lightcurve amplitudes ($P_1, A_1, P_2, A_2$), and a number of observed satellites (bound secondaries) of the primary (Sat.$_1$).
In the last column, we note subsections and/or tables where more data and information are given for a given asteroid pair.  For most asteroid pairs, we also give further information in Electronic Supplementary Information.

In Table~\ref{AstPairsTaxColorstable}, we give the geometric albedos ($p_{V,1}$) of the pair primaries that we refined from the WISE data using our accurate absolute magnitudes or from our thermophysical modeling presented
in Appendix~\ref{AppendTPM}.
In columns Taxon.$_1$ and Taxon.$_2$, we report the primary and secondary taxonomic classifications.  The 6th and 7th columns are their color indices in the Johnson-Cousins VR photometric system that we obtained as by-product
of our lightcurve observations or derived from their Sloan colors using the formula $(V-R) = 1.09 (r-i) + 0.22$ (Jester et al., 2005).
In the last column, we mention where more information on the reported quantities is given.

In Table~\ref{AstPairsPolestable}, we report the ecliptic coordinates (in equinox
J2000) of the spin poles of several paired asteroids for which we derived their
spin vectors or the orbit poles of four binary systems among asteroid pair primaries.
The sidereal periods of the paired asteroids with derived spin vectors
are given in the $P_1$ or $P_2$ columns
in Table~\ref{AstPairsDatatable}.  Their convex shape models or information about the binary systems are given in
subsections mentioned in the last column.

In Table~\ref{AstPairSatstable}, we report the best estimates (nominal values) for several parameters of the binary systems among asteroid pairs.
Uncertainties of the values are given in the text or they are
available in the binary asteroid parameters tables at
http://www.asu.cas.cz/$\sim$asteroid/binastdata.htm (update of the original tables from Pravec and Harris, 2007).
We give the diameter of the primary (main body) of the binary system $D_{1,p}$, the mean diameter ratio between the unbound secondary and the primary $D_2/D_{1,p}$,
the mean diameter ratio between the bound secondary (satellite) and the primary $D_{1,s}/D_{1,p}$,
the ratio of the major semiaxis of the binary system to the primary's mean diameter $a_{\rm orb}/D_{1,p}$,
the constrained or assumed eccentricity $e$ (3-$\sigma$ ranges or upper limits on the eccentricity are given, or zero eccentricity is assumed in cases where the available data is consistent with circular orbit of the satellite but
we did not obtain a full model of the secondary's orbit yet),
the primary and secondary rotation periods $P_{1,p}$ and $P_{1,s}$,
the orbital period $P_{\rm orb}$,
the normalized total angular momentum $\alpha_L$ (see Pravec and Harris, 2007, for its definition),
the observed lightcurve amplitudes of the primary and the secondary $A_{1,p}$ and $A_{1,s}$,
measured at mean solar phase SolPh,
and the estimated primary and secondary equatorial axes ratios $a_{1,p}/b_{1,p}$ and $a_{1,s}/b_{1,s}$.
In the last column, we note the subsections on the paired binary systems where more information is given.

In following subsections, we present the results for selected individual asteroid pairs that are particularly interesting in certain points.

\newpage
\subsection{(1741) Giclas and (258640) 2002 ER36}
\label{1741sect}

\begin{figure}
%\vspace{1cm}
\includegraphics[width=\textwidth]{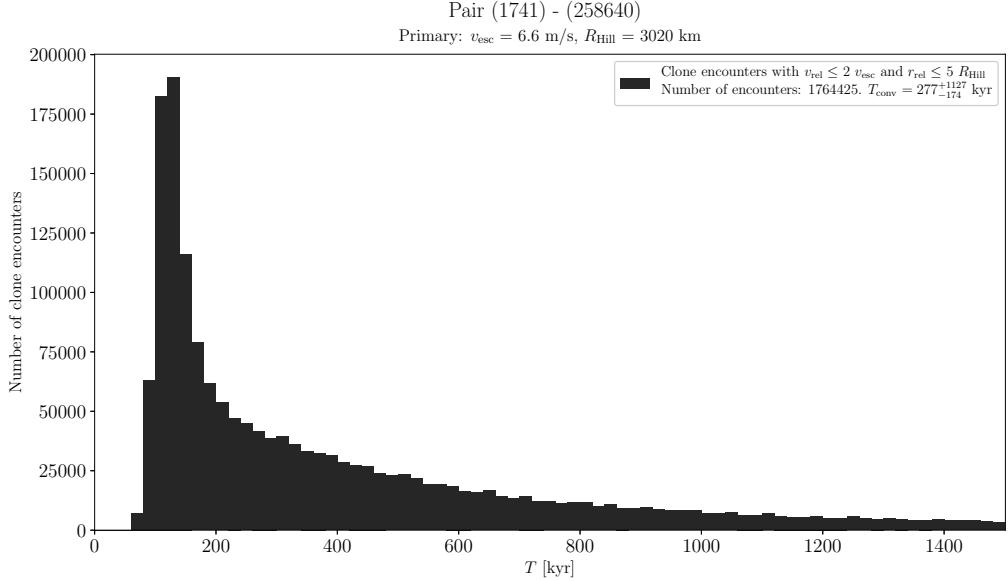}
\caption{\rm Distribution of past times of close and slow primary--secondary clone encounters for the asteroid pair 1741--258640.
}
\label{1741enchist}
\vspace{1cm}
\end{figure}

\begin{figure}
%\vspace{1cm}
\includegraphics[width=\textwidth]{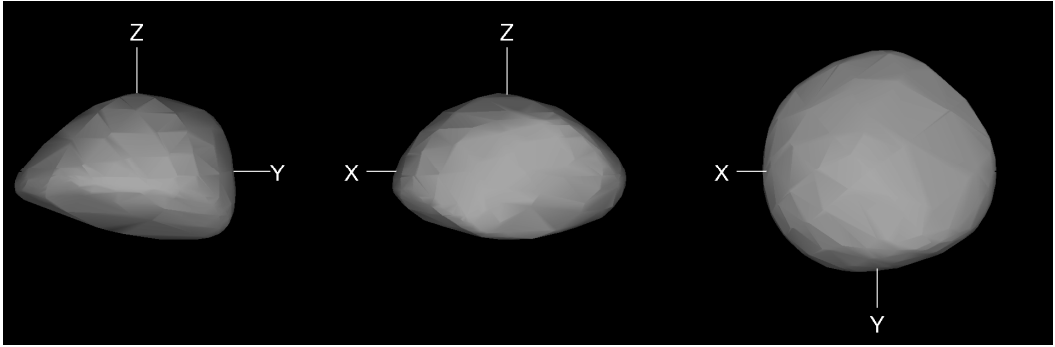}
\caption{\rm Convex shape model of (1741) Giclas for the pole solution $(L_1,B_1) = (105^\circ, +30^\circ)$.
In this and other figures below, the model is shown from two equatorial views 90$^\circ$ apart and pole on. The $Z$-axis is the axis of rotation.
}
\label{1741model_1}
\vspace{1cm}
\end{figure}

\begin{figure}
%\vspace{1cm}
\includegraphics[width=\textwidth]{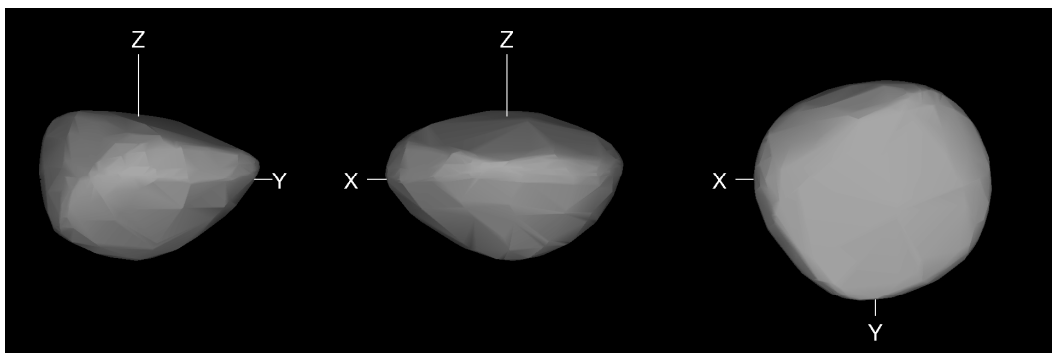}
\caption{\rm Convex shape model of (1741) Giclas for the pole solution $(L_1,B_1) = (288^\circ, +24^\circ)$.
}
\label{1741model_2}
\vspace{1cm}
\end{figure}

This is a secure asteroid pair with an estimated age about 200~kyr (Fig.~\ref{1741enchist}).
From spectral observations of the primary, Polishook et~al.~(2014a) derived that it is S/Sq type.
We obtained its color index $(V-R)_1 = 0.466 \pm 0.010$ as the weighted mean of our measured $(V-R)_1 = 0.471 \pm 0.010$ and the value $0.456 \pm 0.015$ by Slivan et~al.~(2008).
%Using our determined mean absolute magnitude, we refined the WISE data by Masiero et~al.~(2011) and obtained $D_1 = 12.3 \pm 1.3$~km and $p_{V,1} = 0.26 \pm 0.06$.
We derived its prograde spin pole (with two mirror solutions in longitude, see Table~\ref{AstPairsPolestable}).
The best-fit convex shape models for the two pole solutions are shown in Figs.~\ref{1741model_1} and \ref{1741model_2}.
From our thermophysical modeling, we derived its effective volume-equivalent diameter $D_1$ and geometric albedo $p_{V,1}$ (Appendix~\ref{AppendTPM}).
On the nights 2015-06-05.6, -06.6, and 2018-01-25.2, we detected brightness attenuations with depths 0.06--0.07~mag that suggest a presence of satellite (bound secondary).  This suspicion of binary nature of (1741) Giclas
needs to be confirmed with thorough observations in the future.

\clearpage

\subsection{(2110) Moore-Sitterly and (44612) 1999 RP27}
\label{2110sect}

\begin{figure}
%\vspace{1cm}
\includegraphics[width=\textwidth]{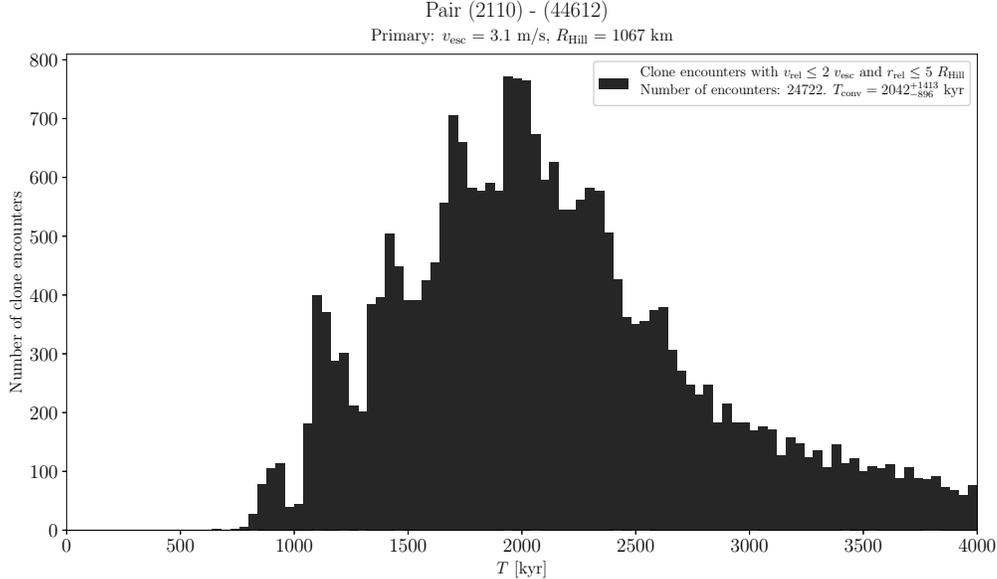}
\caption{\rm Distribution of past times of close and slow primary--secondary clone encounters for the asteroid pair 2110--44612.
}
\label{2110enchist}
\vspace{1cm}
\end{figure}

\begin{figure}
%\vspace{1cm}
\includegraphics[width=\textwidth]{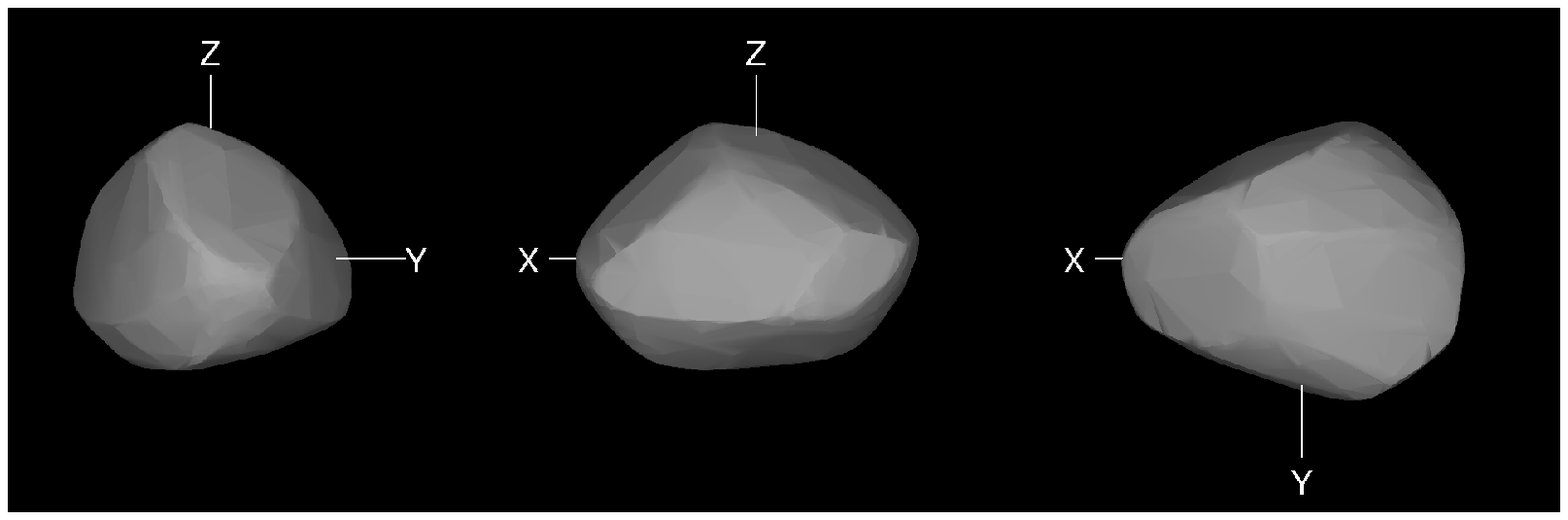}
\caption{\rm Convex shape model of (2110) Moore-Sitterly for the pole solution $(L_1,B_1) = (91^\circ, -75^\circ)$.
}
\label{2110model_1}
\vspace{0.5cm}
\end{figure}

\begin{figure}
%\vspace{1cm}
\includegraphics[width=\textwidth]{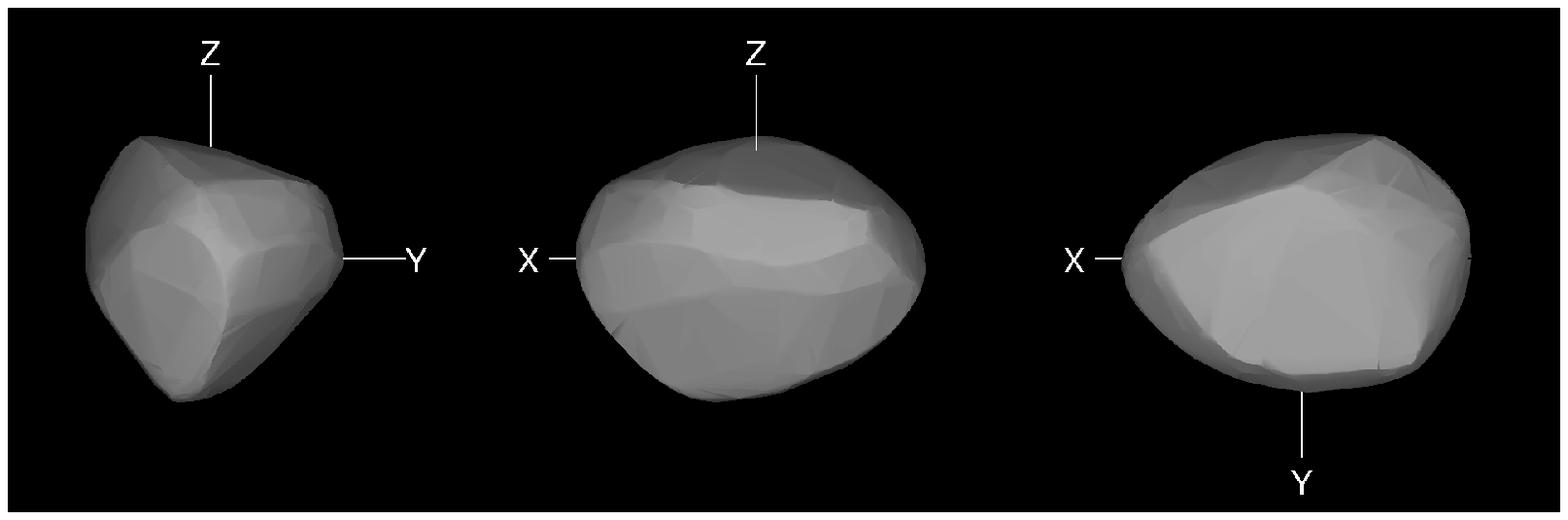}
\caption{\rm Convex shape model of (2110) Moore-Sitterly for the pole solution $(L_1,B_1) = (270^\circ, -77^\circ)$.
}
\label{2110model_2}
\vspace{0.5cm}
\end{figure}

\begin{figure}
%\vspace{1cm}
\includegraphics[width=\textwidth]{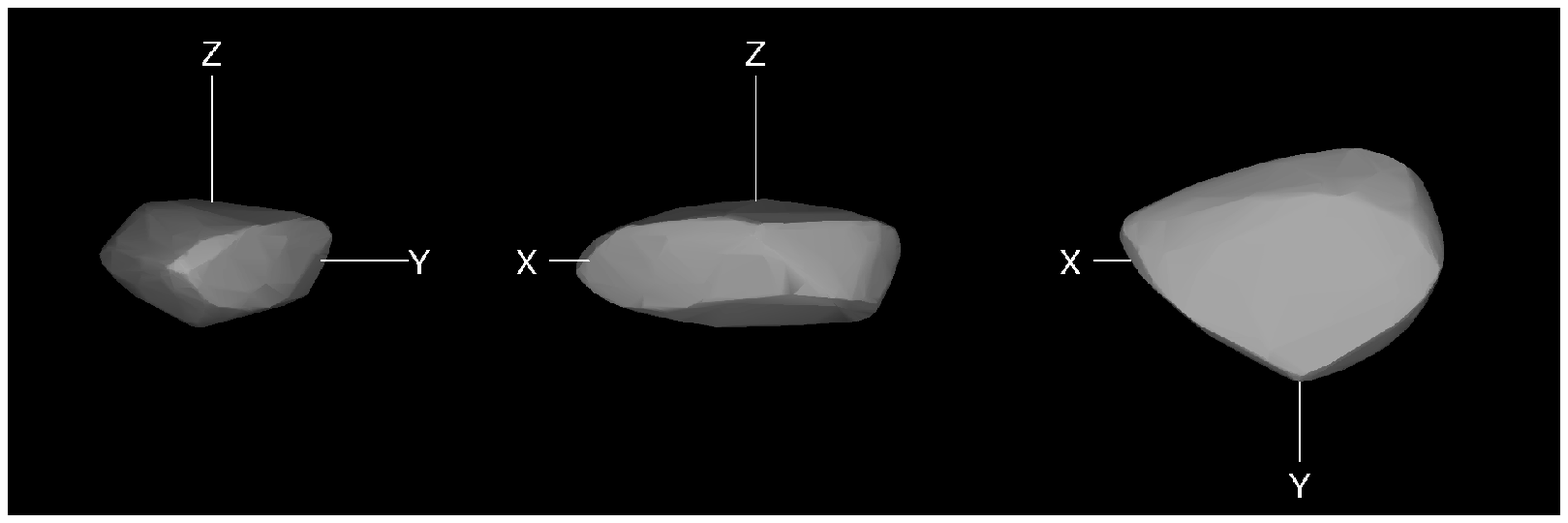}
\caption{\rm Convex shape model of (44612) 1999 RP27 for the pole solution $(L_2,B_2) = (8^\circ, -73^\circ)$.
}
\label{44612model_1}
\vspace{0.5cm}
\end{figure}

\begin{figure}
%\vspace{1cm}
\includegraphics[width=\textwidth]{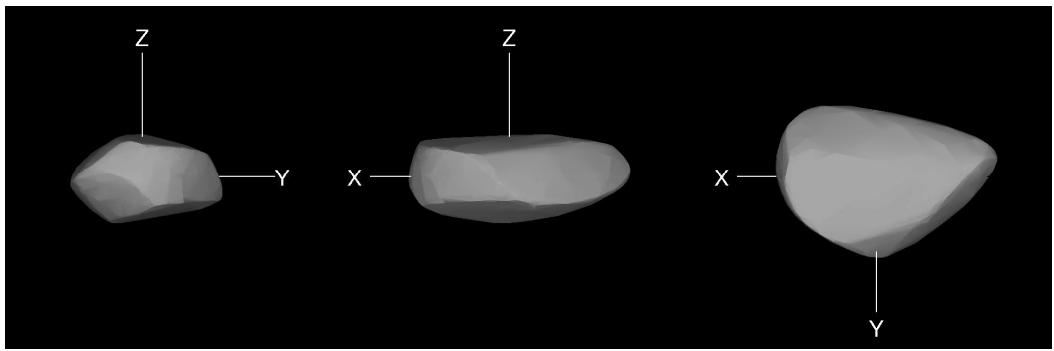}
\caption{\rm Convex shape model of (44612) 1999 RP27 for the pole solution $(L_2,B_2) = (193^\circ, -69^\circ)$.
}
\label{44612model_2}
\vspace{0.5cm}
\end{figure}

This is a secure asteroid pair with an estimated age about 2~Myr (Fig.~\ref{2110enchist}).
From their spectral observations, Polishook et~al.~(2014a) derived S and Sq/Q types for the primary and secondary, respectively.
Their color indices are $(V-R)_1 = 0.45 \pm 0.02$ (Moskovitz 2012) and $(V-R)_2 = 0.444 \pm 0.010$ (our measurement).
Using our derived mean absolute magnitude, we refined the WISE data (Masiero et~al.~2011) for the secondary (44612) and obtained
%$D_1 = 5.8 \pm 0.7$~km and $p_{V,1} = 0.20 \pm 0.05$, and
$D_2 = 2.0 \pm 0.4$~km and $p_{V,2} = 0.22 \pm 0.08$.
For both asteroids, we derived their retrograde spin vectors (with two mirror solutions in longitude, see Table~\ref{AstPairsPolestable}), in agreement with their earlier models by Polishook~(2014a).
The best-fit convex shape models for the two pole solutions for both the primary and the secondary are shown in Figs.~\ref{2110model_1} to \ref{44612model_2}.
Though the best-fit pole positions for the two asteroids are nearly 90\dg distant in longitude, their 3-$\sigma$ pole uncertainty areas overlap, see their plots in the Electronic Supplementary Information.
Finally, from our thermophysical modeling, we derived the primary's volume-equivalent diameter $D_1$ and geometric albedo $p_{V,1}$ (Appendix~\ref{AppendTPM}).

\clearpage

\clearpage

\subsection{(3749) Balam and (312497) 2009 BR60}
\label{3749sect}

\begin{figure}
%\vspace{1cm}
\includegraphics[width=\textwidth]{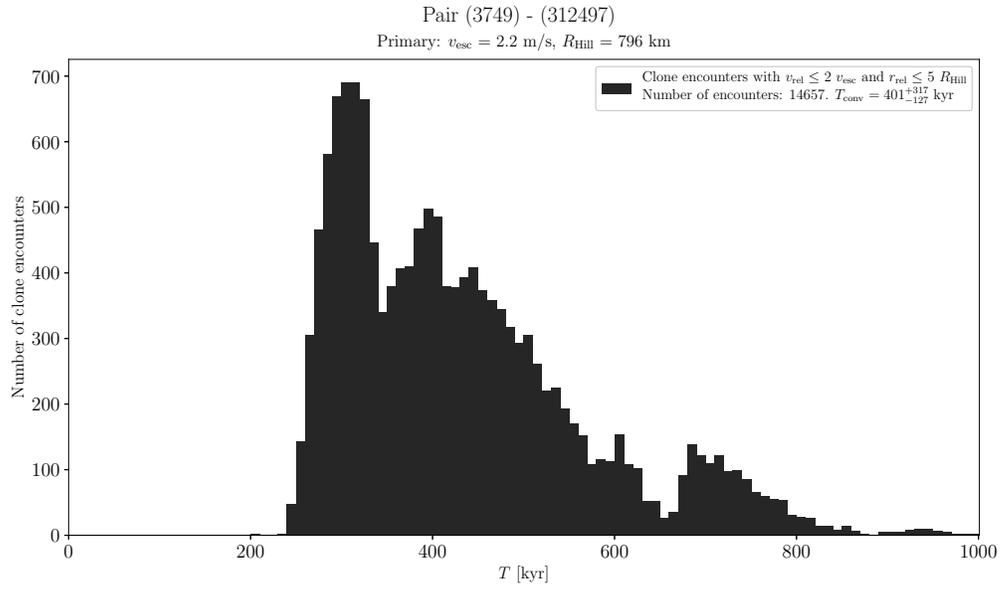}
\caption{\rm Distribution of past times of close and slow primary--secondary clone encounters for the asteroid pair 3749--312497.
}
\label{3749enchist}
\vspace{0.5cm}
\end{figure}

\begin{figure}
%\vspace{1cm}
\includegraphics[width=\textwidth]{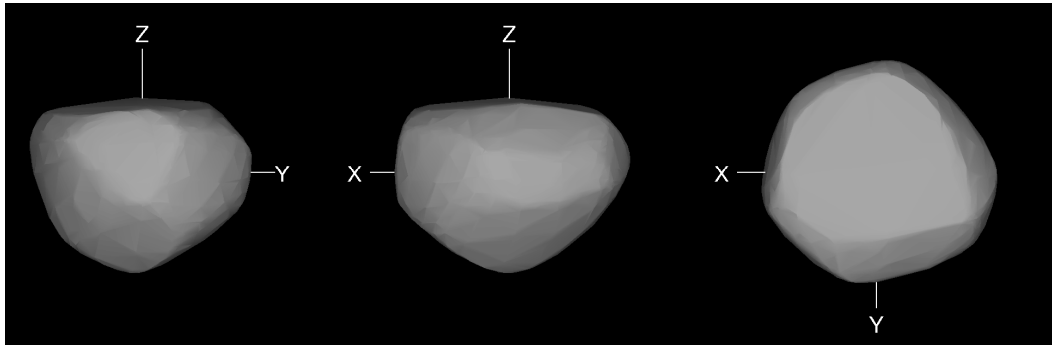}
\caption{\rm Convex shape model of (3749) Balam for the pole solution $(L_1,B_1) = (49^\circ, -69^\circ)$.
}
\label{3749model_1}
\vspace{0.5cm}
\end{figure}

\begin{figure}
%\vspace{1cm}
\includegraphics[width=\textwidth]{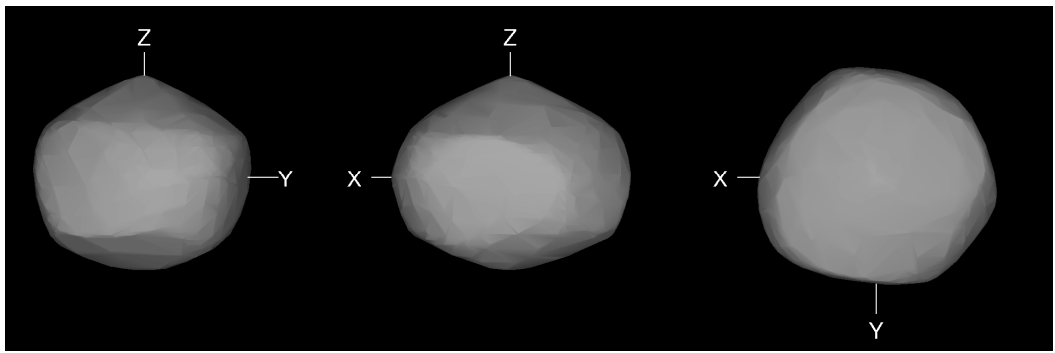}
\caption{\rm Convex shape model of (3749) Balam for the pole solution $(L_1,B_1) = (250^\circ, -71^\circ)$.
}
\label{3749model_2}
\vspace{0.5cm}
\end{figure}

This is a secure asteroid pair with an estimated age about 400~kyr (Fig.~\ref{3749enchist}).
The primary (3749) Balam has two satellites.  The outer satellite was discovered by Merline et~al.~(2002) and the inner satellite was discovered by Marchis et~al.~(2008).
Vokrouhlick\'y~(2009) identified the unbound secondary (312497).
For the outer satellite of the primary, Vachier et~al.~(2012) derived a set of possible orbital solutions with the semi-major axis ranging from 189 to 298~km,
orbital period from 1306 to 3899~h, and eccentricity from 0.35 to 0.77.
From the magnitude difference between their images of the outer satellite and the primary plus inner satellite (which was not resolved in their images) reported by Merline et~al.~(2002) and applying a correction for
the presence of the unresolved inner satellite, we estimate the size ratio $D_{1,s}/D_{1,p} = 0.24$ for the outer satellite.
From the observed total mutual events between the inner satellite and the primary, we derived its size ratio $D_{1,s}/D_{1,p} = 0.46 \pm 0.05$.
Its orbital period is $33.38 \pm 0.02$~h (Marchis et~al.~2008) and it appears synchronous, i.e., its rotational period $P_{1,s}$ appears equal to the orbital period.
Its orbit is slightly eccentric, $e = 0.03$--0.08 (3-$\sigma$ range; Scheirich et al., in preparation).
Refining the WISE thermal measurements (Masiero et~al.~2011) with our accurately determined mean absolute magnitude of the whole system of Balam $H_1 = 13.57 \pm 0.07$,
we obtained the effective diameter $D_1 = 4.7 \pm 0.5$~km and geometric albedo $p_{V,1} = 0.30 \pm 0.07$.
Correcting it for the satellites presence, we derived the primary's mean diameter $D_{1,p} = 4.1 \pm 0.5$~km.
Polishook et~al.~(2014a) derived from their spectral observations that it is an Sq type.
We also derived its retrograde spin pole (with two mirror solutions in longitude, see Table~\ref{AstPairsPolestable}), which confirms its earlier model by Polishook~(2014a).
The best-fit convex shape models for the two pole solutions are shown in Figs.~\ref{3749model_1} and \ref{3749model_2}.

\clearpage

\bigskip
\subsection{(4765) Wasserburg and (350716) 2001 XO105}
\label{4765sect}

\begin{figure}
%\vspace{1cm}
\includegraphics[width=\textwidth]{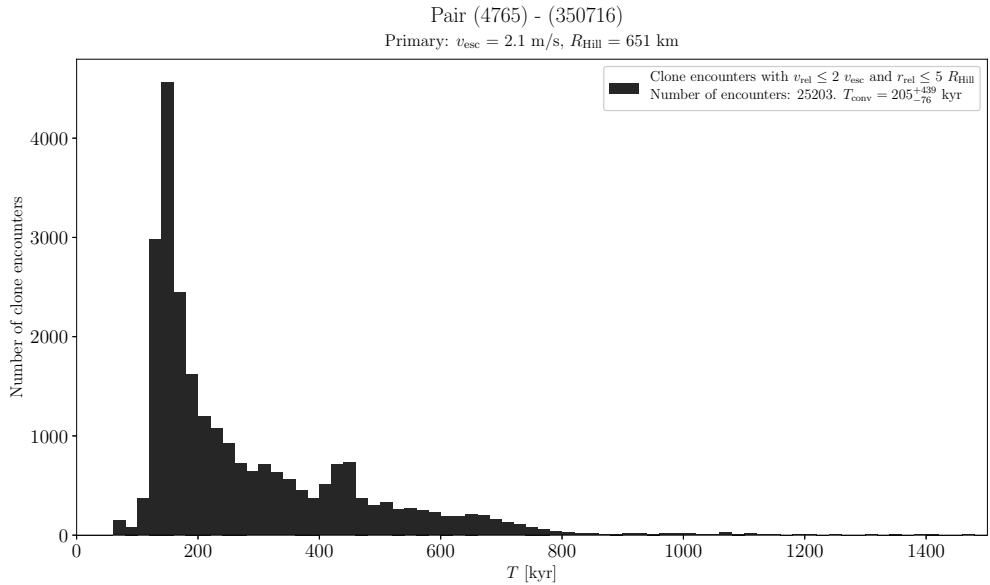}
\caption{\rm Distribution of past times of close and slow primary--secondary clone encounters for the asteroid pair 4765-350716.
}
\label{4765enchist}
\vspace{1cm}
\end{figure}

\begin{figure}
%\vspace{1cm}
\includegraphics[width=\textwidth]{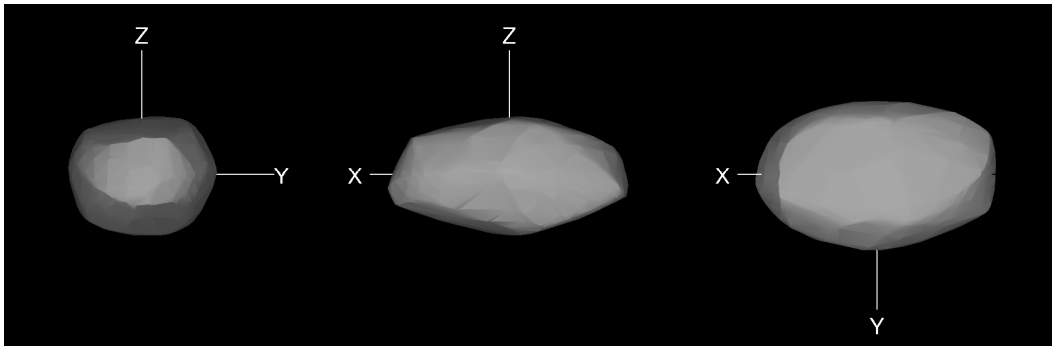}
\caption{\rm Convex shape model of (4765) Wasserburg for the pole solution $(L_1,B_1) = (235^\circ, +8^\circ)$.
}
\label{4765model}
\vspace{1cm}
\end{figure}

Backward integrations of their heliocentric orbits suggest that these two asteroids separated about 200~kyr ago, though higher ages are also possible (Fig.~\ref{4765enchist}).
From spectral observations of the primary (4765), Polishook et~al.~(2014a) found that it is an X type in the near-IR Bus-DeMeo taxonomy system;
it probably belongs to the E class in the Tholen taxonomy, considering its position in the Hungaria asteroid group.
From the SDSS measurements, we obtained its $(V-R)_1 = 0.40 \pm 0.03$.
We also derived its spin vector and constructed a convex shape model (Fig.~\ref{4765model}).
It is interesting that it has an obliquity close to $90^\circ$, i.e., its spin axis is close to the ecliptic plane.
We have found that the small one-opposition asteroid 2016~GL253 is very close to this pair, its distance from the primary in osculating elements is $d = 2.4 \pm 0.2$~m/s only.  We also checked that
their nominal orbits converge in the past.  So, this seems to be actually an asteroid cluster, similar to the clusters studied by Pravec et~al.~(2018), but a final confirmation awaits for backward orbit
integrations after a better orbit is derived for 2016~GL253 from its future observations.

\clearpage

\bigskip
\subsection{(4905) Hiromi and (7813) Anderserikson}
\label{4905sect}

\begin{figure}
%\vspace{1cm}
\includegraphics[width=\textwidth]{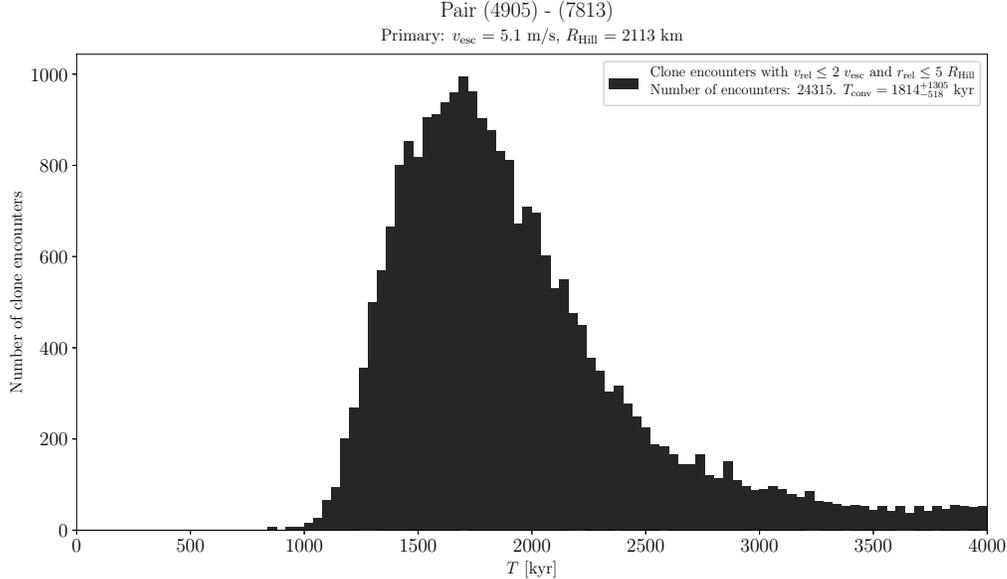}
\caption{\rm Distribution of past times of close and slow primary--secondary clone encounters for the asteroid pair 4905--7813.
}
\label{4905enchist}
\vspace{1cm}
\end{figure}

\begin{figure}
%\vspace{1cm}
\includegraphics[width=\textwidth]{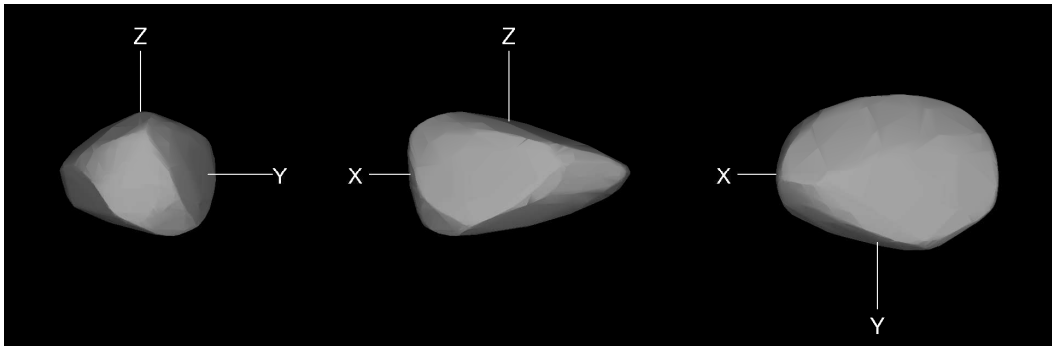}
\caption{\rm Convex shape model of (4905) Hiromi  for the pole solution $(L_1,B_1) = (185^\circ, -87^\circ)$.
}
\label{4905model}
\vspace{1cm}
\end{figure}

Despite the somewhat larger distance between these two asteroids in mean orbital elements than usual, $d_{\rm mean} = 28$~m/s (but $d_{\rm prop} = 0.9$~m/s only!),
this is a secure pair.  They show a good orbital convergence with estimated age about 1.8~Myr, see Fig.~\ref{4905enchist}.
From spectral observations of the primary (4905), Polishook et~al.~(2014a) found that it is an Sw type.
We also derived its retrograde spin pole (see Table~\ref{AstPairsPolestable}).
The best-fit convex shape model is shown in Fig.~\ref{4905model}.
From our thermophysical modeling, we derived its volume-equivalent diameter $D_1$ and geometric albedo $p_{V,1}$ (Appendix~\ref{AppendTPM}).
For the secondary (7813), we derived from our measured Sloan colors that it is an S type\footnote{K type is not entirely ruled out for (7813), but it is a rather rare taxonomic type and it is
much more likely that the asteroid belongs to the S complex.}; from its SDSS colors, Carvano et~al.~(2010) derived an S type as well.
Its period $P_2 = 13.277 \pm 0.002$~h is likely, but values twice or thrice that are not ruled out (Suppl.~Fig.~17).
Using our derived mean absolute magnitudes, we refined the WISE data by Masiero et~al.~(2011) and obtained
%$D_1 = 10.9 \pm 1.1$~km and $p_{V,1} = 0.16 \pm 0.03$, and
$D_2 = 6.3 \pm 0.6$~km and $p_{V,2} = 0.20 \pm 0.04$.

\clearpage

\bigskip
\subsection{(5026) Martes and 2005 WW113}
\label{5026sect}

\begin{figure}
%\vspace{1cm}
\includegraphics[width=\textwidth]{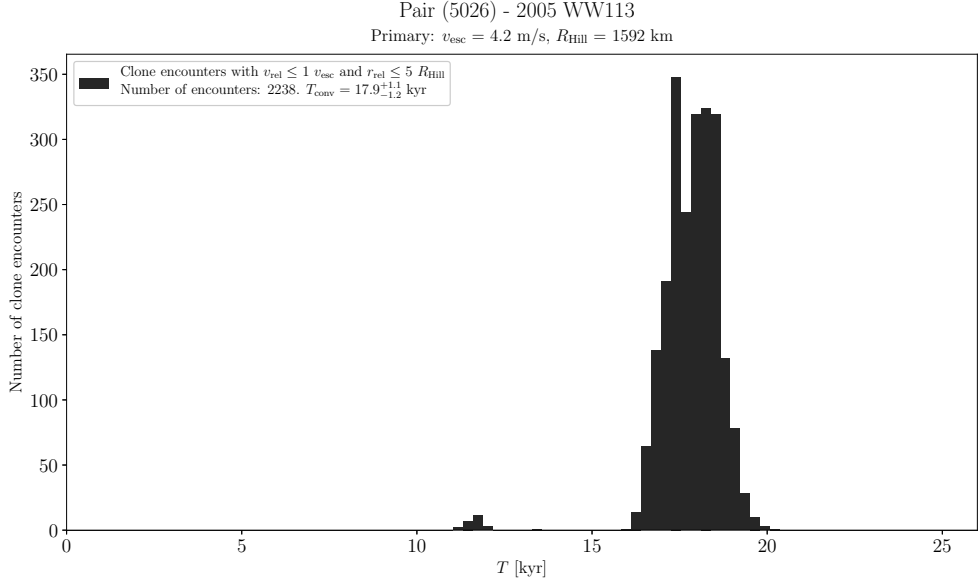}
\caption{\rm Distribution of past times of close and slow primary--secondary clone encounters for the asteroid pair 5026--2005WW113.
}
\label{5026enchist}
\vspace{1cm}
\end{figure}

\begin{figure}
%\vspace{1cm}
\includegraphics[width=\textwidth]{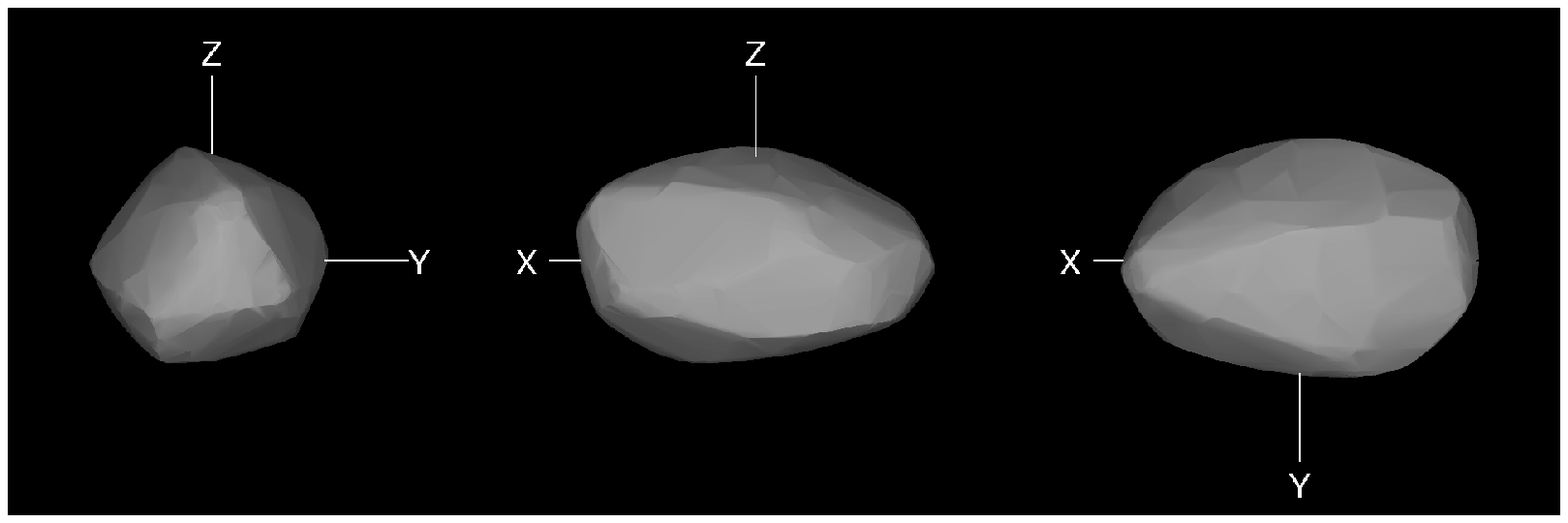}
\caption{\rm Convex shape model of (5026) Martes for the pole solution $(L_1,B_1) = (11^\circ, +62^\circ)$.
}
\label{5026model_1}
\vspace{1cm}
\end{figure}

\begin{figure}
%\vspace{1cm}
\includegraphics[width=\textwidth]{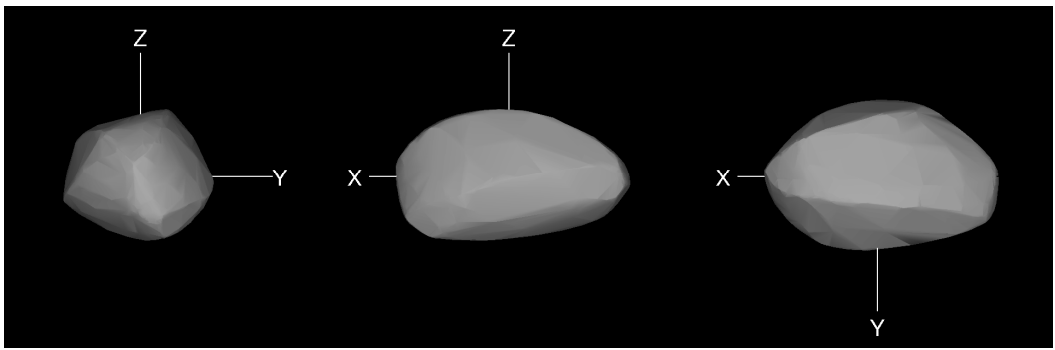}
\caption{\rm Convex shape model of (5026) Martes for the pole solution $(L_1,B_1) = (197^\circ, +47^\circ)$.
}
\label{5026model_2}
\vspace{1cm}
\end{figure}

This is a very young pair.  Backward integrations of their heliocentric orbits suggest that these two asteroids separated about 18~kyr ago (Fig.~\ref{5026enchist}).
%Some even younger clone encounters (at times about 2.5 and about 12~kyr) have higher relative velocities and we consider them to be subsequent approaches between the two asteroids rather than their real separation time.
For the primary (5026), Polishook et~al.~(2014a) found that it is a Ch type.
We also derived its prograde spin pole (with two mirror solutions in longitude, see Table~\ref{AstPairsPolestable}), in agreement with the earlier model by Polishook~(2014a).
The best-fit convex shape models for the two pole solutions are shown in Figs.~\ref{5026model_1} and \ref{5026model_2}.

\newpage

\clearpage

\bigskip
\subsection{(6070) Rheinland and (54827) Kurpfalz}
\label{6070sect}

This young asteroid pair was studied in detail by Vokrouhlick\'y et~al.~(2017).  Their data are summarized in Table~\ref{AstPairsDatatable} to \ref{AstPairsPolestable}.
One of their interesting findings is that the spin vectors of the two asteroids are neither aligned at present nor they were aligned at the time of separation of the two asteroids 16.34~kyr ago, but they were tilted by $38^\circ \pm 12^\circ$.
We will discuss it in Section~\ref{PolesSect}.
The primary (6070) and the secondary (54827) were classified as Sq and Q types, respectively (Polishook et~al.~2014).  They interpreted the spectral difference as the secondary having a fresher, less space weathered surface.

\bigskip
\subsection{(6369) 1983 UC and (510132) 2010 UY57}
\label{6369sect}

\begin{figure}
%\vspace{1cm}
\includegraphics[width=\textwidth]{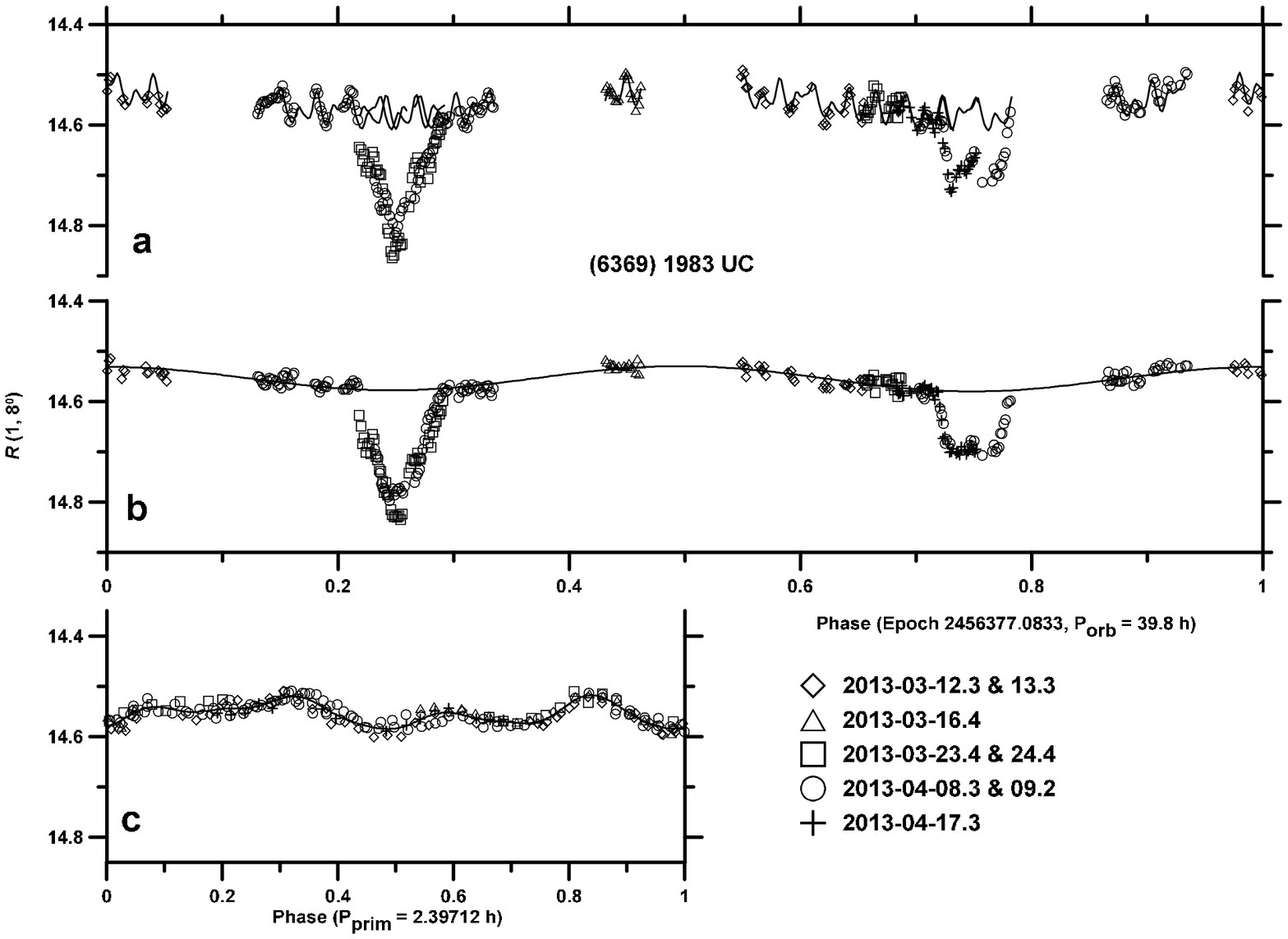}
\caption{\rm Lightcurve data of (6369) 1983 UC from 2013.
(a) The original data showing all the lightcurve components, folded with the orbital period.
(b) The orbital plus secondary rotational lightcurve components, derived after subtraction of the primary lightcurve component, showing the mutual events between components
of the binary system superimposed to the secondary rotational lightcurve. (c) The primary lightcurve component.
}
\label{6369_13lc}
\vspace{1cm}
\end{figure}

Backward integrations of their heliocentric orbits suggest that these two asteroids separated about 700~kyr ago (Suppl.~Fig.~18).
From our observations taken in March-April 2013, we found that the primary (6369)~1983~UC is a binary system.
The satellite (bound secondary) has a secondary-to-primary mean diameter ratio of $D_{1,s}/D_{1,p} = 0.37 \pm 0.02$,
an orbital period of $39.80 \pm 0.02$~h, and it is synchronous, i.e., its rotational period $P_{1,s}$ is equal to the orbital period (Fig.~\ref{6369_13lc}).
The fact that we did not observe mutual eclipse events in the system during its 2nd, return apparition in February 2016 indicates that the satellite's orbit has a significant obliquity, i.e., the orbital pole is
not close to the north or south pole of the ecliptic.
The primary's rotational period $P_{1,p} = 2.39712 \pm 0.00005$~h is likely.  Though we cannot formally rule out a period twice as long, it would be a lightcurve with 8 pairs of maxima and minima per rotation, which is unlikely.
%The color index of (6369) is $(V-R)_1 = 0.472 \pm 0.010$.
We will discuss it, together with other multiple (paired-binary) asteroid systems, in Section~\ref{P1qdistrSect}.

\clearpage

\bigskip
\subsection{(7343) Ockeghem and (154634) 2003 XX38}
\label{7343sect}

\begin{figure}
%\vspace{1cm}
\includegraphics[width=\textwidth]{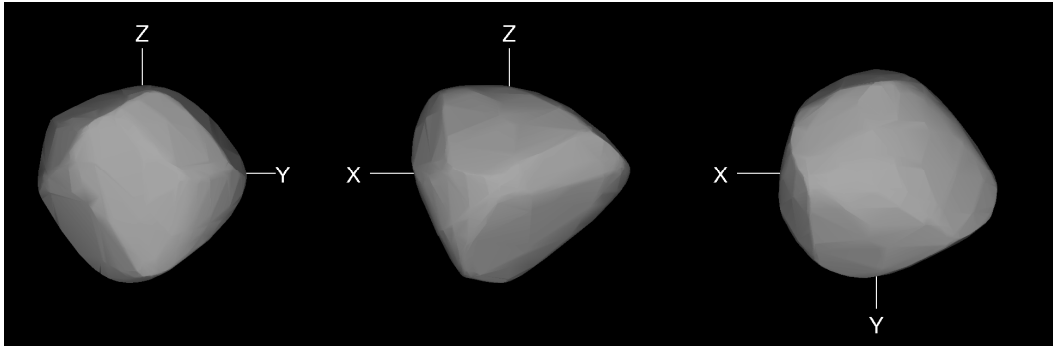}
\caption{\rm Convex shape model of (7343) Ockeghem for the pole solution $(L_1,B_1) = (39^\circ, +57^\circ)$.
}
\label{7343model_1}
\vspace{1cm}
\end{figure}

\begin{figure}
%\vspace{1cm}
\includegraphics[width=\textwidth]{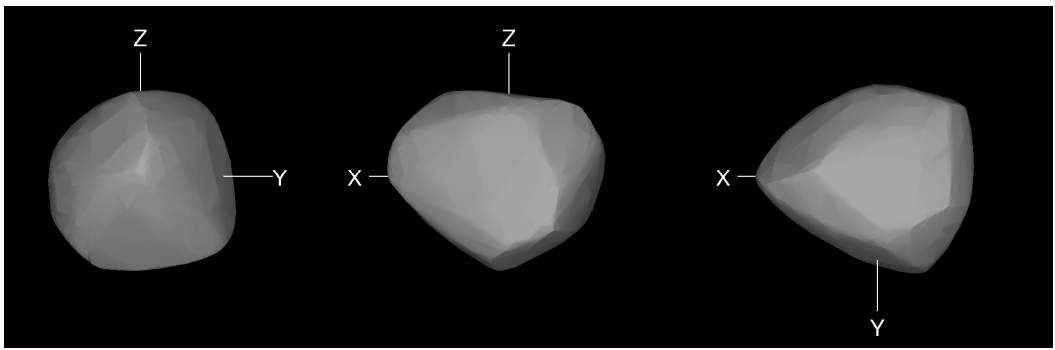}
\caption{\rm Convex shape model of (7343) Ockeghem for the pole solution $(L_1,B_1) = (231^\circ, +52^\circ)$.
}
\label{7343model_2}
\vspace{1cm}
\end{figure}

Backward integrations of their heliocentric orbits suggest a lower limit on the age of this pair of 382~kyr (Suppl.~Fig.~20).
Duddy et~al.~(2012) found that these two asteroids have very similar spectra belonging to the S class.
They also derived the primary's effective diameter $D_1 = 4.1 \pm 0.6$~km.
With our determined mean absolute magnitude $H_1 = 14.31 \pm 0.11$, we obtain the geometric albedo $p_{V,1} = 0.20 \pm 0.06$.
For the primary (7343), we derived its prograde spin pole (with two mirror solutions in longitude, see Table~\ref{AstPairsPolestable}).
The best-fit convex shape models for the two pole solutions are shown in Figs.~\ref{7343model_1} and \ref{7343model_2}.

\newpage

\clearpage

\bigskip
\subsection{(8306) Shoko and 2011 SR158}
\label{8306sect}

Backward integrations of their heliocentric orbits suggest that these two asteroids separated about 400~kyr ago (Suppl.~Fig.~21).
Pravec et~al.~(2013) found that the primary (8306) Shoko is a binary, possibly ternary system, from their observations taken during September--December 2013
(see also Pravec et~al.~2016).
The satellite (bound secondary) has a secondary-to-primary mean diameter ratio of $D_{1,s}/D_{1,p} \ge 0.40$,
an orbital period of $36.20 \pm 0.04$~h, and it is synchronous, i.e., its rotational period $P_{1,s}$ is equal to the orbital period.
The fact that we did not observe mutual eclipse events in the system during its 2nd, return apparition in January--February 2015 indicates that the satellite's orbit has a significant obliquity, i.e., the orbital pole is
not close to the north or south pole of the ecliptic.
The primary's rotational period $P_{1,p} = 3.35015 \pm 0.00005$~h is likely, though we cannot formally rule out a period twice as long with 4 pairs of maxima/minima per rotation.
%The color index of (8306) is $(V-R)_1 = 0.47 \pm 0.01$.
Note that the normalized total angular momentum of the system $\alpha_L = 1.19 \pm 0.17$ given in Table~\ref{AstPairSatstable} was computed without accounting for the possible second satellite (see the references above).
Polishook et~al.~(2014a) found from their spectral observations that it is an Sq type.
We will discuss this asteroid system, together with other multiple (paired-binary) asteroid systems, in Section~\ref{P1qdistrSect}.

\bigskip
\subsection{(9783) Tensho-kan and (348018) 2003 SF334}
\label{9783sect}

This is a secure asteroid pair, showing a good orbital convergence about 600~kyr ago (Suppl.~Fig.~24).
We found that the primary (9783) Tensho-kan is a binary system.  The satellite (bound secondary) has a secondary-to-primary mean diameter ratio of $D_{1,s}/D_{1,p} = 0.24 \pm 0.02$,
an orbital period of $29.5663 \pm 0.0006$~h and a retrograde orbit with pole near the south ecliptic pole (Scheirich et al., in preparation).
The primary's rotational period $P_{1,p} = 3.0108 \pm 0.0003$ is likely, though a period twice that is not entirely ruled out formally.
Refining the WISE thermal measurements (Masiero et~al.~2011) with our accurately determined mean absolute magnitude of the whole system, $H_1 = 14.06 \pm 0.02$,
we obtained the effective diameter $D_1 = 5.3 \pm 0.6$~km and geometric albedo $p_{V,1} = 0.15 \pm 0.03$.
Correcting it for the satellite presence, we derived the primary's mean diameter $D_{1,p} = 5.1 \pm 0.6$~km.
%The color index of (9783) is $(V-R)_1 = 0.471 \pm 0.010$.
We will discuss it, together with other multiple (paired-binary) asteroid systems, in Section~\ref{P1qdistrSect}.

\bigskip
\subsection{(10123) Fide\"oja and (117306) 2004 VF21}
\label{10123sect}

\begin{figure}
%\vspace{1cm}
\includegraphics[width=\textwidth]{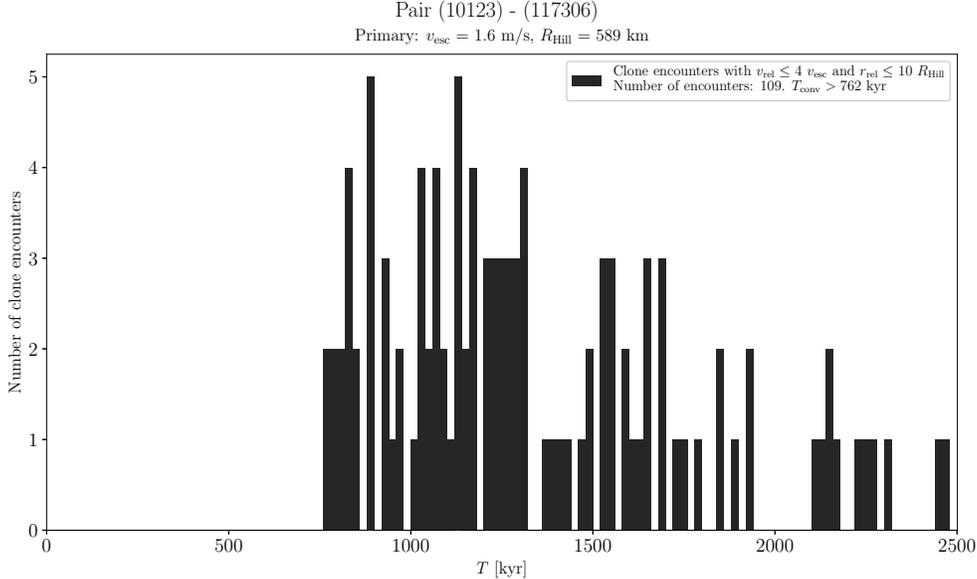}
\caption{\rm Distribution of past times of close and slow primary--secondary clone encounters for the asteroid pair 10123--117306.
}
\label{10123enchist}
\vspace{1cm}
\end{figure}

Backward orbital integrations of these two asteroids show a modest numbers of clone encounters about 1-2~Myr ago (Fig.~\ref{10123enchist}).
This is because the orbits undergo irregular jumps over the 7:2 mean motion resonance with Jupiter.
Despite this drawback, we consider this pair to be real, as it is supported also by a good past convergence of the nominal orbits.
It is further supported by that the two asteroids have the same colors, as found by Moskovitz (2012), who also classified (10123)
as a Ld type, and by our measured $(V-R)_1 = 0.468 \pm 0.010$ and $(V-R)_2 = 0.464 \pm 0.018$.

We found that the primary (10123) Fide\"oja is a binary system.
The satellite (bound secondary) has a secondary-to-primary mean diameter ratio $D_{1,s}/D_{1,p} = 0.36 \pm 0.02$ and
an orbital period of $56.46 \pm 0.02$~h.
The primary's rotational period $P_{1,p} = 2.8662 \pm 0.0001$~h is likely, but a period twice as long is also formally possible.
In the best data taken in February--March~2013, there is also apparent a second rotational lightcurve with period of $38.8 \pm 0.2$~h and
an amplitude in the combined primary plus secondary lightcurve of 0.04~mag.  Whether it belongs to the observed eclipsing secondary or
to a third body (second satellite) in the system, remains to be seen from future studies.
Refining the WISE thermal measurements (Masiero et~al.~2011) with our accurately determined mean absolute magnitude of the whole system, $H_1 = 14.55 \pm 0.03$,
we obtained the effective diameter $D_1 = 3.4 \pm 0.6$~km and geometric albedo $p_{V,1} = 0.24 \pm 0.09$.
Correcting it for the satellite presence, we derived the primary's mean diameter $D_{1,p} = 3.2 \pm 0.6$~km.
The unbound secondary (117306) 2004 VF21 has a period $P_2 = 14.462 \pm 0.010$~h (Suppl.~Fig.~25).
We will discuss the system of Fide\"oja, together with other multiple (paired-binary) asteroid systems, in Section~\ref{P1qdistrSect}.

\bigskip
\subsection{(17198) Gorjup and (229056) 2004 FC126}
\label{17198sect}

This is a secure asteroid pair, showing a good orbital convergence about 300~kyr ago (Suppl.~Fig.~47).
Wolters et~al.~(2014), following spectral observations by Duddy et~al.~(2013), found that the primary is Sa while the secondary is Sr type;
the secondary has a deeper 1-$\mu$m absorption band.  They suggested that it could be due to a more weathered surface of the primary.
Polishook et~al.~(2014a) reported an Sw type for the primary.

\bigskip
\subsection{(21436) Chaoyichi and (334916) 2003 YK39}
\label{21436sect}

\begin{figure}
%\vspace{1cm}
\includegraphics[width=\textwidth]{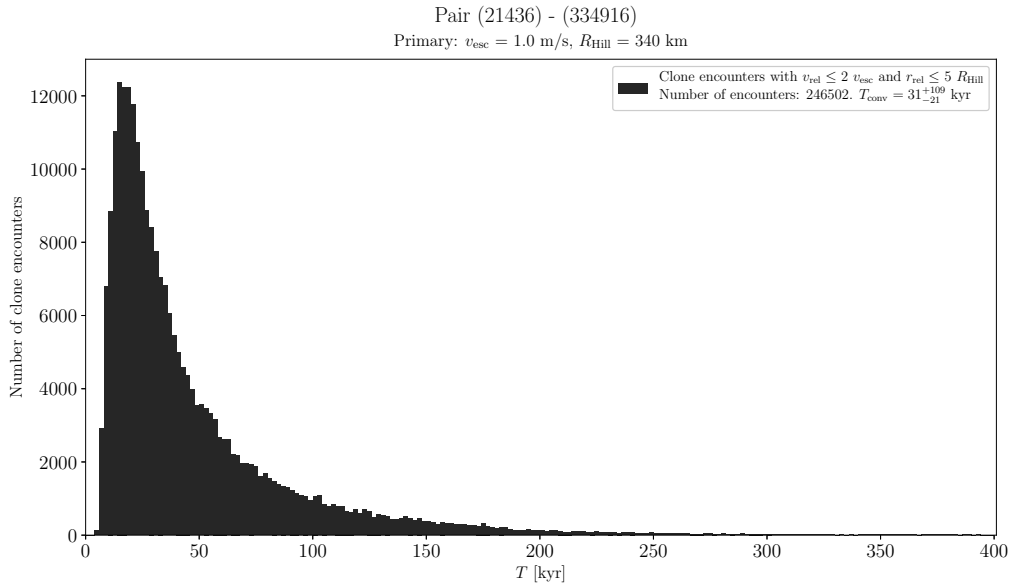}
\caption{\rm Distribution of past times of close and slow primary--secondary clone encounters for the asteroid pair 21436--334916.
}
\label{21436enchist}
\vspace{1cm}
\end{figure}

This is a young asteroid pair, showing an orbital convergence about 30~kyr ago (Fig.~\ref{21436enchist}).
We found that the primary (21436) Chaoyichi is a binary system.  It has the size ratio $D_{1,s}/D_{1,p} = 0.36 \pm 0.02$ and
orbital period $81.19 \pm 0.02$ (Scheirich et al., in preparation).
A particularly interesting feature is that it has a non-zero eccentricity of $0.19 \pm 0.03$ (3-$\sigma$ uncertainty).
Refining the WISE thermal measurements (Masiero et~al.~2011) with our accurately determined mean absolute magnitude of the whole system, $H_1 = 15.62 \pm 0.05$,
we obtained the effective diameter $D_1 = 2.0 \pm 0.3$~km and geometric albedo $p_{V,1} = 0.26 \pm 0.09$.
Correcting it for the satellite presence, we derived the primary's mean diameter $D_{1,p} = 1.9 \pm 0.3$~km.
We will discuss it, together with other multiple (paired-binary) asteroid systems, in Section~\ref{P1qdistrSect}.

\bigskip
\subsection{(25021) Nischaykumar and (453818) 2011 SJ109}
\label{25021sect}

Backward integrations of their heliocentric orbits suggest that these two asteroids separated about 900~kyr ago (Suppl.~Fig.~59).
We found that the primary (25021) Nischaykumar is a binary system.
It has the size ratio $D_{1,s}/D_{1,p} = 0.28 \pm 0.03$ and
orbital period of $23.4954 \pm 0.0004$ (Scheirich et al., in preparation).
The data
suggests that the satellite is synchronous, i.e., its rotational period $P_{1,s}$ is equal to the orbital period.
The primary's rotational period $P_{1,p} = 2.5344 \pm 0.0012$~h is likely, but we cannot rule out some longer periods for its low amplitude.
Refining the WISE thermal measurements (Masiero et~al.~2011) with our accurately determined mean absolute magnitude of the whole system, $H_1 = 15.94 \pm 0.03$,
we obtained the effective diameter $D_1 = 2.1 \pm 0.6$~km and geometric albedo $p_{V,1} = 0.16 \pm 0.08$.
Correcting it for the satellite presence, we derived the primary's mean diameter $D_{1,p} = 2.0 \pm 0.6$~km.
We will discuss it, together with other multiple (paired-binary) asteroid systems, in Section~\ref{P1qdistrSect}.

\bigskip
\subsection{(25884) Asai and (48527) 1993 LC1}
\label{25884sect}

\begin{figure}[b]
%\vspace{1cm}
\includegraphics[width=\textwidth]{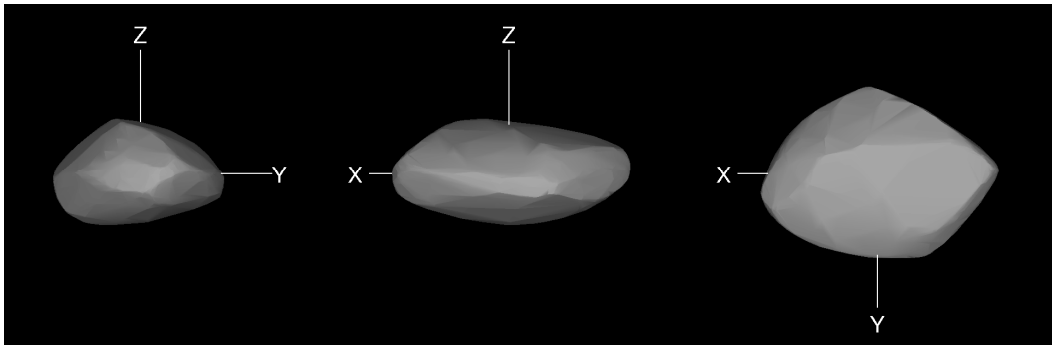}
\caption{\rm Convex shape model of (25884) Asai for the pole solution $(L_1,B_1) = (159^\circ, -57^\circ)$.
}
\label{25884model}
\vspace{1cm}
\end{figure}

Backward integrations of their heliocentric orbits suggest that these two asteroids separated about 700~kyr ago (Suppl.~Fig.~60).
For the primary (25884), using our derived mean absolute magnitude $H_1 = 15.05 \pm 0.12$, we refined the WISE data by Masiero et~al.~(2011) and obtained $D_1 = 1.9 \pm 0.5$~km and $p_{V,1} = 0.48 \pm 0.26$.
Polishook et~al.~(2014a) found it to be an X type in the near-IR Bus-DeMeo taxonomy system.  Considering its high albedo and position in the Hungaria asteroid family, it probably belongs to the E class in the Tholen taxonomy.
We also derived its retrograde spin pole (see Table~\ref{AstPairsPolestable}).
The best-fit convex shape model is shown in Fig.~\ref{25884model}.

\bigskip
\subsection{(26416) 1999 XM84 and (214954) 2007 WO58}
\label{26416sect}

This is a secure asteroid pair, showing a good orbital convergence about 270~kyr ago (Suppl.~Fig.~61).
Polishook~(2014b) suggested from observations taken from Wise in 2011 that the primary (26416) 1999~XM84 is a binary system, and we confirmed it
with observations from La Silla in 2015 (Pravec et~al.~2015).
The satellite (bound secondary) has a secondary-to-primary mean diameter ratio $D_{1,s}/D_{1,p} \ge 0.25$,
an orbital period of $20.7805 \pm 0.0002$~h and a retrograde orbit with pole within 10\dg of the south ecliptic pole (Scheirich et al., in preparation).
The satellite is synchronous, i.e., its rotational period $P_{1,s}$ is equal to the orbital period.
The primary's rotational period $P_{1,p} = 2.9660 \pm 0.0001$~h is likely, though a period twice that is not formally ruled out.
%The color index of (26416) is $(V-R)_1 = 0.495 \pm 0.011$.
The rotational period of the unbound secondary (214954), $P_2 = 2.7689 \pm 0.0002$~h is likely (Suppl.~Fig.~62), though a period twice that with four pairs of lightcurve maxima and minima per rotation is also possible.
We will discuss it, together with other multiple (paired-binary) asteroid systems, in Section~\ref{P1qdistrSect}.

\bigskip
\subsection{(26420) 1999 XL103 and 2012 TS209}
\label{26420sect}

This is a secure asteroid pair, showing an orbital convergence about 250~kyr ago (Suppl.~Fig.~63).
We found that the primary (26420) 1999 XL103 is a binary system.
The satellite (bound secondary) has a secondary-to-primary mean diameter ratio $D_{1,s}/D_{1,p} \ge 0.34$ and
an orbital period of $23.90 \pm 0.02$~h or $47.80 \pm 0.05$~h (Suppl.~Figs.~64 and 65).
The primary's rotational period $P_{1,p}$ has not been uniquely determined, there are a few possible solutions from 2.2 to 4.2~h.
%The color index of (26420) is $(V-R)_1 = 0.482 \pm 0.019$.
From our Sloan color measurements, we derived that it is a V type.
We will discuss it, together with other multiple (paired-binary) asteroid systems, in Section~\ref{P1qdistrSect}.

\bigskip
\subsection{(42946) 1999 TU95 and (165548) 2001 DO37}
\label{42946sect}

Backward integrations of their heliocentric orbits suggest that these two asteroids separated about 700~kyr ago (Suppl.~Fig.~77).
From our measured Sloan colors, we found that both asteroids are S types.
Polishook et~al.~(2014a) reported that the primary is an Sr or Sw type.
Using our derived weighted mean absolute magnitude $H_1 = 14.05 \pm 0.04$, we refined the WISE data by Masiero et~al.~(2011) and obtained $D_1 = 4.8 \pm 0.5$~km and $p_{V,1} = 0.19 \pm 0.04$; it is
an albedo typical for S types.

\bigskip
\subsection{(43008) 1999 UD31 and (441549) 2008 TM68}
\label{43008sect}

\begin{figure}
%\vspace{1cm}
\includegraphics[width=\textwidth]{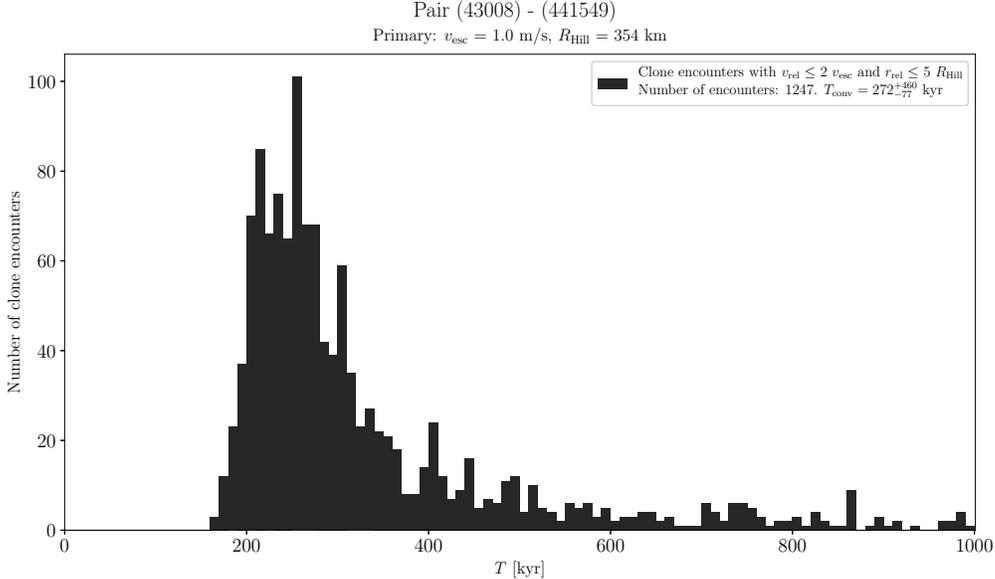}
\caption{\rm Distribution of past times of close and slow primary--secondary clone encounters for the asteroid pair 43008--441549.
}
\label{43008enchist}
\vspace{1cm}
\end{figure}

\begin{figure}
%\vspace{1cm}
\includegraphics[width=\textwidth]{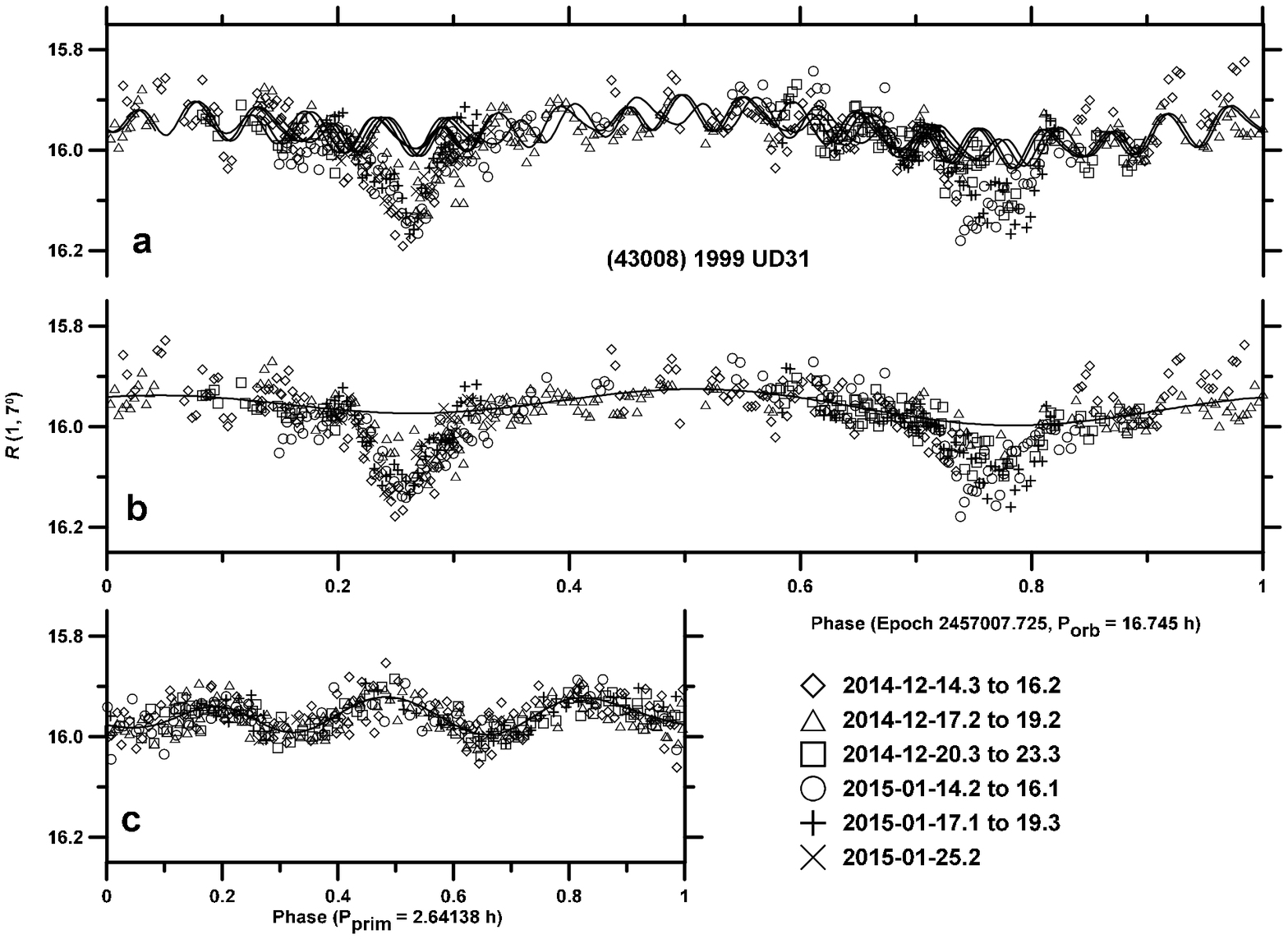}
\caption{\rm Lightcurve data of (43008) 1999 UD31 from 2014--2015.
(a) The original data showing all the lightcurve components, folded with the orbital period.
(b) The orbital plus secondary rotational lightcurve components, derived after subtraction of the primary lightcurve component, showing the mutual events between components
of the binary system superimposed to the secondary rotational lightcurve. (c) The primary lightcurve component.
}
\label{43008_14lc}
\vspace{1cm}
\end{figure}

\begin{figure}
%\vspace{1cm}
\includegraphics[width=\textwidth]{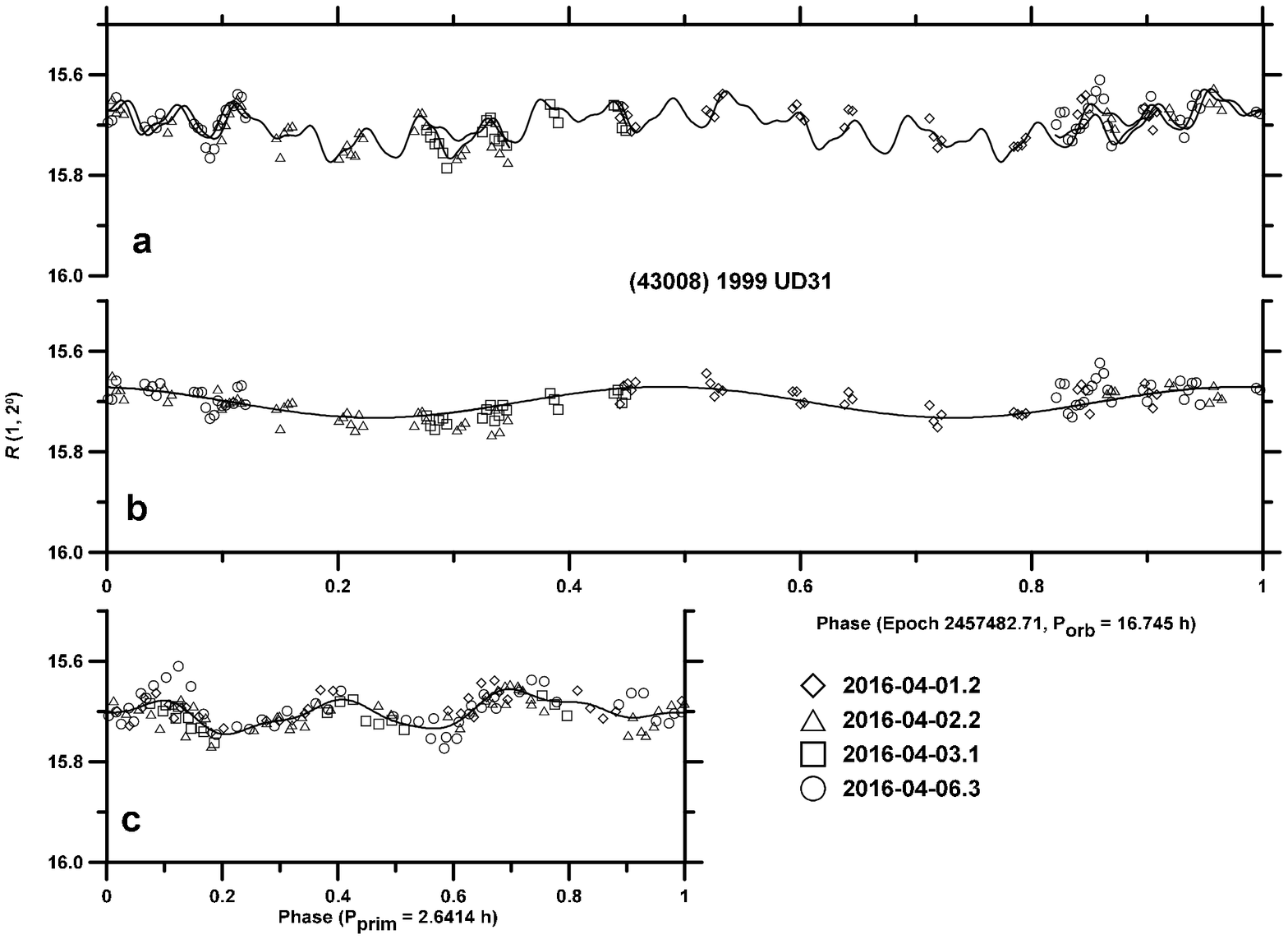}
\caption{\rm Lightcurve data of (43008) 1999 UD31 from 2016.
(a) The original data showing all the lightcurve components, folded with the orbital period.
(b) The secondary rotational lightcurve component, derived after subtraction of the primary lightcurve component.  Mutual events did not occur.
(c) The primary lightcurve component.
}
\label{43008_16lc}
\vspace{1cm}
\end{figure}

\begin{figure}
%\vspace{1cm}
\includegraphics[width=\textwidth]{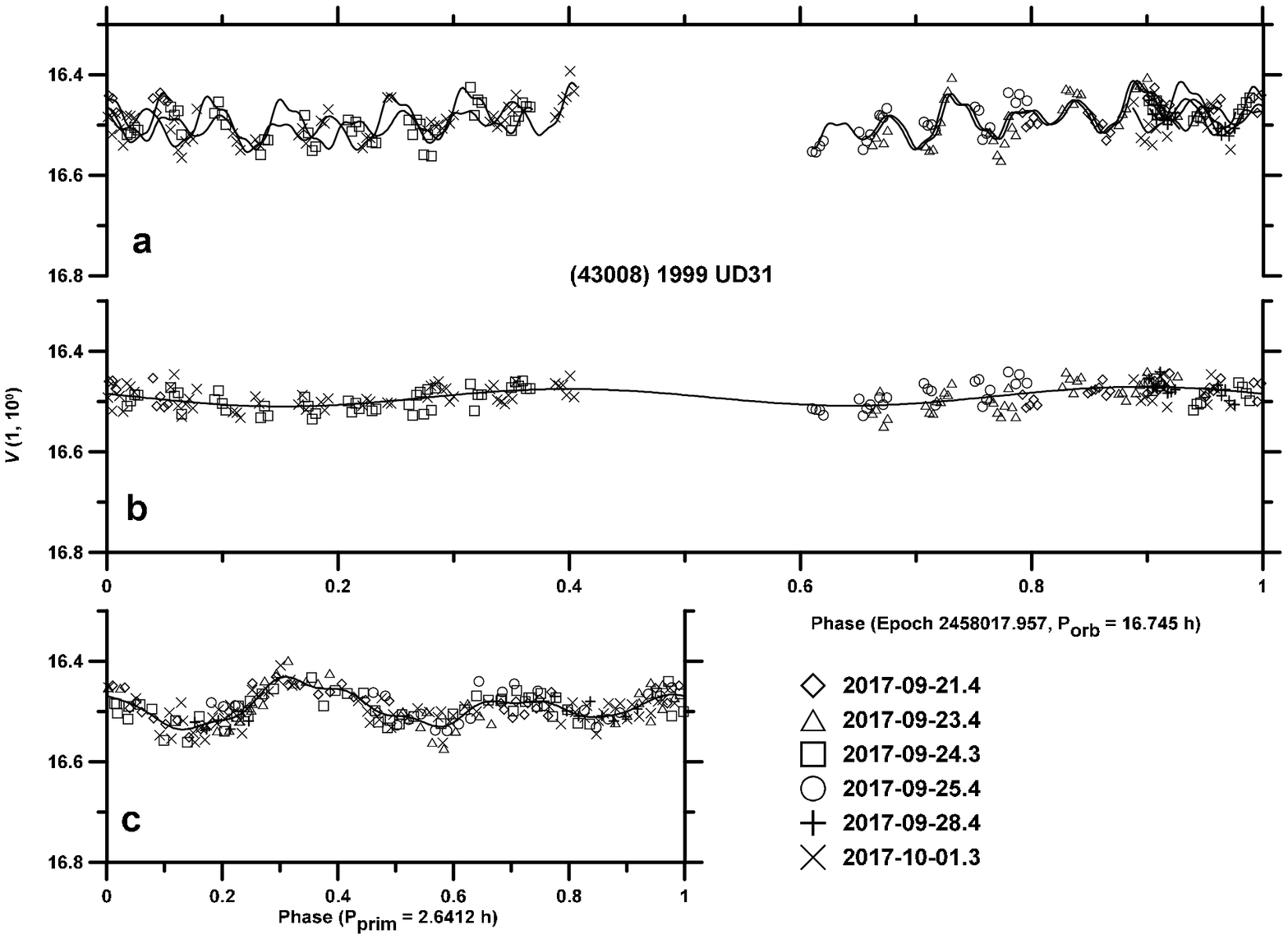}
\caption{\rm Lightcurve data of (43008) 1999 UD31 from 2017.
(a) The original data showing all the lightcurve components, folded with the orbital period.
(b) The secondary rotational lightcurve component, derived after subtraction of the primary lightcurve component.  Mutual events did not occur.
(c) The primary lightcurve component.
}
\label{43008_17lc}
\vspace{1cm}
\end{figure}

Backward integrations of their heliocentric orbits suggest that these two asteroids separated about 270~kyr ago (Fig.~\ref{43008enchist}).
From our observations taken during December 2014-January 2015, we found that the primary (43008) 1999~UD31 is a binary system.
The satellite (bound secondary) has a secondary-to-primary mean diameter ratio $D_{1,s}/D_{1,p} \ge 0.35 \pm 0.02$ and
an orbital period of $16.745 \pm 0.005$~h (Fig.~\ref{43008_14lc}).
The fact that we did not observe mutual eclipse events in the system during its 2nd and 3rd, return apparitions in April~2016 and September--October~2017 (Figs.~\ref{43008_16lc} and \ref{43008_17lc})
indicates that the satellite's orbit has a significant obliquity, i.e., the orbital pole is
not close to the north or south pole of the ecliptic.
The satellite is synchronous, i.e., its rotational period $P_{1,s}$ is equal to the orbital period.
The primary's rotational period $P_{1,p} = 2.64138 \pm 0.00007$~h is likely, but we cannot rule a period twice that with six pairs of lightcurve maxima/minima per rotation.
%The color index of (43008) is $(V-R)_1 = 0.458 \pm 0.020$.
The unbound secondary (441549) has a likely period $P_2 = 7.96 \pm 0.01$~h, but other nearby periods are not ruled out (Suppl.~Fig.~81).
We will discuss the system of (43008), together with other multiple (paired-binary) asteroid systems, in Section~\ref{P1qdistrSect}.

\clearpage

\bigskip
\subsection{(44620) 1999 RS43 and (295745) 2008 UH98}
\label{44620sect}

Backward integrations of their heliocentric orbits suggest that these two asteroids separated about 700~kyr ago (Suppl.~Fig.~82).
We found that the primary (44620) 1999 RS43 is a binary system.
The satellite (bound secondary) has the secondary-to-primary mean diameter ratio $D_{1,s}/D_{1,p} = 0.39 \pm 0.03$,
an orbital period of $33.6455 \pm 0.0003$~h and a prograde orbit with pole within 7\dg of the north ecliptic pole (Scheirich et al., in preparation).
The satellite is synchronous, i.e., its rotational period $P_{1,s}$ is equal to the orbital period.
The primary's rotational period $P_{1,p} = 3.1393 \pm 0.0003$~h is likely.  Though a period twice that is not formally ruled out, it would be a complex lightcurve with numerous maxima/minima per rotation, which is unlikely.
From our measured Sloan colors, we derived that the primary is an S type.
We will discuss it, together with other multiple (paired-binary) asteroid systems, in Section~\ref{P1qdistrSect}.

\bigskip
\subsection{(46829) McMahon and 2014 VR4}
\label{46829sect}

\begin{figure}
%\vspace{1cm}
\includegraphics[width=\textwidth]{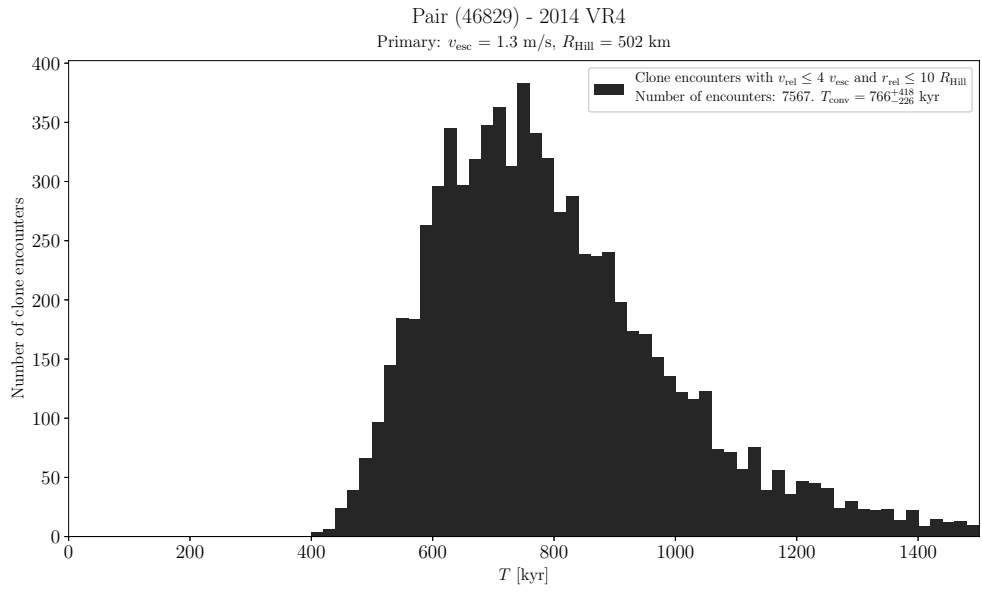}
\caption{\rm Distribution of past times of close and slow primary--secondary clone encounters for the asteroid pair 46829--2017VR4.
}
\label{46829enchist}
\vspace{1cm}
\end{figure}

\begin{figure}
%\vspace{1cm}
\includegraphics[width=\textwidth]{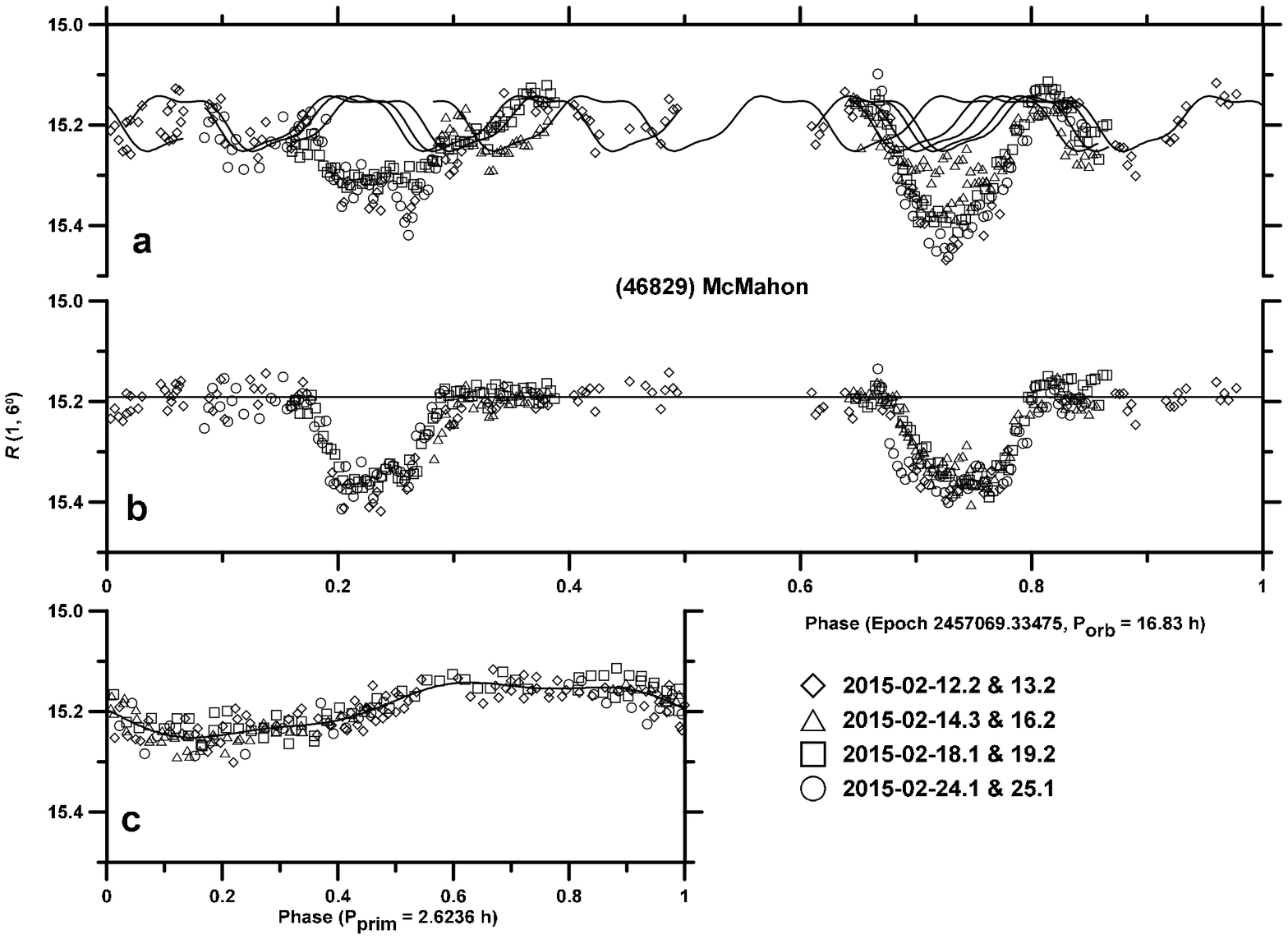}
\caption{\rm Lightcurve data of (46829) McMahon from February 2015.
(a) The original data showing all the lightcurve components, folded with the orbital period.
(b) The orbital lightcurve component, derived after subtraction of the primary lightcurve component, showing the mutual events between components
of the binary system. (c) The primary lightcurve component.
}
\label{43008_15alc}
\vspace{1cm}
\end{figure}

\begin{figure}
%\vspace{1cm}
\includegraphics[width=\textwidth]{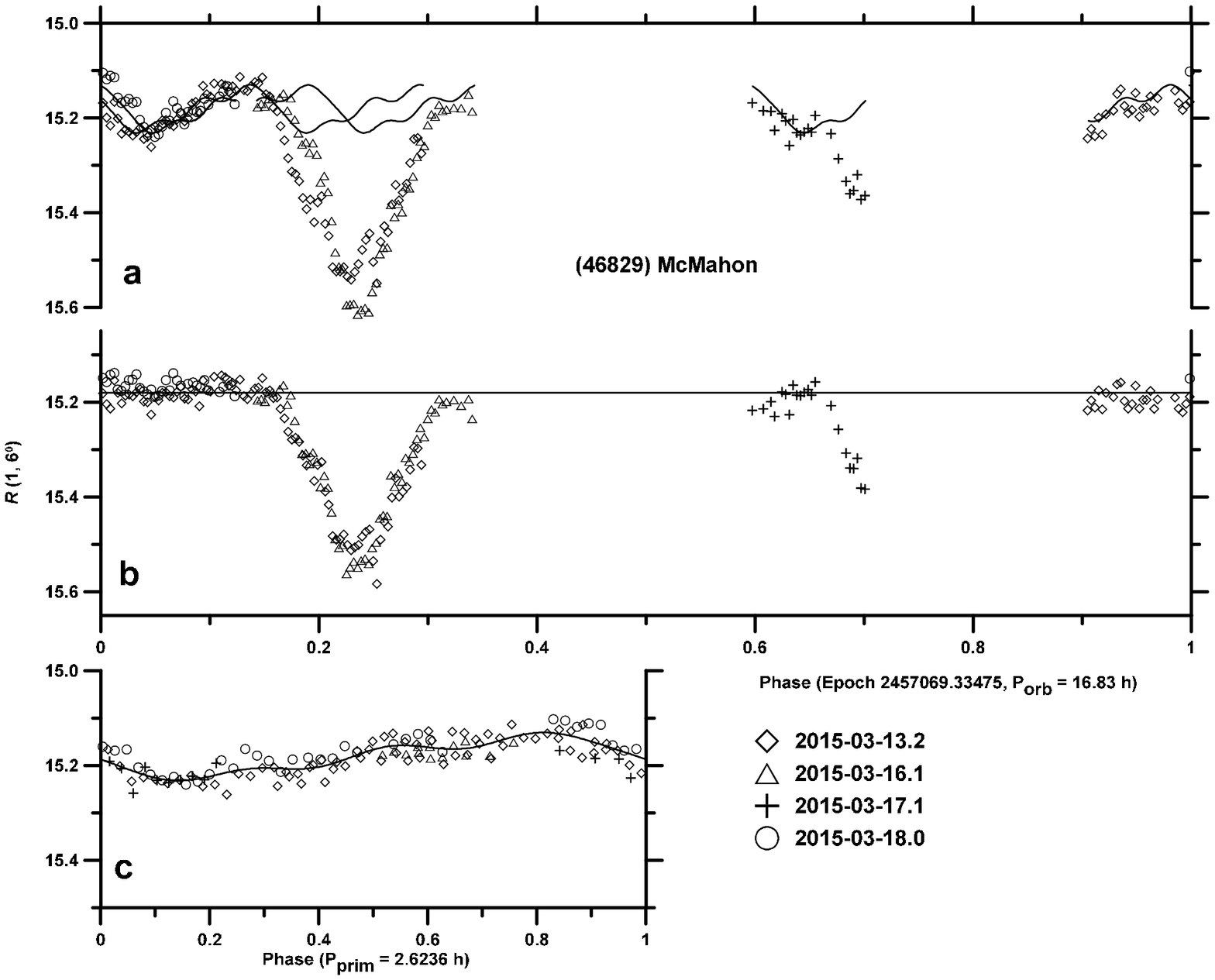}
\caption{\rm Lightcurve data of (46829) McMahon from March 2015.
(a) The original data showing all the lightcurve components, folded with the orbital period.
(b) The orbital lightcurve component, derived after subtraction of the primary lightcurve component, showing the mutual events between components
of the binary system. (c) The primary lightcurve component.
}
\label{43008_15blc}
\vspace{1cm}
\end{figure}

This asteroid pair is probably real; we calculated the probability that it is a random orbital coincidence of two independent asteroids $P_2/N_p = 0.0012$.
For backward integrations of their heliocentric orbits, we used 3000 orbital clones for the primary and 10000 clones for the secondary.  We chose that because
the orbit of 2014~VR4 is not very accurately determined yet so we sampled its large uncertainty hyperellipsoid with ten times more clones than for other asteroid pairs.
The orbital clones show a convergence about 800~kyr ago (Fig.~\ref{46829enchist}).
We found that the primary (46829) McMahon is a binary system.
The satellite (bound secondary) has a secondary-to-primary mean diameter ratio of $D_{1,s}/D_{1,p} = 0.40 \pm 0.02$ and
an orbital period of $16.833 \pm 0.002$~h (Figs.~\ref{43008_15alc} and \ref{43008_15blc}).
The primary's rotational period $P_{1,p} = 2.6236 \pm 0.0003$~h or twice that.
%The color index of (46829) is $(V-R)_1 = 0.470 \pm 0.016$.
We will discuss it, together with other multiple (paired-binary) asteroid systems, in Section~\ref{P1qdistrSect}.

\clearpage

\bigskip
\subsection{(49791) 1999 XF31 and (436459) 2011 CL97}
\label{49791sect}

\begin{figure}
%\vspace{1cm}
\includegraphics[width=\textwidth]{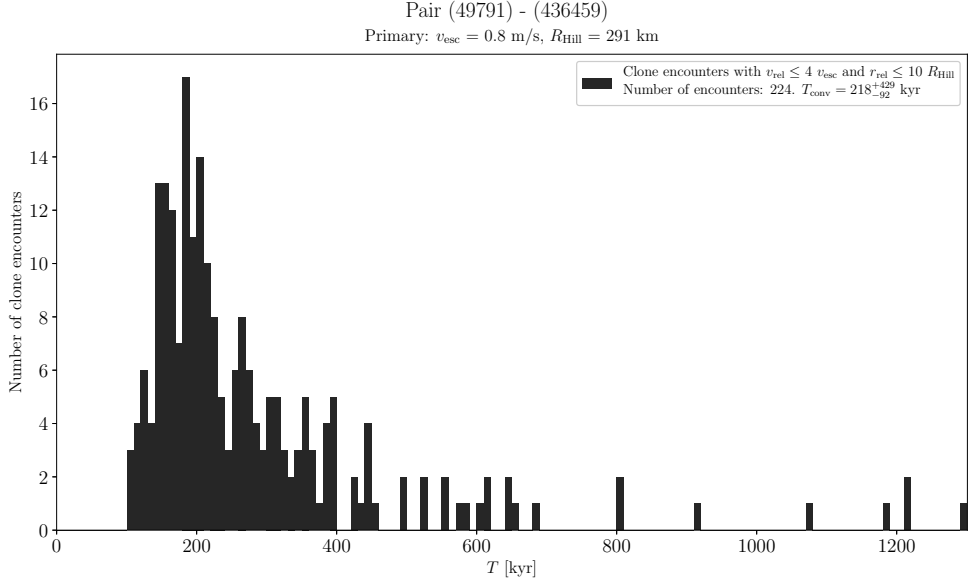}
\caption{\rm Distribution of past times of close and slow primary--secondary clone encounters for the asteroid pair 49791--436459.
}
\label{49791enchist}
\vspace{1cm}
\end{figure}

Backward orbital integrations of these two asteroids showed a relatively low number of clone encounters (Fig.~\ref{49791enchist}).  This is probably because
of a strong Yarkovsky effect for the small secondary, for which our coverage with 1000 orbital clones is not very dense.  Contributing to it may be also a strong chaoticity
of their orbits (probably due to the 15:8 mean motion resonance with Mars).
We calculated that the probability that this pair is a random coincidence of two independent asteroids in the space of mean orbital elements is 3\%.
We consider this pair to be probably real, but it will have to be confirmed with future studies.
For the primary (49791), V type is a likely classification from the SDSS measurements, though our measured $(V-R)_1 = 0.428 \pm 0.020$ differs from the mean $(V-R) = 0.516 \pm 0.037$ for V types (Pravec et~al.~2012b)
by more than $2 \sigma$; a confirmation of the taxonomic classification with further observations will be needed.
The period of (49791), $P_1 = 13.822 \pm 0.002$~h is likely, but a value twice that with 4 pairs of lightcurve maxima/minima per rotation is not entirely ruled out (Suppl.~Fig.~86).
With the $\Delta H = 2.7 \pm 0.3$, this primary rotation is too slow for formation of this asteroid pair by rotational fission.
We will discuss this anomalous case, together with other two similar ones, in Section~\ref{P1qdistrSect}.

\bigskip
\subsection{(52852) 1998 RB75 and (250322) 2003 SC7}
\label{52852sect}

Backward integrations of their heliocentric orbits suggest that these two asteroids separated about 500~kyr ago (Suppl.~Fig.~93).
Polishook et~al.~(2014a) found that the primary (52852) is a V type.
Its rotational period is estimated $P_1 = 5.4348 \pm 0.0005$h, but half-integer multiples of this value are also possible.
In our observations taken in September-October 2015, there appeared several brightness attenuations about 0.06~mag deep that might be mutual
events due to a satellite of the primary, but we did not obtain a satisfactory solution for period of the suspect events.
A confirmation with future high-quality observations is needed.

\bigskip
\subsection{(53537) 2000 AZ239 and (503955) 2004 ED107}
\label{53537sect}

\begin{figure}
%\vspace{1cm}
\includegraphics[width=\textwidth]{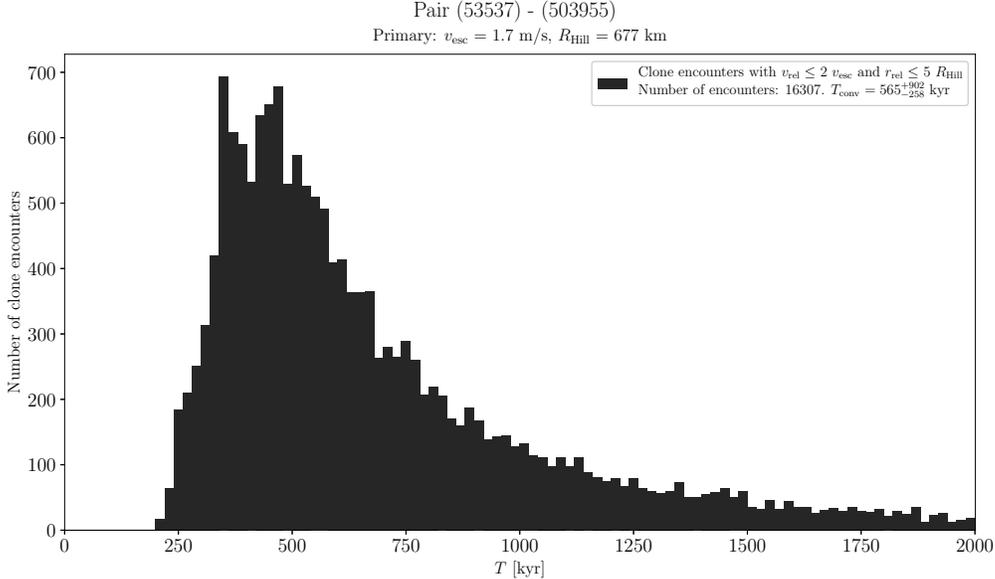}
\caption{\rm Distribution of past times of close and slow primary--secondary clone encounters for the asteroid pair 53537--503955.
}
\label{53537enchist}
\vspace{1cm}
\end{figure}

\begin{figure}
%\vspace{1cm}
\includegraphics[width=\textwidth]{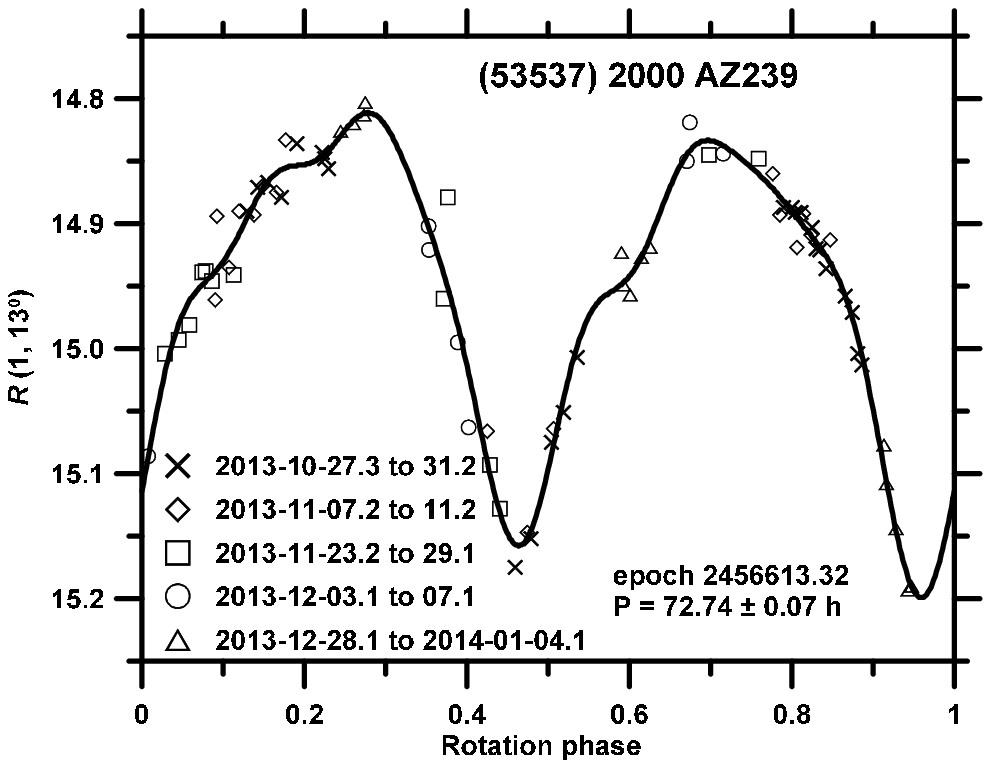}
\caption{\rm Composite lightcurve of (53537) 2000 AZ239.
}
\label{53537_13lc}
\vspace{1cm}
\end{figure}

This is a secure pair.
Backward integrations of their heliocentric orbits suggest that these two asteroids separated about 500~kyr ago (Fig.~\ref{53537enchist}).
However, the pair has an anomalously low angular momentum content.  The primary's spin period was uniquely determined $P_1 = 72.74 \pm 0.07$~h,
there is no ambiguity or significant uncertainty in it (Fig.~\ref{53537_13lc}).  In particular, all shorter periods are ruled out.
With the pair's $\Delta H = 3.3 \pm 0.3$, it is a too slow primary rotation to be explained by the theory of pair formation by rotational fission.
We will discuss this and other two similar anomalous cases in Section~\ref{P1qdistrSect}.
%For its $H_1$ determination and $D_1$ estimation, reported in Table~\ref{AstPairsDatatable}, we assumed $(V-R)_1 = 0.49 \pm 0.05$ and $p_{V,1} = 0.20$, which are mean values for S-type asteroids (Pravec et~al.~2012b).

\bigskip
\subsection{(54041) 2000 GQ113 and (220143) 2002 TO134}
\label{54041sect}

This is a secure pair.  The two asteroids are close one to each other ($d_{\rm mean} = 0.72$~m/s, $P_2/N_p < 10^{-4}$) and they show a good orbital convergence with
estimated age about 200~kyr (Suppl.~Fig.~96).
The period of (54041) is ambiguous, it is either $P_1 = 6.610$~h or twice that.
The period of (220143) $P_2 = 3.4987 \pm 0.0007$~h is likely.  Though a period twice as long cannot be formally ruled out, it would be a complex lightcurve with numerous maxima and minima, which seems implausible.
See Suppl.~Figs.~97 to 102.
% Polishook et~al.~(2014) found that the primary (54041) is a V type.
With the mean $H_1$ value (see Electronic Supplementary Information), we refined the WISE effective diameter and geometric albedo (Masiero et~al.~2011):
$D_1 = 2.6 \pm 0.8$~km and $p_{V,1} = 0.25 \pm 0.14$.
We measured the color indices $(V-R)_1 = 0.492 \pm 0.010$ and $(V-R)_2 = 0.447 \pm 0.018$; the former value is consistent with the V type found for the primary by Polishook et~al.~(2014a).\footnote{From SDSS measurements,
(54041) is suggested to be an S type.}
These two color indices differ by $0.045 \pm 0.021$, i.e., the difference is significant at 2-$\sigma$ level.
It will be good to do a spectral study of this asteroid pair in the future.

\bigskip
\subsection{(56232) 1999 JM31 and (115978) 2003 WQ56}
\label{56232sect}

\begin{figure}
%\vspace{1cm}
\includegraphics[width=\textwidth]{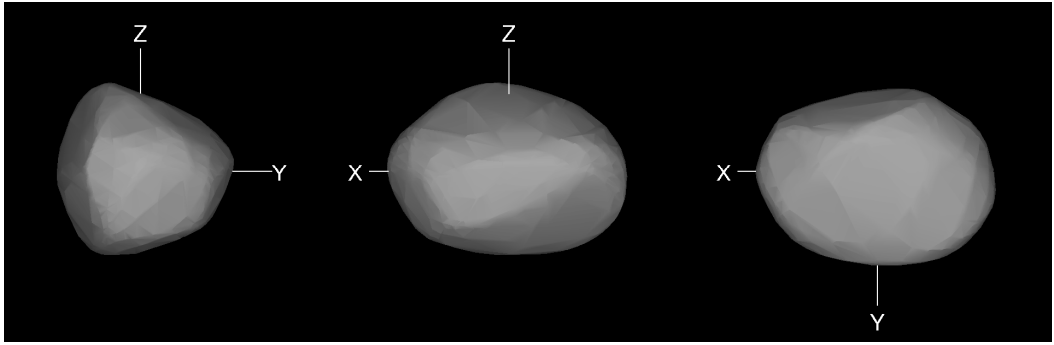}
\caption{\rm Convex shape model of (56232) 1999 JM31 for the pole solution $(L_1,B_1) = (190^\circ, -80^\circ)$.
}
\label{56232model}
\vspace{1cm}
\end{figure}

Backward integrations of their heliocentric orbits suggest that these two asteroids separated about 130~kyr ago (Suppl.~Fig.~109).
For the primary (56232), we derived its retrograde spin vector (see Table~\ref{AstPairsPolestable}).
The best-fit convex shape model is shown in Fig.~\ref{56232model}.

\bigskip
\subsection{(57202) 2001 QJ53 and (276353) 2002 UY20}
\label{57202sect}

This is a secure pair.  Backward integrations of their heliocentric orbits suggest that these two asteroids separated about 130~kyr ago (Suppl.~Fig.~116).
The rotational period of the primary (57202) is likely $P_1 = 2.44482 \pm 0.00007$h; periods twice or thrice as long are not entirely ruled out, but appear implausible (Suppl.~Figs.~117 and 118).
From the SDSS measurements, it is likely an S type, though an L type cannot be entirely ruled out.
%We measured its $(V-R)_1 = 0.483 \pm 0.012$.
In our observations taken in January 2017, there occurred two brightness attenuations 0.08--0.11~mag deep that could be mutual
events due to a satellite of the primary.  This probable paired binary needs to be confirmed with future observations.

\bigskip
\subsection{(60677) 2000 GO18 and (142131) 2002 RV11}
\label{60677sect}

\begin{figure}
%\vspace{1cm}
\includegraphics[width=\textwidth]{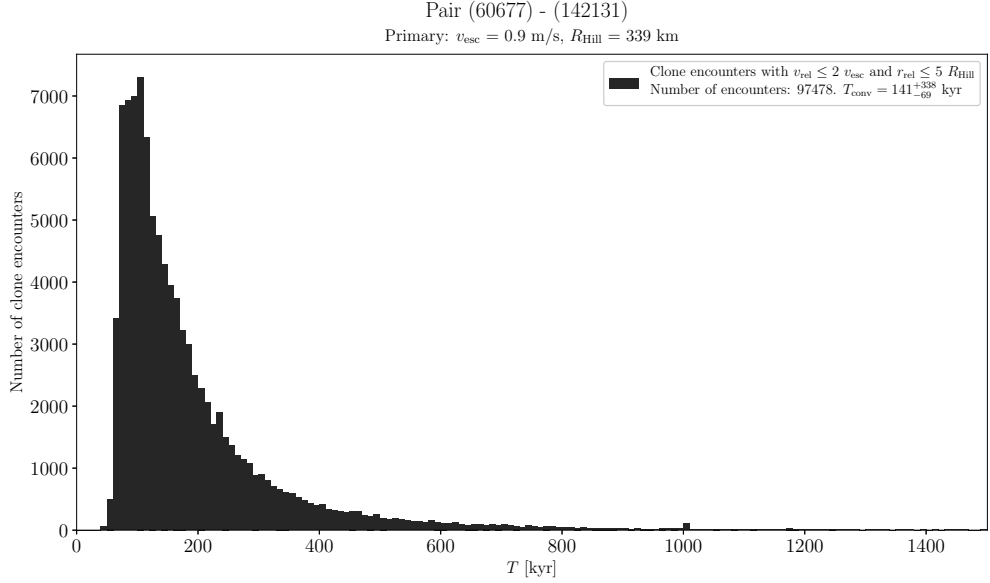}
\caption{\rm Distribution of past times of close and slow primary--secondary clone encounters for the asteroid pair 60677--142131.
}
\label{60677enchist}
\vspace{1cm}
\end{figure}

\begin{figure}
%\vspace{1cm}
\includegraphics[width=\textwidth]{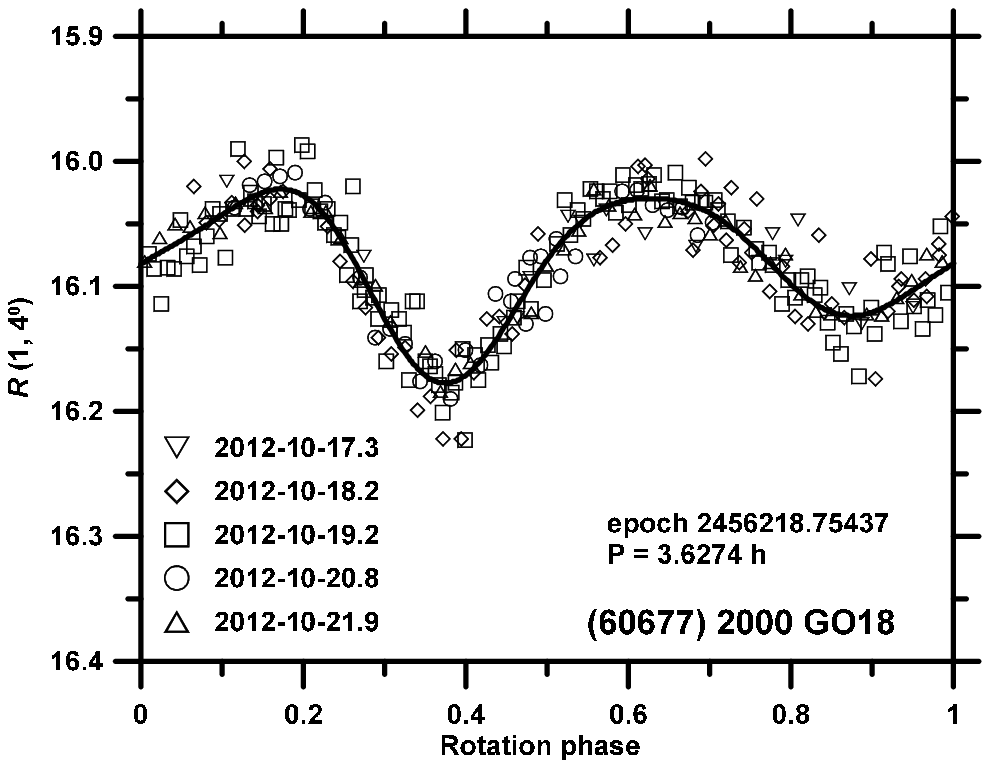}
\caption{\rm Composite lightcurve of (60677) 2000 GO18 from 2012.
}
\label{60677_12lc}
\vspace{1cm}
\end{figure}

\begin{figure}
%\vspace{1cm}
\includegraphics[width=\textwidth]{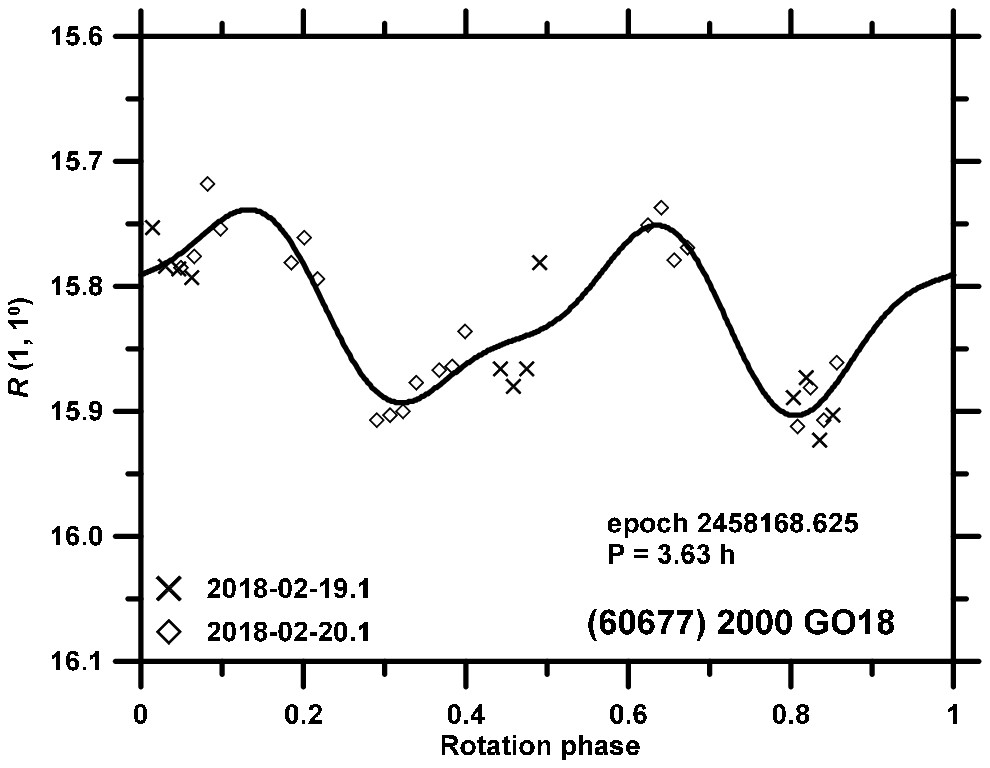}
\caption{\rm Composite lightcurve of (60677) 2000 GO18 from 2018.
}
\label{60677_18lc}
\vspace{1cm}
\end{figure}

\begin{figure}
%\vspace{1cm}
\includegraphics[width=\textwidth]{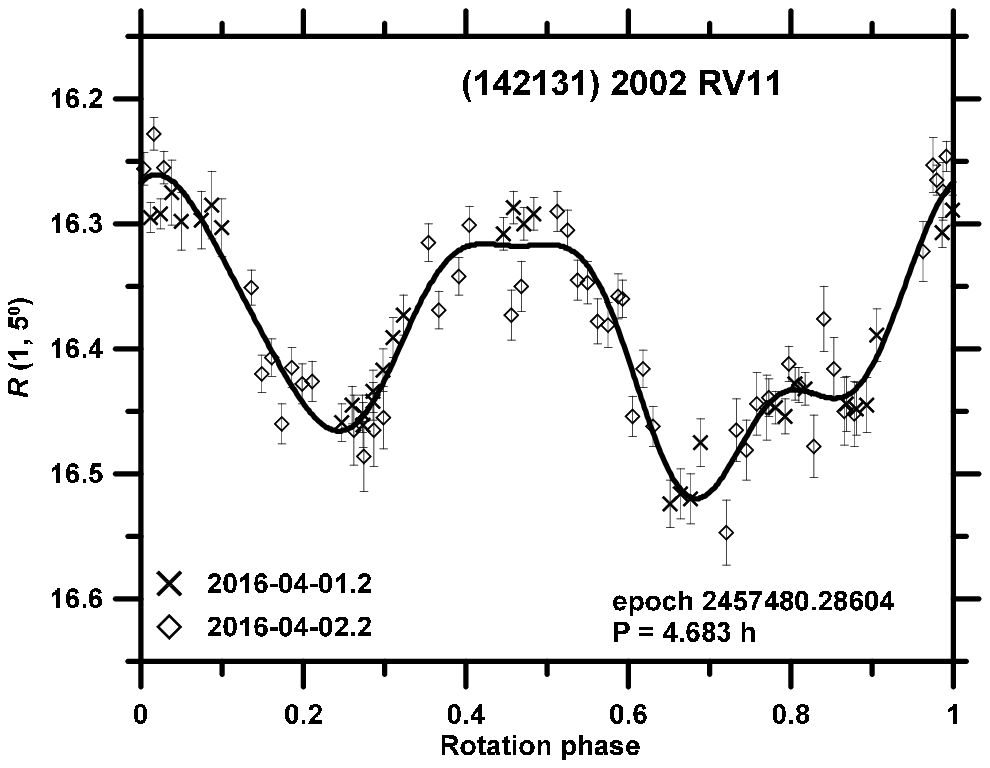}
\caption{\rm Composite lightcurve of (142131) 2002 RV11 from 2016.
}
\label{142131_16lc}
\vspace{1cm}
\end{figure}

This is a secure pair.  Backward integrations of their heliocentric orbits suggest that these two asteroids separated about 140~kyr ago (Fig.~\ref{60677enchist}).
The period of (60677), $P_1 = 3.6274 \pm 0.0008$~h is likely, but a value twice that with 4 pairs of lightcurve maxima/minima per rotation is not entirely ruled out (Figs.~\ref{60677_12lc} and \ref{60677_18lc}).
The period of (142131), $P_2 = 4.683 \pm 0.008$~h is likely; a value twice that does not appear plausible (Fig.~\ref{142131_16lc}).
From the SDSS colors, we derived that the secondary belongs to S, A or L class.
However, a particularly interesting feature of this pair is that it has a low $\Delta H = 0.27 \pm 0.05$, i.e., an anomalously high mass ratio $q = 0.69 \pm 0.05$.
This is not predicted for an asteroid pair with fast rotating primary by the theory of rotational fission.
We will discuss this anomalous case, together with other three similar ones, in Section~\ref{P1qdistrSect}.

\clearpage

\bigskip
\subsection{(60744) 2000 GB93 and (218099) 2002 MH3}
\label{60744sect}

\begin{figure}
%\vspace{1cm}
\includegraphics[width=\textwidth]{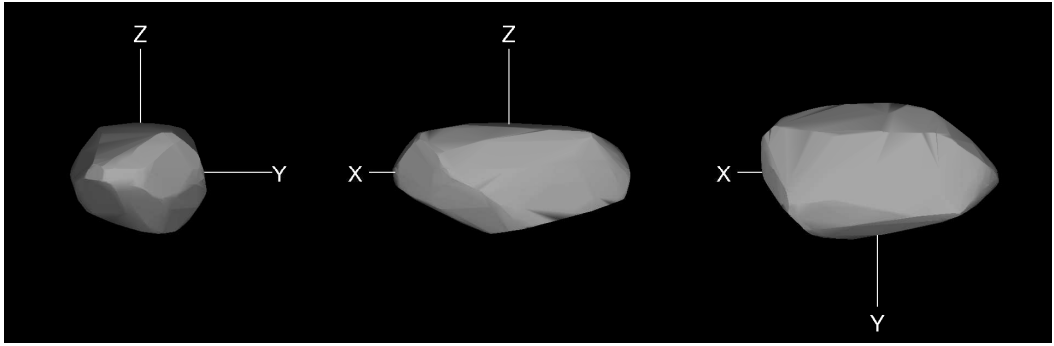}
\caption{\rm Convex shape model of (60744) 2000 GB93 for the pole solution $(L_1,B_1) = (202^\circ, -69^\circ)$.
}
\label{60744model}
\vspace{1cm}
\end{figure}

Backward integrations of their heliocentric orbits suggest that these two asteroids separated about 350~kyr ago (Suppl.~Fig.~121).
%(Fig.~\ref{60744enchist}).
We measured the color indices $(V-R)_1 = 0.480 \pm 0.010$ and $(V-R)_2 = 0.485 \pm 0.010$.
From the SDSS measurements, we classify the primary as an S type.
%We measured the primary's mean absolute magnitudes $H_1 = 15.44 \pm 0.04, 15.40 \pm 0.04, 15.42 \pm 0.03$ and $15.48 \pm 0.04$ in 2012--2013, 2014, 2015 and 2018, respectively, with the phase relation slope parameter $G = 0.27 \pm 0.04$ measured in 2012--2013.
%We measured the secondary's mean absolute magnitudes $H_2 = 16.46 \pm 0.03, 16.46 \pm 0.03$ and $16.41 \pm 0.03$ in 2013--2014, 2015 and 2016, respectively, with the phase relation slope parameter $G = 0.21 \pm 0.02$ measured in 2016.
We also derived its retrograde spin vector (see Table~\ref{AstPairsPolestable}).
The best-fit convex shape model is shown in Fig.~\ref{60744model}.

\clearpage

\bigskip
\subsection{(69142) 2003 FL115 and (127502) 2002 TP59}
\label{69142sect}

\begin{figure}
%\vspace{1cm}
\includegraphics[width=\textwidth]{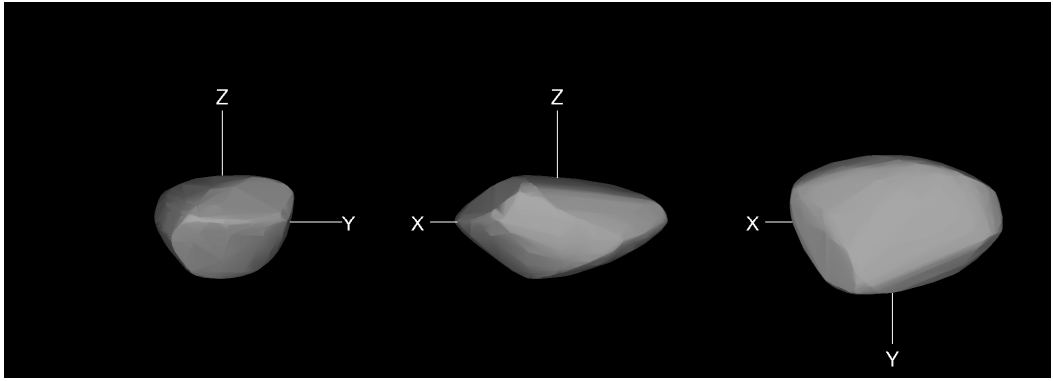}
\caption{\rm Convex shape model of (69142) 2003 FL115 for the pole solution $(L_1,B_1) = (90^\circ, +55^\circ)$.
}
\label{69142model}
\vspace{1cm}
\end{figure}

Despite the somewhat increased distance between these two asteroids ($d_{\rm mean} = 17.1$~m/s) and its relatively high estimated about 1~Myr (Suppl.~Fig.~134), it appears to be a real pair.
For the primary (69142), we derived its prograde spin vector (see Table~\ref{AstPairsPolestable}).
The best-fit convex shape model is shown in Fig.~\ref{69142model}.

\clearpage

\bigskip
\subsection{(76148) 2000 EP17 and (56048) 1998 XV39}
\label{56048sect}

\begin{figure}
%\vspace{1cm}
\includegraphics[width=\textwidth]{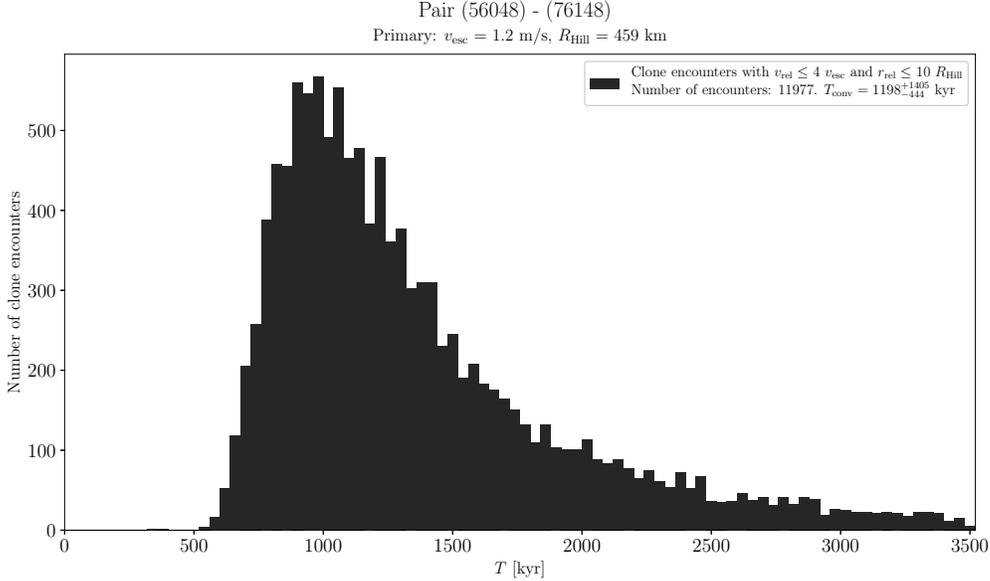}
\caption{\rm Distribution of past times of close and slow primary--secondary clone encounters for the asteroid pair 76148--56048.
}
\label{56048enchist}
\vspace{1cm}
\end{figure}

\begin{figure}
%\vspace{1cm}
\includegraphics[width=\textwidth]{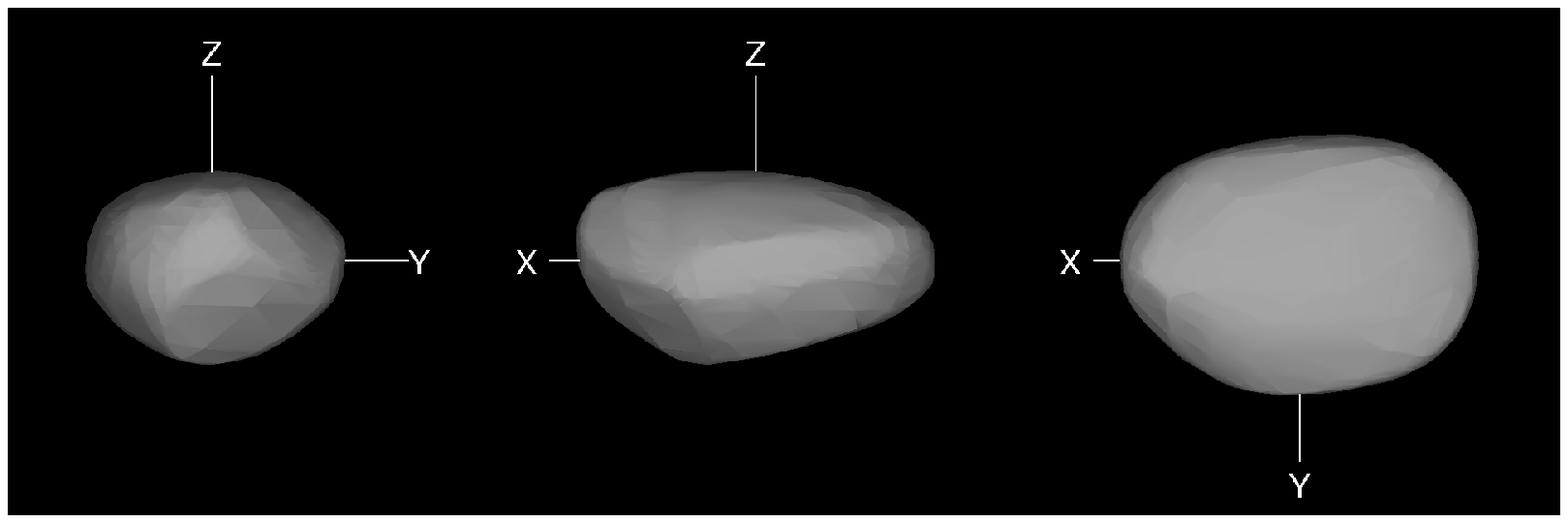}
\caption{\rm Convex shape model of (56048) 1998 XV39 for the pole solution $(L_1,B_1) = (83^\circ, +31^\circ)$.
}
\label{56048model_1}
\vspace{1cm}
\end{figure}

\begin{figure}
%\vspace{1cm}
\includegraphics[width=\textwidth]{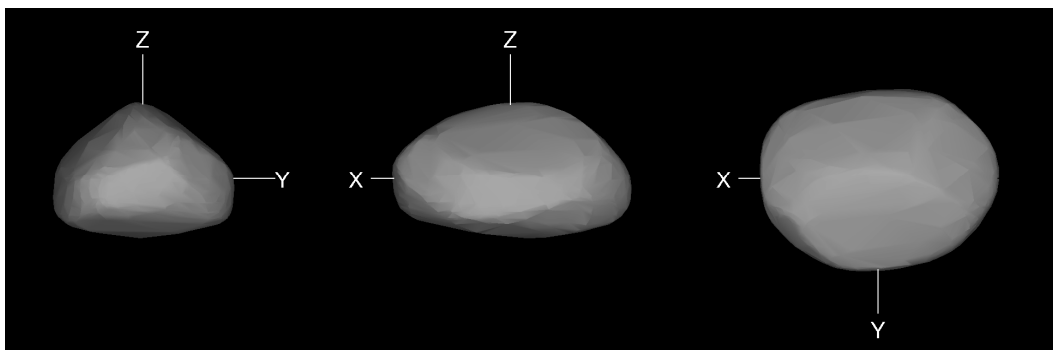}
\caption{\rm Convex shape model of (56048) 1998 XV39 for the pole solution $(L_1,B_1) = (267^\circ, +10^\circ)$.
}
\label{56048model_2}
\vspace{1cm}
\end{figure}

This is a secure pair.
Backward integrations of their heliocentric orbits suggest that these two asteroids separated about 1~Myr ago (Fig.~\ref{56048enchist}).
For (76148), we found a long period with the formal best fit $P_1 = 65.33 \pm 0.09$~h, but it is not an unique solution and its exact period has to be derived from future observations (Suppl.~Fig.~144).
For (56048), we derived its prograde spin pole (with two mirror solutions in longitude, see Table~\ref{AstPairsPolestable}).
The best-fit convex shape models for the two pole solutions are shown in Figs.~\ref{56048model_1} and \ref{56048model_2}.
We also measured their color indices $(V-R)_1 = 0.502 \pm 0.017$ and $(V-R)_2 = 0.481 \pm 0.010$.
What is, however, particularly interesting is that these two asteroids have about the same absolute magnitudes; $\Delta H = 0.08 \pm 0.20$, i.e., the mass ratio $q = 0.90^{+0.28}_{-0.22}$.
This is not predicted for an asteroid pair formed by rotational fission.
We will discuss this anomalous case, together with other three similar ones, in Section~\ref{P1qdistrSect}.

\newpage

\newpage

\clearpage

\subsection{(80218) 1999 VO123 and (213471) 2002 ES90}
\label{80218sect}

\begin{figure}
%\vspace{1cm}
\includegraphics[width=\textwidth]{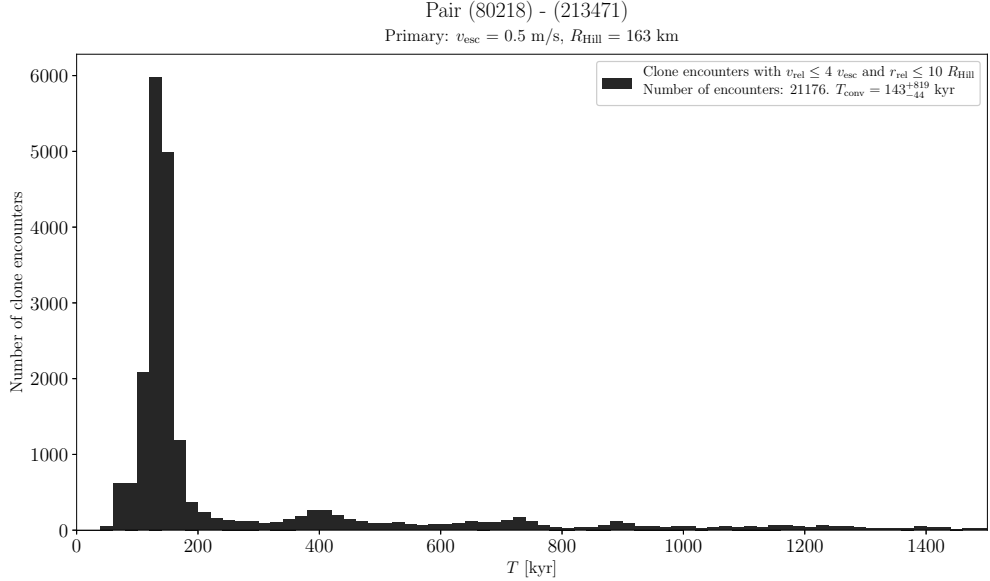}
\caption{\rm Distribution of past times of close and slow primary--secondary clone encounters for the asteroid pair 80218--213471.
}
\label{80218enchist}
\vspace{1cm}
\end{figure}

\begin{figure}
%\vspace{1cm}
\includegraphics[width=\textwidth]{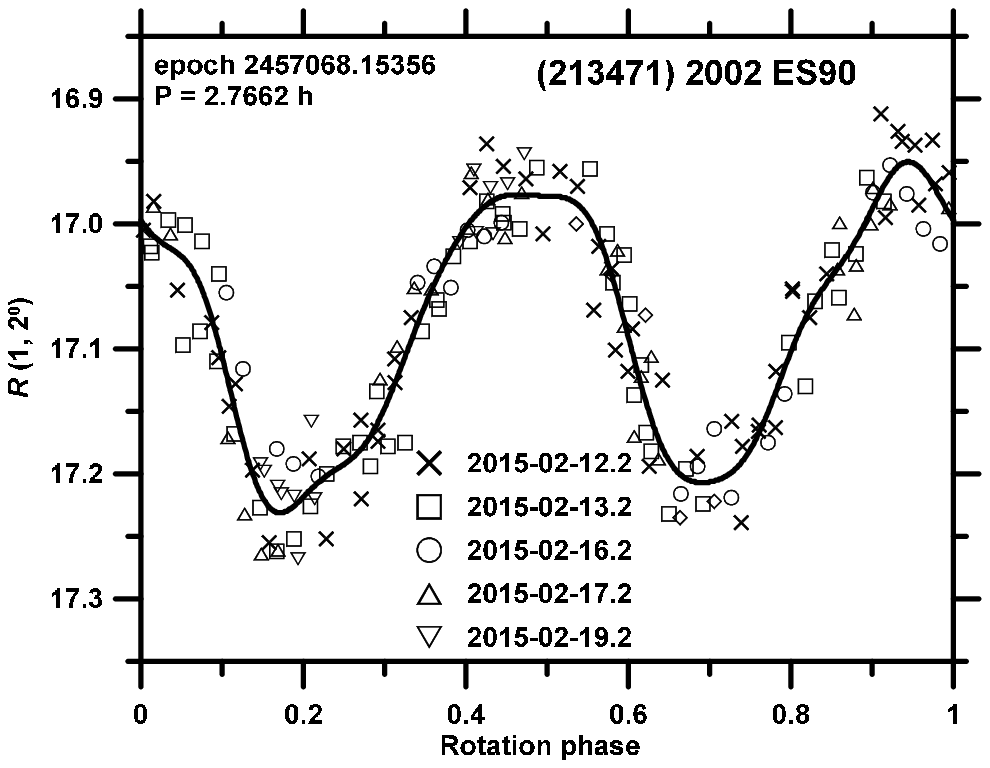}
\caption{\rm Composite lightcurve of (213471) 2002 ES90 from 2015.
}
\label{213471_15lc}
\vspace{1cm}
\end{figure}

This is a secure asteroid pair, showing an orbital convergence about 140~kyr ago (Fig.~\ref{80218enchist}).
We found that the primary (80218) 1999 VO123 is a binary system.
The satellite (bound secondary) has a secondary-to-primary mean diameter ratio $D_{1,s}/D_{1,p} = 0.32 \pm 0.02$,
an orbital period of $33.10 \pm 0.05$~h, and it is synchronous, i.e., its rotational period $P_{1,s}$ is equal to the orbital period (see Pravec et~al.~2016).
The primary's rotational period $P_{1,p} = 3.1451 \pm 0.0002$~h is likely; a period twice as long with 4 pairs of lightcurve maxima and minima per rotation is formally not ruled out, but it appears unlikely.
The period of the unbound secondary (213471), $P_2 = 2.7662 \pm 0.0003$~h is well established (Fig.~\ref{213471_15lc}).
We measured their color indices $(V-R)_1 = 0.403 \pm 0.010$ and $(V-R)_2 = 0.410 \pm 0.023$; an excellent agreement.
However, it is even more interesting, as this asteroid pair has an anomalously low $\Delta H = 0.28 \pm 0.06$.  After a correction of $H_1$ for the satellite presence, it is $\Delta H = 0.17 \pm 0.06$, i.e., it is a high mass ratio $q = 0.79 \pm 0.07$.
We will discuss this very interesting asteroid system in Section~\ref{P1qdistrSect}.

\clearpage

\subsection{(101703) 1999 CA150 and (142694) 2002 TW243}
\label{101703sect}

\begin{figure}
%\vspace{1cm}
\includegraphics[width=\textwidth]{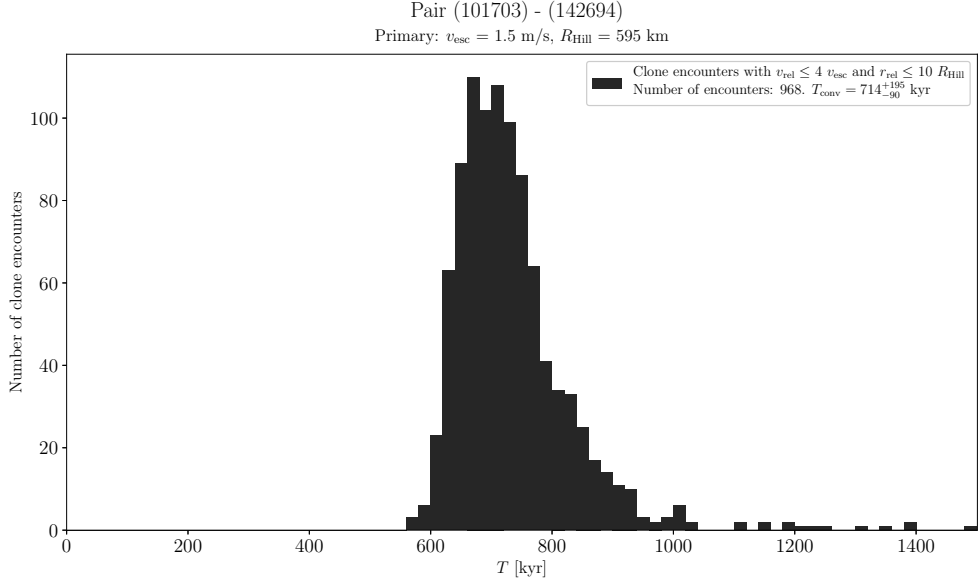}
\caption{\rm Distribution of past times of close and slow primary--secondary clone encounters for the asteroid pair 101703--142694.
}
\label{101703enchist}
\vspace{1cm}
\end{figure}

\begin{figure}
%\vspace{1cm}
\includegraphics[width=\textwidth]{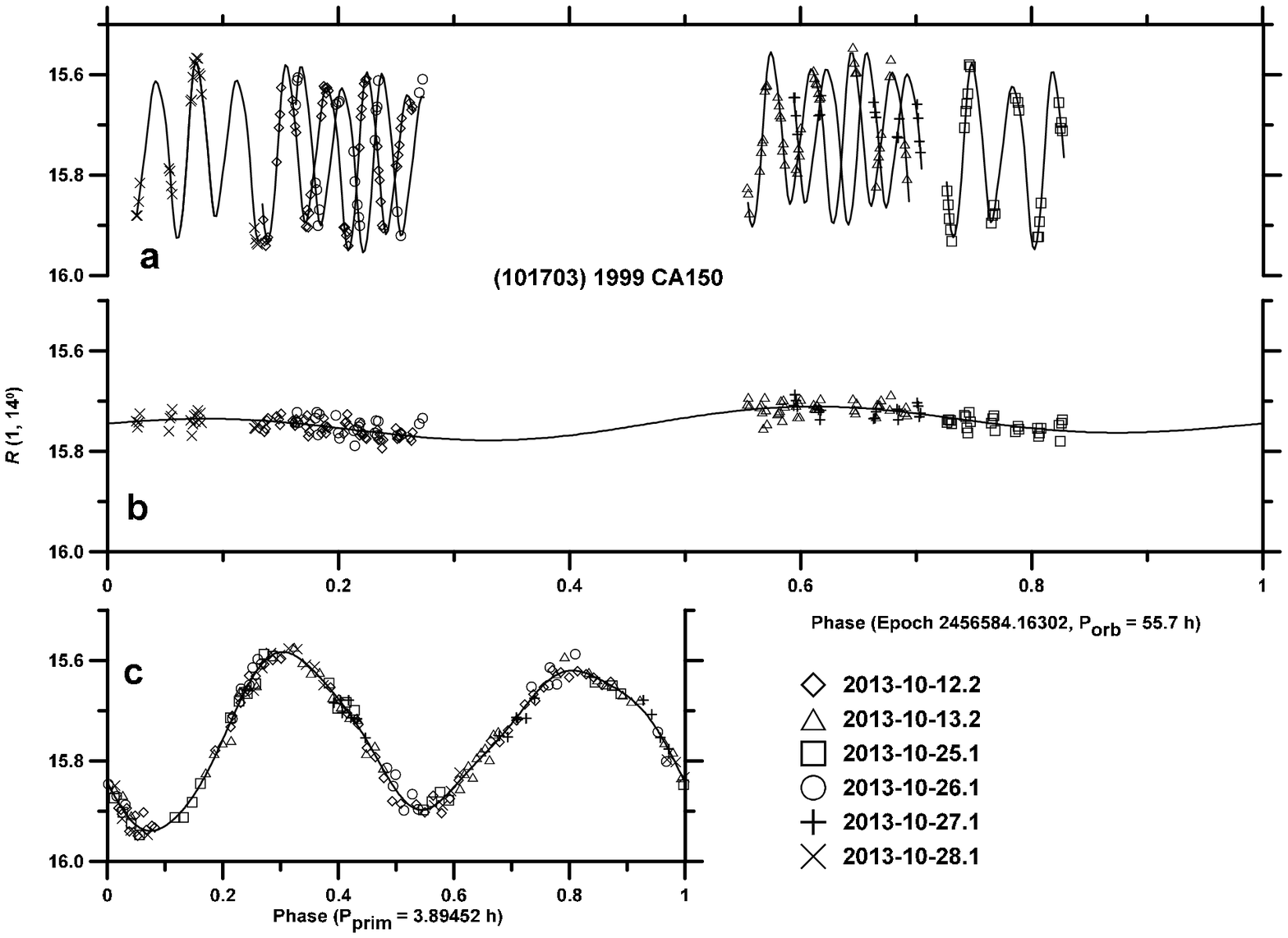}
\caption{\rm Lightcurve data of (101703) 1999 CA150 from 2013.
(a) The original data showing all the lightcurve components, folded with the orbital period.
(b) The secondary rotational lightcurve component, derived after subtraction of the primary lightcurve component.  Mutual events were not detected.
(c) The primary lightcurve component.
}
\label{101703_13lc}
\vspace{1cm}
\end{figure}

Despite the large distance between these two asteroids in mean elements ($d_{\rm mean} = 63.5$~m/s), caused by interaction with the $g + g_5 - 2g_6$ secular resonance, this appears to be a real pair.
%YOUR ASSESSMENT, DAVID?
Backward integrations of their heliocentric orbits suggest its age about 700~kyr (Fig.~\ref{101703enchist}).
The primary (101703) was found to be an Sw or Q type (Polishook et~al.~2014).
Its rotation period $P_1 = 3.8948 \pm 0.0004$~h is well established.
%Its $(V-R)_1 = 0.482 \pm 0.010$.
However, in the data of 2009-09-20.2, there appeared a brightness attenuation that could be a mutual event due to a satellite of the primary (see Suppl.~Fig.~38 in Pravec et~al.~2010).
Moreover, in our accurately calibrated data taken in October 2013, there appears to be a second lightcurve component with a period on an order of 50--60~h that might be a rotational lightcurve of the suspected satellite (Fig.~\ref{101703_13lc}).
The suggested binary nature of (101703) will have to be confirmed with future observations.

\clearpage

\subsection{(122173) 2000 KC28 and (259585) 2003 UG220}
\label{122173sect}

\begin{figure}
%\vspace{1cm}
\includegraphics[width=\textwidth]{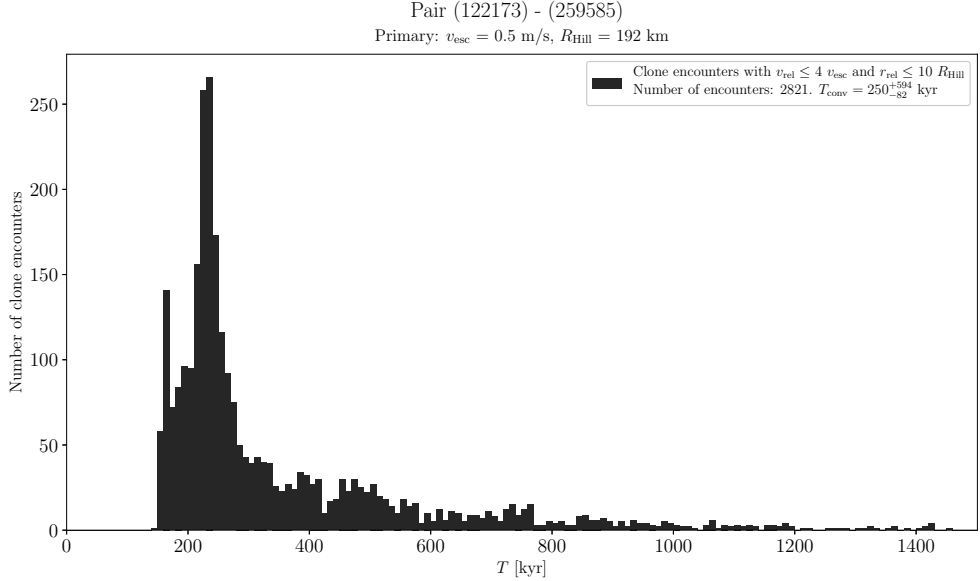}
\caption{\rm Distribution of past times of close and slow primary--secondary clone encounters for the asteroid pair 122173--259585.
}
\label{122173enchist}
\vspace{1cm}
\end{figure}

\begin{figure}
%\vspace{1cm}
\includegraphics[width=\textwidth]{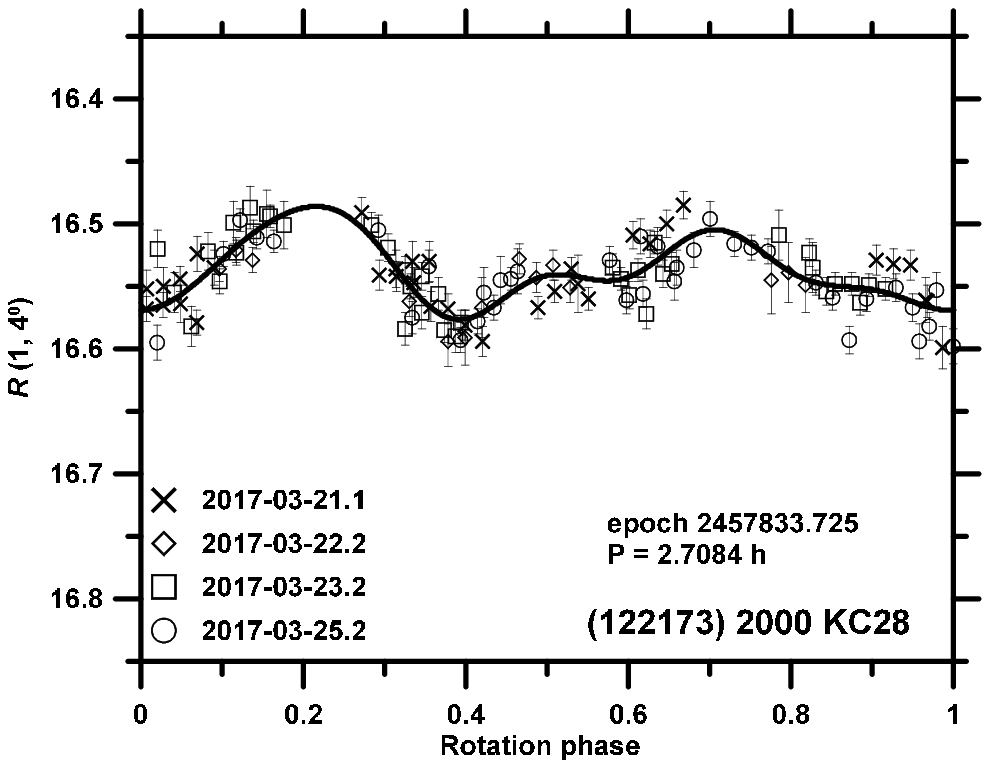}
\caption{\rm Composite lightcurve of (122173) 2000 KC28 from 2017.
}
\label{122173_17lc}
\vspace{1cm}
\end{figure}

\begin{figure}
%\vspace{1cm}
\includegraphics[width=\textwidth]{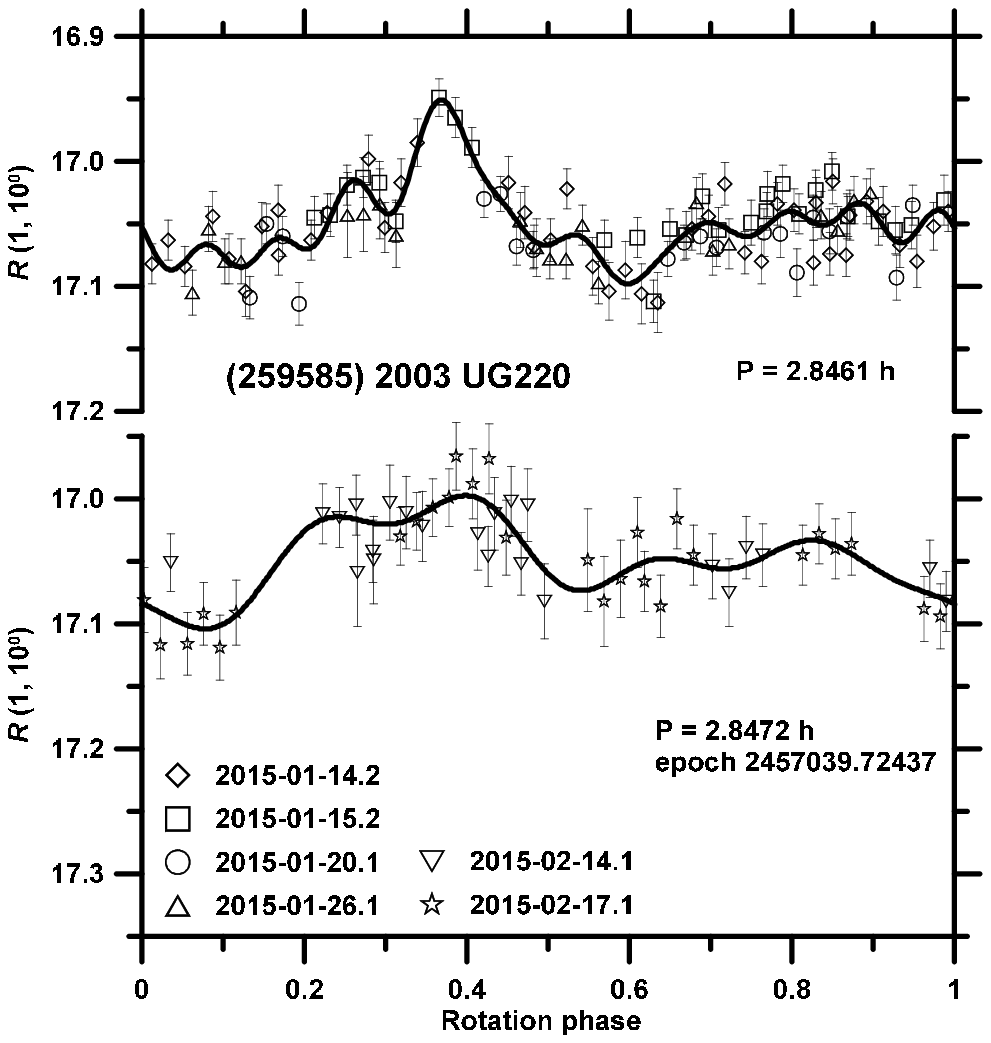}
\caption{\rm Composite lightcurves of (259585) 2003 UG220 from 2015.
}
\label{259585_15lc}
\vspace{1cm}
\end{figure}

This is a secure pair.  Backward integrations of their heliocentric orbits suggest that these two asteroids separated about 250~kyr ago (Fig.~\ref{122173enchist}).
The period of (122173), $P_1 = 2.7084 \pm 0.0009$~h is likely, but half-integer multiples of it are not ruled out (Fig.~\ref{122173_17lc}).
The period of (259585) was not accurately established, a value of $P_2 = 2.8461 \pm 0.0003$~h appears possible (Fig.~\ref{259585_15lc}), but some longer periods are possible as well.
We measured their color indices $(V-R)_1 = 0.443 \pm 0.016$ and $(V-R)_2 = 0.446 \pm 0.016$; an excellent agreement.
However, a particularly interesting feature of this pair is that it has a low $\Delta H = 0.35 \pm 0.04$, i.e., an anomalously high mass ratio $q = 0.62 \pm 0.04$.
This is not predicted for an asteroid pair with fast rotating primary by the theory of rotational fission.
We will discuss this anomalous case, together with other three similar ones, in Section~\ref{P1qdistrSect}.

\clearpage
\section{Albedos, colors and taxonomic classes}
\label{AlbedoColorTaxonSect}

\begin{figure}
%\vspace{1cm}
\includegraphics[width=\textwidth]{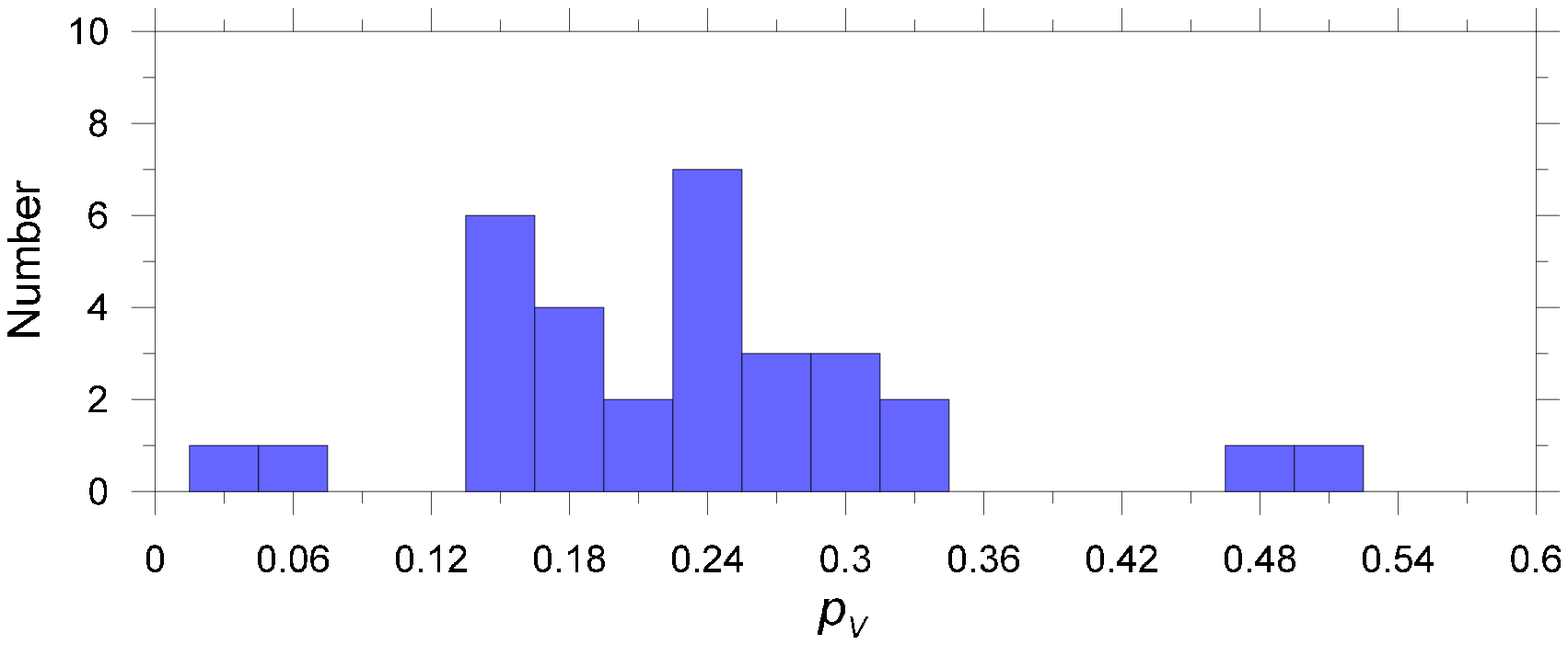}
\caption{\rm Geometric albedos $p_V$ of asteroid pairs.
}
\label{pVfig}
\vspace{1cm}
\end{figure}

\begin{figure}
%\vspace{1cm}
\includegraphics[width=\textwidth]{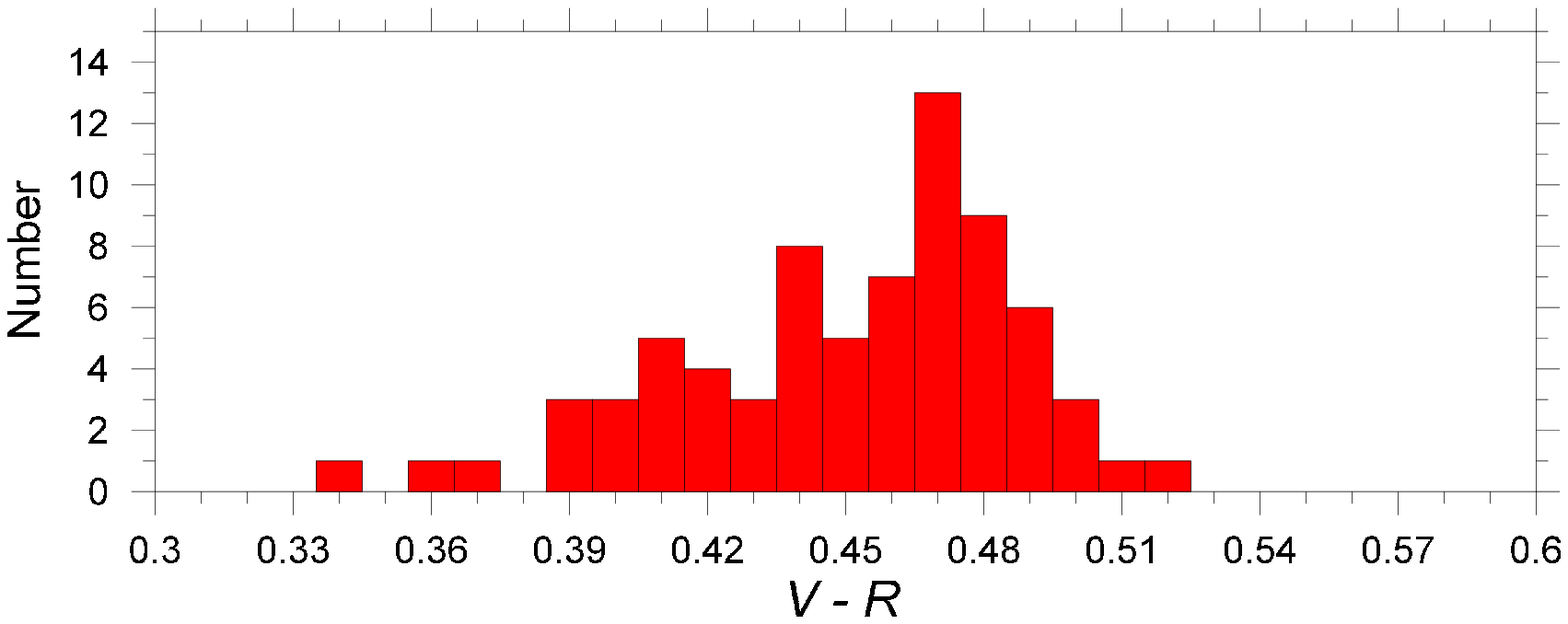}
\caption{\rm Color indices $(V - R)$ of asteroid pairs.
}
\label{VmRfig}
\vspace{1cm}
\end{figure}

\begin{figure}
%\vspace{1cm}
\includegraphics[width=\textwidth]{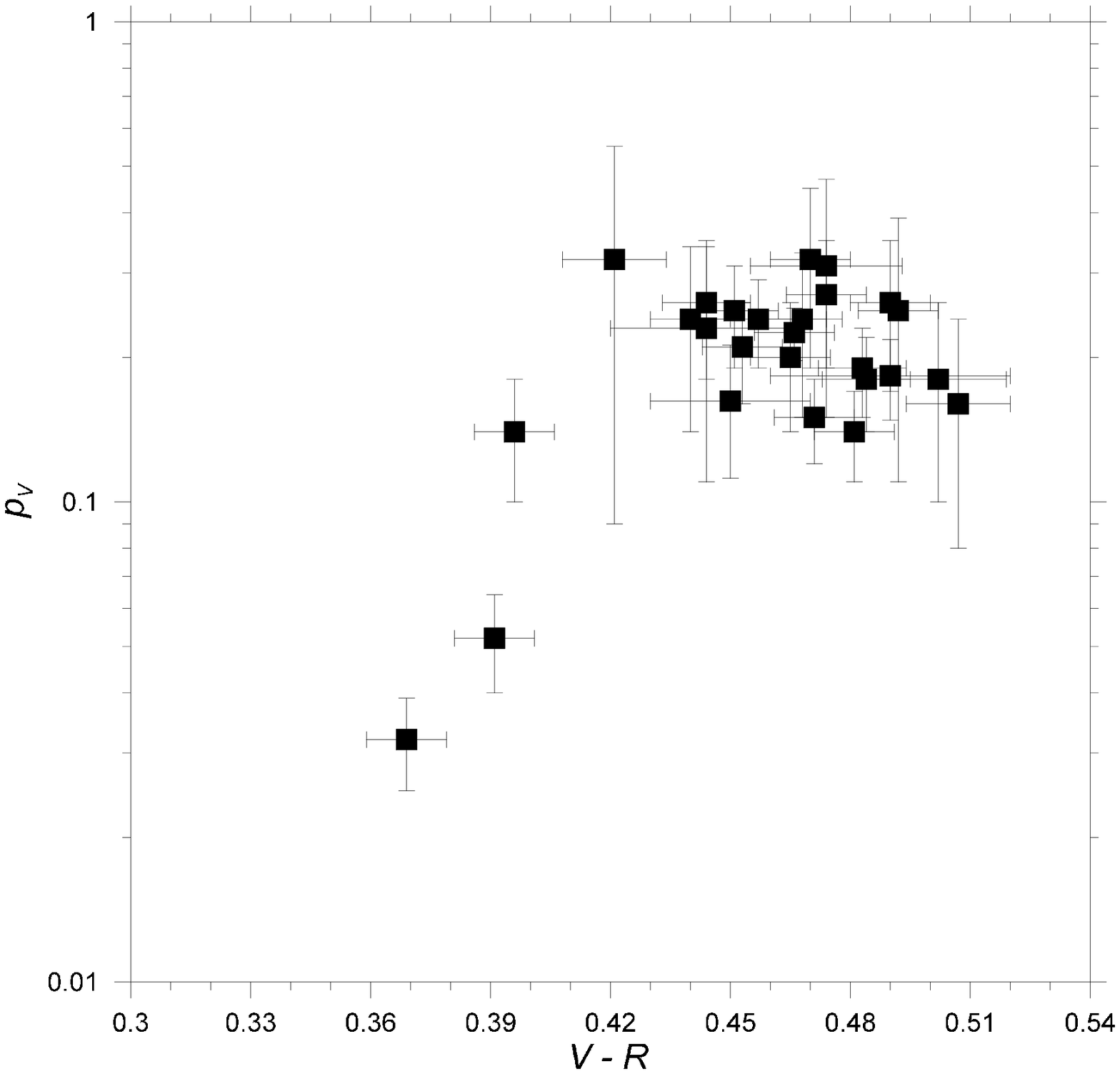}
\caption{\rm Geometric albedos $p_V$ versus color indices $(V - R)$ of asteroid pairs.
}
\label{pV_VmRfig}
\vspace{1cm}
\end{figure}

We obtained geometric albedos $p_{V,1}$, refined from the WISE data (Masiero et~al.~2011) with our accurate absolute magnitudes $H_1$ using the method described in Pravec et~al.~(2012b), for 31 asteroid pair primaries.  The data are given in the 3rd column of
Table~\ref{AstPairsTaxColorstable} and they are plotted in Fig.~\ref{pVfig}.  27 of the 31 pairs have medium albedos between 0.14 and 0.32, while 2 and 2 are low- and high-albedo asteroids ($p_{V,1}$ about 0.04 and about 0.49), respectively.
The scarcity of low-albedo asteroid pairs in our sample is suspected to be due to a bias against their detection, as the pair secondaries are mostly small (on an order of 1 km) and the known population of main belt asteroids
at these diameters is heavily biased towards higher albedo objects due to the magnitude-limited sky surveys in the visual spectral range.  For 74 asteroid pairs, we obtained $(V-R)$ color indices for one or both components.
The data are given in the 6th and 7th columns of Table~\ref{AstPairsTaxColorstable} and plotted in Fig.~\ref{VmRfig}.  Nearly 3/4 (53 of the 74) of the asteroid pairs have $(V-R)$ in the range 0.44--0.52, which is a range
predominated by the S complex and where also Q, V and L type asteroids are.
%The tail towards higher $(V-R)$ values, up to 0.52, is likely a mixture of red-end S types and V types,
The tail towards lower $(V-R)$ values, down to 0.34, is likely a mixture of blue-end members of the SQ complex, X types, and neutral reflectance (solar-like color)\footnote{The solar $(V-R)$ is $0.367 \pm 0.006$.}
primitive (C and C-like) types.  From Fig.~\ref{pV_VmRfig} where we plot the $p_{V,1}$ vs $(V-R)$ data for 26 asteroid pairs where we got
both, it is apparent that most (or all, for our specific sample) asteroid pairs with the $(V-R)$ values from 0.42 to 0.51 have medium albedos, consistent with them being (mostly) S/Q/L types (see below).  Two of the three points with neutral to slightly red $(V-R)$ values
from 0.36 to 0.40 are low-albedo objects and one has a medium albedo; the former are probably primitive (C/C-like) types and the latter, (11286) is an X/M type.  Though unique taxonomic classifications cannot
be given from the single-color data, the observed distribution of the albedos and colors is consistent with what we see in the general population of the main asteroid belt if we consider the observational bias against small low-albedo
asteroids.

The taxonomic classifications that we obtained from spectral or color measurements for one or both components of 42 asteroid pairs (given in the 4th and 5th columns of Table~\ref{AstPairsTaxColorstable})
generally confirm the picture suggested from the albedos and $(V-R)$ color indices above.  28 of the 42 are S, Q or L types, 8 are X (mostly E) types,
4 or 5 (one classification is uncertain) are V types, and 1 is a Ch type.  In 9 cases, we obtained taxonomic classifications for both components.  One of the 9 is an X type and 8 belong to the SQ complex.
It is significant that in all the 9 cases, both components of a given pair belong to the same taxonomic complex, with no or only moderate difference between them.  In three cases (see Table~\ref{AstPairsTaxColorstable}),
we see that the secondary is apparently less space weathered, having stronger absorption features (Q, Sq, Sr) than the primary (Sq, S, Sa).  It suggests that in at least some asteroid pairs, the secondary has a fresher surface
than the primary.

\clearpage

\clearpage
\section{Spin and orbit poles}
\label{PolesSect}

We determined spin or orbit pole positions for 19 asteroids or asteroid binaries in 17 asteroid pairs.  The data are presented in Table~\ref{AstPairsPolestable} and the referenced subsections in Section~\ref{IndivPairsSect}.
We analyse them in below.

Of the 17 asteroid pairs with determined poles (for at least one pair member), 7 are prograde and 10 retrograde.
The slight preference for retrograde poles is not statistically significant.

There is an apparent concentration of the determined asteroid pair poles to the north and south poles of the ecliptic.
However, its reality needs to be confirmed with simulations of selection effects affecting the sample; there is present an observational/modeling bias against asteroids with high obliquities.
Note that for asteroids with low obliquities, their poles can be uniquely (in latitude at least) derived with less amount of data typically than for asteroids with obliquities close to 90\dg (for that the latter have lower mean observed lightcurve amplitudes on average,
due to the projection effect).
The selection effect is even stronger for asteroid binaries for which we need to observe typically 3 apparitions with mutual events in order to derive an unique orbital pole.
For binaries with high obliquities that are in eclipse/occultation geometry and therefore show mutual events in only some apparitions, it means that we need to observe them in many more apparitions
than for binaries with low obliquities which show mutual events most or all the time.
Such multi-apparition data are not available for the binary systems in our asteroid pairs sample yet.  Hence all the four binaries
in Table~\ref{AstPairsPolestable} have $B_1$ close to $+90^\circ$ or $-90^\circ$.
For three other paired binaries that we re-observed in their 2nd or 3rd~apparition (after the discovery of their binary nature in the 1st~apparition),
namely (6369), (8306) and (43008) (see their subsections), there were not present mutual events in the return apparitions, indicating that their obliquities are not low.  Note that for general main-belt asteroid binaries,
we simulated the observational selection effects and found a real strong concentration of their poles towards the ecliptic poles (Pravec et~al.~2012a), but we will need to do the de-biasing for paired binaries when sufficient data are available in future.

For two pairs, 2110--44612 and 6070--54827, we determined poles of both pair members.  The important finding is that in each of the two pairs, both members have the same sense of rotation (retrograde in both cases).
The pair 6070--54827 was studied in detail by Vokrouhlick\'y et~al.~(2017) who found that the original spin vectors of the two asteroids were not co-linear but tilted by $38^\circ \pm 12^\circ$
at the time of their separation.  In the case of 2110-44612, we found that though the 3-$\sigma$ uncertainty areas of their pole positions overlap (see Suppl.~Figs.~1 and 2), their nominal pole positions (for both mirror solutions)
differing by about 90\dg in longitude call for a more detailed look into the long-term evolution of asteroid spin poles.  We present it in following.

From our observations, we determine an asteroid spin vector at the present epoch.
In the optimum situation, we have
information on the spin axes of both components of an asteroid pair.
However, we want to know what were their spin axes right after formation of the pair, as such data
can be confronted with predictions from the theories of asteroid pair formation.
A question arises to what degree the current spin configuration preserves
the original configuration.  Here we give a brief look to
this problem, having in mind the sample of paired asteroids
for which we derived their spin poles (Table~\ref{AstPairsPolestable}). These bodies
are large enough, and their estimated ages short enough, so that we may neglect
non-gravitational effects, such as YORP, in the first approximation. However,
the spin axis orientation
is affected by gravitational torques, primarily from the Sun,
which may tilt the original spin vectors on a timescale comparable to
or shorter than the pair ages. Therefore, we must consider
their effect for the pairs.

Breiter et~al.~(2005) presented an efficient numerical implementation of
the secular spin dynamics due to the solar torques. Because this problem
is tightly coupled with orbital dynamics, in particular changes in orientation
of the orbital plane in the inertial space, we also need to propagate the
heliocentric orbit over the same interval of time for which we seek the spin
evolution. To infer this information, we embedded the secular model of
Breiter et~al.~(2005) into the symplectic orbit-evolution package {\tt swift}
(e.g., Levison and Duncan 1994).
The initial data for the orbit integration are taken from the {\tt AstDyS}
web site. We first integrate the nominal orbit of the asteroid backward in time
to reach the inferred epoch of the pair formation. At that moment, we set
up initial data for the spin integration and propagate both the orbit and spin
forward in time to the current date. The spin integration requires the
value of the precession constant $\alpha$ to be known. Apart from the orbital
data, $\alpha$ depends also on the spin rate $\omega$ and the dynamical
flattening $\Delta$ of the asteroid. The spin rate is known well from our
observations, but the flattening
is determined only approximately with our convex shape model. In
particular, $\Delta=(2C-A-B)/2C$ where $(A,B,C)$ are the principal values of the
inertia tensor. We assume a homogeneous density distribution in the asteroid when estimating
these values. However, an analysis of shape variants that also satisfy the
photometric data implies that $\Delta$ is known with only $\simeq
15-30$\% accuracy typically for asteroid convex shape models.
The situation is even more complicated when the asteroid
has a satellite. The effective $\Delta$ value then depends also on
the orbital and physical parameters of the satellite (see Sec.~5.1.1 of
Pravec et~al.~2012a). The necessary data are taken from our observations
(Table~\ref{AstPairsPolestable}) that allow to estimate the required quantities. As above, the
value of $\Delta$ is known with poor accuracy, contributing by the largest
value to the uncertainty in the precession constant $\alpha$. In order to
sample possible outcomes of the spin evolution, we propagate different variants
for which the precession constant has been fractionally changed by
$15-30$\% from the nominal value.

\begin{figure}
\begin{center}
 \includegraphics[width=10cm]{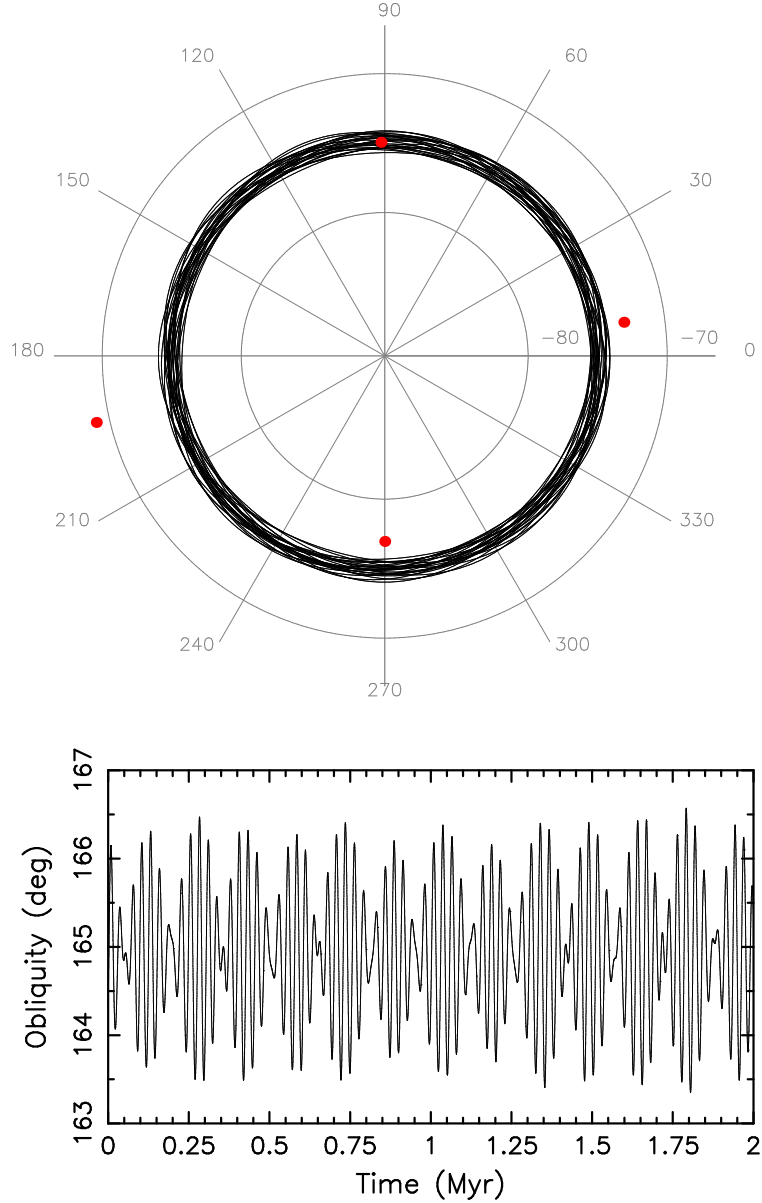}
\end{center}
\caption{An example of spin axis evolution for (2110) Moore-Sitterly
 over $2$~Myr interval of time (comparable to the estimated age of the
 2110--44612 pair). Top: Motion of the spin axis in the ecliptic longitude and latitude
 in polar projection (the south ecliptic pole in the center).   The four
 red symbols indicate the nominal pole solutions (two mirror solutions each) for (2110) and (44612).
%The gray lines show isolines of constant ecliptic longitude (in $30^\circ$ steps), and isolines of constant ecliptic latitudes ($-80^\circ$,  inner circle, and $-70^\circ$, outer circle).
 Bottom: Time evolution of the obliquity. Initial data is close to the first nominal pole solution of (2110) Moore-Sitterly (Table~\ref{AstPairsPolestable}).
 \label{2110_spin}}
\vspace{1cm}
\end{figure}

The simplest situation occurs for very young pairs, for which the today's pole
orientation may still be relatively close to its original value. For instance, in the
case of the 6070--54827 pair, the relative configuration of the primary and
secondary asteroid poles did not change much. This is because (6070) has
a nearly stationary spin axis directed at the south ecliptic pole, while the
spin pole of (54827) performed only a $\sim 90^\circ$ displacement
along a constant ecliptic latitude. When the age is higher, and the rotation
sense is retrograde, the regular precession may smear the pole location to
all possible ecliptic longitudes. As an example, we note that it is important
to consider this effect
when interpreting our result for the 2110--44612 asteroid pair. In this case,
our solution gives the nominal poles of both components having about the
same ecliptic latitude and the ecliptic longitudes about $90^\circ$ different.
This result is, however, entirely compatible with
an initially co-linear configuration of the spins of the primary and the
secondary. Figure~\ref{2110_spin} shows an example of the evolution of the
pole orientation
for the primary (2110)~Moore-Sitterly, assuming an age of $2$~Myr and
the nominal value $\Delta=0.33$. Note that the
obliquity performs only small oscillations (keeping the ecliptic latitude
nearly constant), while the ecliptic longitude of the asteroid spin axis
undergoes regular precession around the south ecliptic pole with a period of
about $75$~kyr. The evolutionary track of the secondary's spin pole is
similar with only a different value of the precession period
(because of its different value of the rotation period and dynamical
flattening). So their poles diverge in the ecliptic
longitude to acquire any longitudinal difference at a future time. That said, we obviously
cannot prove initial co-linearity of the rotation poles in the
2110--44612 pair, but the data are consistent with that.
In any case, the poles are never more than $\simeq 25^\circ-30^\circ$
apart during their evolution. This is because of their proximity to the south
ecliptic pole and the low inclination of their heliocentric orbits.

However, things may get much wilder for prograde rotating asteroids residing on
high-inclination heliocentric orbits. This is because precession rate
for prograde-rotating asteroids may occur to be in resonance with
precession rate of the orbital plane (e.g., Colombo~1966, Henrard and
Murigande~1987). Specific examples of secular spin axis evolution for
asteroids in the inner main belt, where most of our studied pairs reside,
can be found in Vokrouhlick\'y et~al.~(2006) or Vra\v{s}til and
Vokrouhlick\'y~(2015).

\begin{figure}
\begin{center}
 \includegraphics[width=10cm]{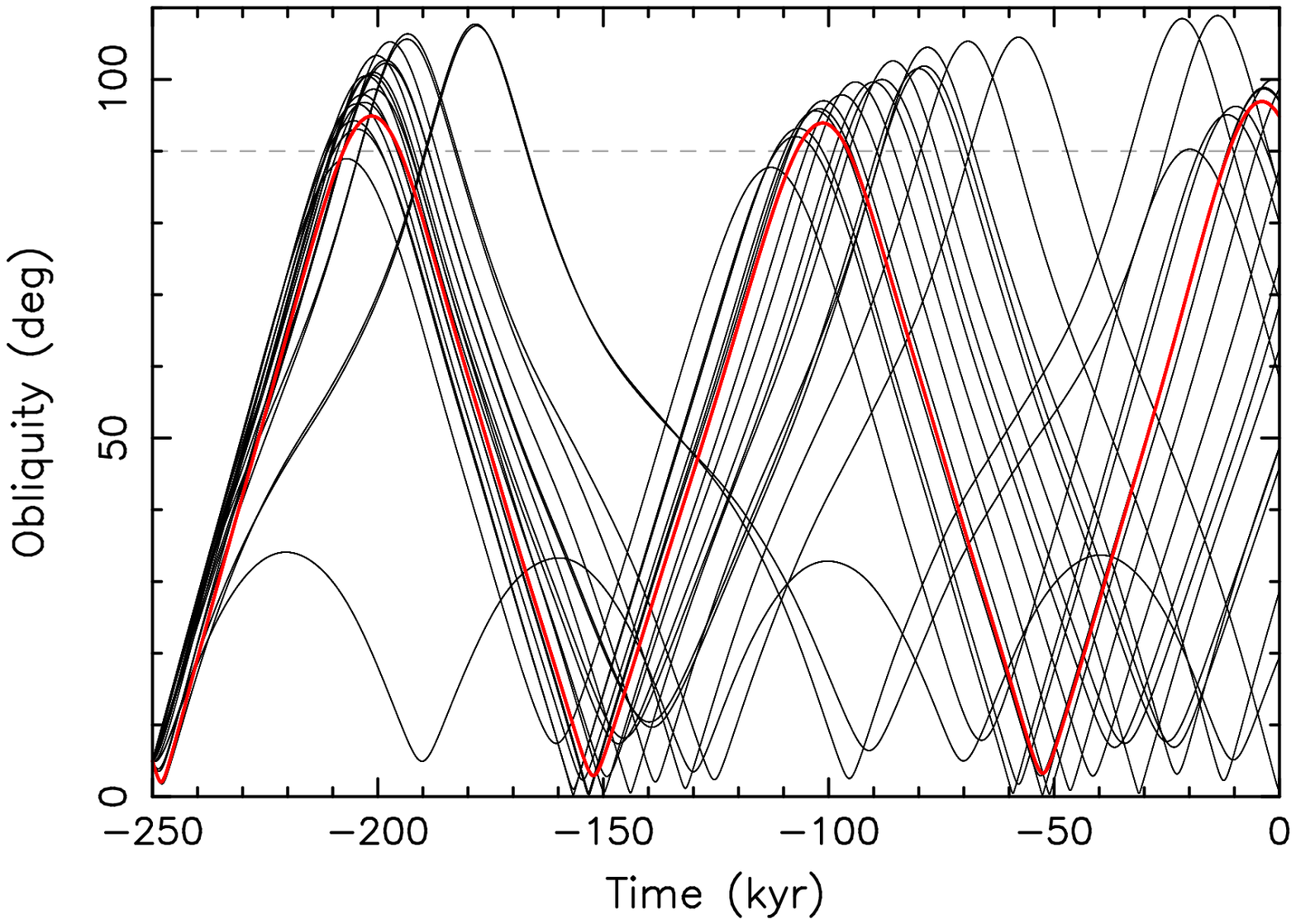}
\end{center}
\caption{Twenty variants (black curves) of possible obliquity
 evolution for (4765)~Wasserburg.
%The abscissa is time in the past, the ordinate shows the obliquity.
 All cases
 were assumed to be formed 250~kyr ago with an obliquity of $5^\circ$.
 The evolutions are different because each of the variant was
 given a slightly different value of the precession constant $\alpha$,
 by assuming a different value of the dynamical flattening $\Delta$, and
 a slightly different longitude of the initial pole. The solution
 highlighted in red terminates close to the currently observed obliquity
 $91^\circ$ of this asteroid. Other variants may, however,
 achieve different today's obliquities in the range $0^\circ-110^\circ$,
 in spite of their common initial obliquity value.
 \label{4675_spin}}
\vspace{1cm}
\end{figure}

The most significant perturbations occur for pairs among Hungaria
asteroids, whose orbital inclinations are high.
To demonstrate the effects, we consider the case of (4765)
Wasserburg, the primary component of the 4765--350715 pair (Section~\ref{4765sect}).
We found that (4765) has a peculiarly large obliquity of $91^\circ$,
which seemed to be at odds with the expected low obliquities of paired
asteroids after YORP-induced fission, close to the YORP asymptotic obliquity values.
For sake of definiteness, we assume
an age of $250$~kyr for the 4765--350715 pair, consistent with its estimated age,
and consider the value $\Delta=0.33$ from
the (4765)'s nominal convex shape model. Most importantly, we assume that the initial
obliquity at the formation of the pair was $5^\circ$ only, very different
from the today's value. With these data, we propagated the nominal model plus 19 clone variants by
taking slightly different $\Delta$ values and initial longitudes of
the spin axis. Evolution of the obliquity for all these cases is
shown in Fig.~\ref{4675_spin}. There are two main features to be
noted: (i) the obliquities oscillate up to $\simeq 110^\circ$, and
(ii) the different cases
quickly diverge, achieving at the current epoch any value
between $0^\circ$ and $110^\circ$ obliquity. This is because the nominal
value of the precession constant $\alpha\simeq 36$~arcsec/yr is close
to the principal frequencies $s\simeq -22.6$~arcsec/yr and $s_6\simeq
-26.3$~arcsec/yr with which the orbital plane precesses in space.
The proper inclination is high, $\simeq 21.2^\circ$, and it implies a
stationary point (Cassini state~2) of the spin-orbit resonance with
the $s$-frequency at about $60^\circ$. A similar stationary point for the
$s_6$-frequency is at about $45^\circ$, enclosed with a resonance zone
spanning from $25^\circ$ to $60^\circ$ obliquity. Interaction of the
two phenomena forms a large chaotic zone for the obliquity evolution
extending to nearly $110^\circ$. As a result, we find that the current position
of the spin pole of (4765) cannot be used to determine its initial spin vector
orientation. We may also imagine that the pole position
of the secondary (350715) could follow one of the tracks shown in
Fig.~\ref{4675_spin} that end at a near-zero obliquity. Therefore, a potential
large angular distance of current poles of the members of
this pair would still be perfectly compatible with them being co-linear right after the pair formation.
The same analysis applies also to the asteroid (69142) 2003~FL115,
the primary component of the 69142--127502 pair.

\begin{figure}
\begin{center}
 \includegraphics[width=10cm]{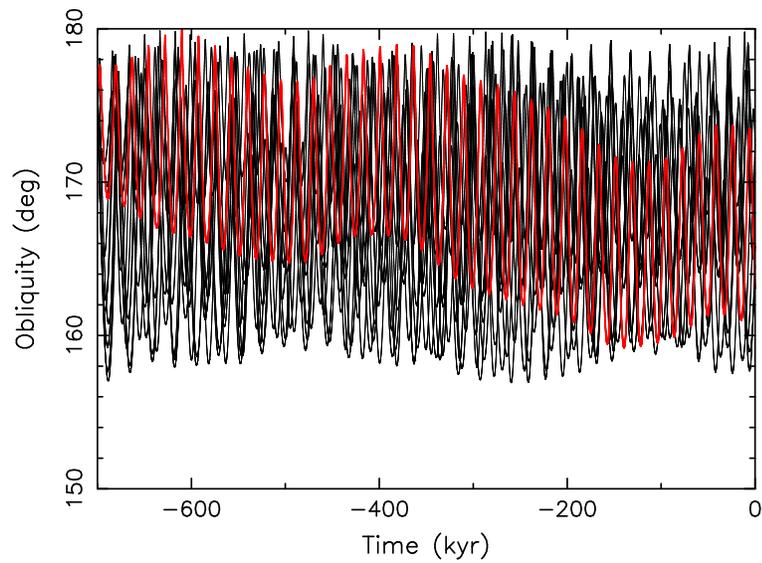}
\end{center}
\caption{Ten variants (black curves) of possible obliquity
 evolution for (25884)~Asai.
% The abscissa is time in the past, the ordinate shows the obliquity.
 All cases
 were assumed to be formed 700~kyr ago with an obliquity of $175^\circ$.
 The individual evolutions are different because each of the variant was
 given a slightly different value of the precession constant $\alpha$,
 by assuming different value of the dynamical flattening $\Delta$, and
 a slightly different longitude of the initial pole. The evolution
 highlighted in red terminates close to the today's observed obliquity
 $162^\circ$ of this asteroid. Other variants may achieve
 different today's obliquities in a relatively narrow range $160^\circ-180^\circ$.
\label{25884_spin}}
\end{figure}

The huge obliquity variation discussed above for (4765) is closely
related to the prograde sense of its rotation. Retrograde rotators
among Hungarias show much smaller effects, principally produced by the
large orbital inclination. We tested this conclusion in the case of
asteroid (25884) Asai, for which we determined the current obliquity
of $\simeq 162^\circ$ (Section~\ref{25884sect} and Table~\ref{AstPairsPolestable}). Adopting an age of $700$~kyr,
compatible with the estimated age of the pair 25884--48527, we repeated our
numerical experiment by propagating 10 possible variants of pole
evolution for (25884), starting with an initial obliquity of $175^\circ$.
The results are shown in Fig.~\ref{25884_spin}. The obliquity oscillations
are now limited to a relatively narrow range between $160^\circ$ and
$180^\circ$. Note that their amplitudes are even smaller than the proper
orbital inclination of $\simeq 20.8^\circ$ and the oscillations have
now a short period of $\simeq 17-20$~kyr. This is because for retrograde
rotators the regular spin precession is opposite to the orbital plane
precession and may not constitute a resonant configuration. As a
result, the today's obliquity should preserve the initial value to within
about $10^\circ$ (which is smaller than the 3-$\sigma$ uncertainty of the pole
solution). Obviously, the ecliptic longitudes of spin poles of members of
a pair older than a half million years can be very different.

Using the tools described above, we analysed all the other paired asteroids with determined spin poles reported in Table~\ref{AstPairsPolestable}.
We found no significant secular
effects that would surpass the solution uncertainty with possibly an
exception of (56048) 1998~XV39. This is the secondary of the anomalous high-mass ratio pair
76148--56048 (see Sections~\ref{56048sect} and \ref{OutlierPairsSect}). We find that its current $\simeq 65^\circ$ to $70^\circ$
obliquity may be acquired from an initially low obliquity through
chaotic evolution due to overlap of the $s$ and $s_6$ secular spin-orbit
resonances.

Our analyses above can be briefly summed up
as follows: The present-epoch pole orientation of a paired asteroid
typically does not represent its original state.  Effects of the
spin evolution over the age of an asteroid pair need to be taken into account.
In the simplest case, the spin dynamics represents only the regular
precession in ecliptic longitude. For low- and mid-latitude pole positions,
this effect may cause a large angular separation of the primary and
secondary poles at the current epoch. In more complicated cases,
generally those of prograde-rotating asteroids residing on high-inclination
heliocentric orbits, the pole evolution may be chaotic, even to a degree preventing
a deterministic connection of the present pole position with its
initial value.  We conclude that each asteroid pole solution needs to be
analysed individually.  Overall, of the 17 pairs in this sample, we found that 13 show low-amplitude
oscillations of their obliquities (smaller than their uncertainties), other 2 ---(25884) and (56048)--- show moderate oscillations (comparable to the uncertainties of their obliquities), and
the last 2 ---the prograde-rotating Hungarias (4765) and (69142)--- show high-amplitude oscillations.

\clearpage

%\newpage
\section{Primary period vs mass ratio distribution, and bound secondaries (paired binaries)}
\label{P1qdistrSect}

\begin{figure}
%\vspace{1cm}
\includegraphics[width=\textwidth]{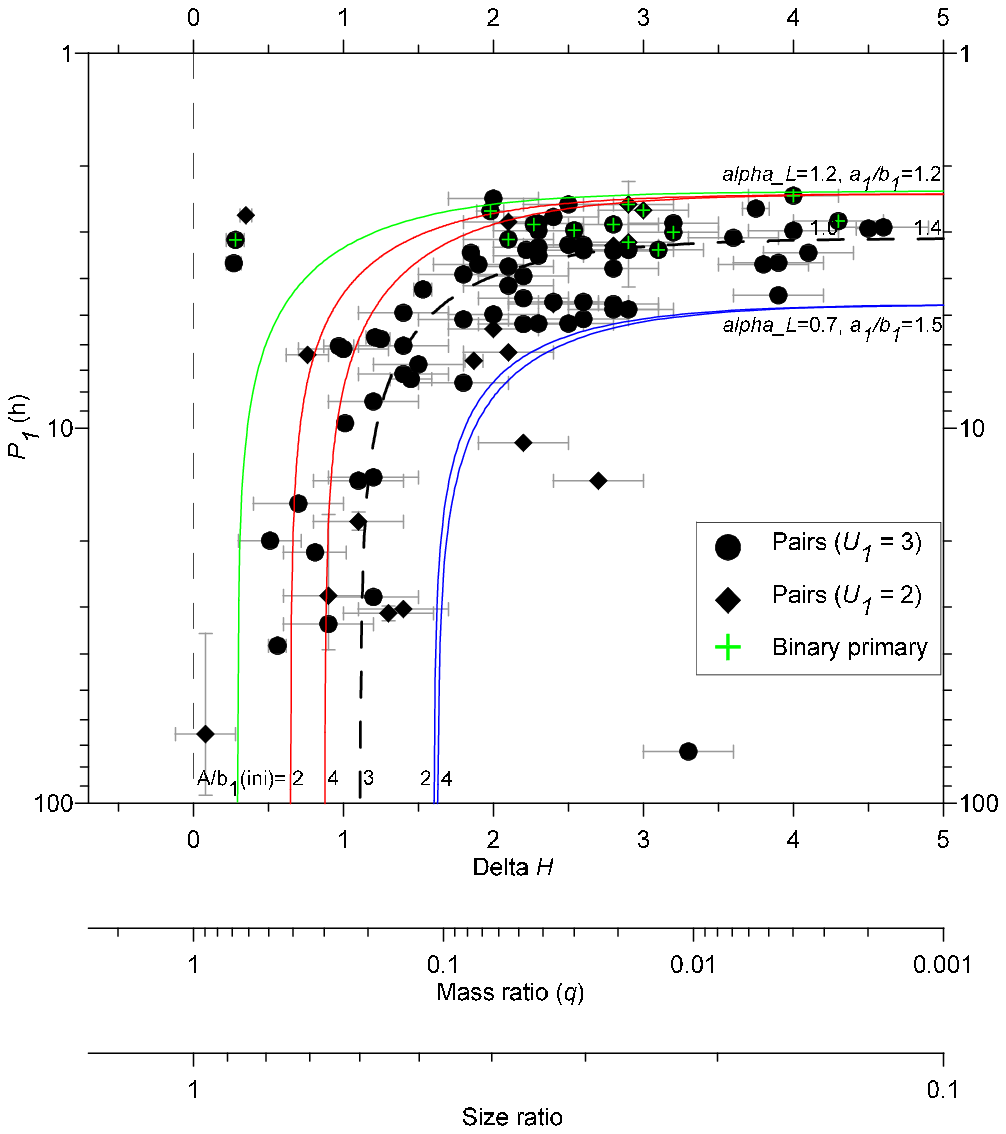}
\caption{\rm Primary rotation periods vs mass ratios of asteroid pairs.
}
\label{P1vsDeltaHfig}
\vspace{1cm}
\end{figure}

In Fig.~\ref{P1vsDeltaHfig}, we plot the primary period vs pair mass ratio data for the 93 studied asteroid pairs.  The filled circles are data with securely determined primary periods (with the period determination quality code $U = 3$),
while the diamonds are data where the primary periods are uncertain by a factor of 1.5 typically, up to 2, usually due to uncertainty of a number of observed lightcurve maxima/minima per rotation ($U = 2$).
The black dashed curve is the nominal relation between the primary period and mass ratio computed from the theory of formation of asteroid pairs by rotational fission, and
the blue, red and green solid curves are theoretical limits (lower or upper) on the primary rotation frequency as derived in Pravec et~al.~(2010, 2018).
Specifically, the black dashed curve
is for the normalized total angular momentum of the system $\alpha_L = 1.0$, the primary's equatorial elongation $a_1/b_1 = 1.4$, $b_1/c_1 = 1.2$,
and the initial relative semi-major axis $A_{\rm ini}/b_1 = 3$.  This set of parameters can be considered as the best representation of pair
parameters.
In particular, the total angular momentum content of 1.0 is about the mean of the distribution of $\alpha_L$ values in small asteroid binaries (Pravec and Harris 2007),
and the axial ratio of 1.4 is about a mean of equatorial elongations of pair primaries suggested by their observed amplitudes.
The red and blue curves represent upper and lower limit cases.
The upper curves are for the system's normalized total angular mometum  $\alpha_L = 1.2$, primary's axial ratio $a_1/b_1 = 1.2$, and
initial orbit's normalized semi-major axis $A_{\rm ini}/b_1 = 2$ and 4.
The lower curves are for  $\alpha_L = 0.7, a_1/b_1 = 1.5$ and $A_{\rm ini}/b_1 = 2$ and 4.
The choice of $a_1/b_1 = 1.2$ for the upper limit cases is because the asteroid pair primaries closest to the upper limit curve have
low amplitudes $\mathcal{A}_1 = 0.1$--0.2~mag.  Similarly, the choice of $a_1/b_1 = 1.5$ for the lower limit cases is because the highest amplitudes
of the points close to the lower limit curve are $\mathcal{A}_1 = 0.4$--0.5~mag, suggesting the equatorial elongations $\sim 1.4$--1.5.
For completness, the green curve gives the theoretical hard upper limit on the final primary spin rate (i.e., lower limit on the period)
as derived in Pravec et~al.~(2018).  As discussed in that paper, the theoretical hard limit was derived involving certain
idealizations that are probably not fulfilled in real asteroids.  In particular, it assumes spherical component shapes while real asteroids are non-spherical.
Thus, real asteroid pairs formed by spin-up fission may stay well below the theoretical hard limit.  Indeed, we see in the plot that the observed
asteroid pairs do not extend to the green curve at $q$ about 0.6, but they are well to the right of the curve
(except for the four anomalous high-mass ratio pairs ---the four leftmost points in Fig.~\ref{P1vsDeltaHfig}---, which we will discuss below).

\newpage
\subsection{Outliers to the $P_1$--$q$ relation}
\label{OutlierPairsSect}

Of the 93 asteroid pairs in our sample, 86 follow the $P_1$--$q$ relation derived from the theory of asteroid pair formation by rotational fission.
Of the 7 outliers, 3 have too slow primary rotations (too low total angular momentum content; they lie below the blue curves in Fig.~\ref{P1vsDeltaHfig}) and 4 have too high mass ratios (they are to the left of the upper limit curves in the plot).

The 3 low-angular momentum pairs are
49791--436459 (Sect.~\ref{49791sect}),
53537--503955 (Sect.~\ref{53537sect}) and
69298--2012FF11.  In the last case, it may be due to a possible error in its $P_1$ or $q$ value; we cannot entirely rule out that the period of (69298) is in error by a factor of 2 if it has a monomodal lightcurve,
or the absolute magnitude of the secondary 2012\,FF11 taken from MPC may be in error.  Thus, the anomalously low angular momentum content of this pair has to be confirmed with accurate measurements of the two uncertain
parameters in the future.  The other two asteroid pairs appear to be real outliers; it is unlikely that their $P_1$ and $q$ values could have big errors.  However, in the case of 49791--436459, we have to consider
the possibility that it may not be a real pair, but that it may be a random orbital coincidence of two unrelated asteroids, see Sect.~\ref{49791sect}.  The case of 53537--503955 is, however, a securely established pair (see Sect.~\ref{53537sect})
and it is a true outlier --- the lowest point at $\Delta H = 3.3$ in the $P_1$--$q$ plot.  We do not have a physical theory for how it could be formed with such slowly rotating primary and very small secondary.  Just, we consider a possibility
that it might not be actually an asteroid pair, but an asteroid cluster, similar to the clusters studied in Pravec et~al.~(2018), with its more secondaries with sizes similar to (503955) waiting to be discovered in the future.
With a few more secondaries with sizes of 0.5--1~km, its mass ratio would rise to $q > 0.1$ that would be in agreement with the theory of asteroid cluster formation.
We note that the known main belt asteroid population is highly
incomplete at asteroid diameters $< 1$~km ---we estimate a diameter of $\sim 0.8$~km for (503955)--- so there may exist yet-to-be-discovered secondaries of the putative cluster of (53537).  We conclude that
all the three anomalous pairs with apparently too low angular momentum content may be just due to our uncertain or incomplete knowledge and they do not represent a real challenge to the theory of asteroid pair
formation by rotational fission.

The 4 anomalous high-mass ratio pairs are
60677--142131 (Sect.~\ref{60677sect}),
76148--56048 (Sect.~\ref{56048sect}),
80218--213471 (Sect.~\ref{80218sect}) and
122173--259585 (Sect.~\ref{122173sect}).
All the four pairs were securely identified, there is no doubt about their reality.
%(see their respective subsections above).
Three of the four form a compact group in the upper left of Fig.~\ref{P1vsDeltaHfig}.
They share a number of common properties.  All the 6 members (3 primaries and 3 secondaries) of the three pairs rotate fast, with periods from 2.7 to 4.7~h and they have low lightcurve amplitudes $<0.28$~mag that suggest
that they have nearly spheroidal shapes with the equatorial axis ratios $<1.3$.  They are small; we estimate their diameters 1--2~km.  The three pairs are relatively young, their likely ages are between 100 and 300~kyr.
They were not taxonomically classified yet; the $(V-R)$ color indices and positions in the inner main belt suggest that the pairs 60677--142131 and 122173--259585 could belong to the SQ complex, while 80218--213471
might be more likely an X type.  The fourth pair, 76148--56048 is somewhat separated in certain properties from the three of the compact group: Its members rotate slower ($\sim 65$ and 7.0~h) and it is older (estimated age $\approx 1200$~kyr),
but, like the other three pairs, it is small asteroids ($D_2 = 2.4 \pm 0.5$~km; see Electronic Supplementary Information) with low lightcurve amplitudes, and it likely belongs to the S complex (based on its $(V-R)$ color indices,
geometric albedo, and position in the inner main belt).  And, to make it even more interesting, the primary of the pair 80218--213471 has a bound, orbiting synchronous secondary, see Sect.~\ref{80218sect} and Table~\ref{AstPairSatstable}.

The 4 outlier asteroid pairs present a challenge for the fission theory that seems to explain the other population of asteroid pairs.
Definitive results on how they formed will require additional information and observations of these systems.
In their absence, we can analyze the extreme energetics of fissioning and reshaping systems to understand what the general theory can say about the limits.
First, we note that the fission theory applied to the other asteroid pairs assumes that the bodies themselves do not undergo reshaping following fission.
At most, we may assume that they fission again, but remain rigid. If we relax this assumption and allow for the fissioned components to reshape
themselves into more spherical shapes, then additional energy can be liberated from the system and can allow asteroid pairs with mass ratios approaching unity.

We note that the 4 outlier asteroid pair systems have relatively unelongated shapes
that may be consistent with this scenario, and thus we attempt to apply this generalized fission theory under the assumption
that the bodies reshaped themselves after fission.
The main elements of the calculations we carry out are given in Scheeres~(2004).
Here we assume that the parent body is an ellipsoid with a given self-potential energy and kinetic energy from rotation.
If it rotates fast enough, it can fission and split into two bodies with masses that sum to the total mass of the initial body.
If we assume that these two bodies take on nearly spherical shapes, this releases additional energy which the system can use to disrupt.

This process can be simply represented if we only allow the different shapes to take on ellipsoidal shapes, as we can evaluate 
the self potentials in closed form.  The general fission process for a single body yields the balance equation
\begin{equation}
E_0 = \frac{1}{2} I_0 \omega_0^2 + {\mathcal U}_{00} = \frac{1}{2} I_1 \omega_1^2 + {\mathcal U}_{11} + \frac{1}{2} I_2 \omega_2^2 + {\mathcal U}_{22} + \frac{1}{2} f (1 - f) M_0 v_{12}^2 + {\mathcal U}_{12},
\end{equation}
where $M_0$ is the mass of the parent body, $f$ is the mass fraction of the two bodies, $I_i$ is the moment of inertia of the $i$th body (assumed to be its maximum moment of inertia), $\omega_i$ is its spin rate,
${\mathcal U}_{ii}$ is the self-potential of the body,
and $v_{12}$ and ${\mathcal U}_{12}$ are the relative velocity and the mutual potential between the fissioned bodies.
Here $i=0$ is the parent body which splits into bodies 1 and 2. The free energy of the fissioned system is
\begin{equation}
E_{\rm free} = E_0 - {\mathcal U}_{11} - {\mathcal U}_{22}
\end{equation}
and is normally taken as a constant. We note that the free energy must be positive for the two bodies to be able to escape from each other (Scheeres, 2002),
and is the fundamental property that the fission theory of asteroid pairs is founded on.
If the two bodies escape from each other, then ${\mathcal U}_{12} \rightarrow 0$ and $v_{12} \rightarrow v_\infty$.
The limiting case is for $v_\infty \sim 0$ and it provides a constraint on the final spin rates of the separated bodies,
\begin{equation}
E_0 - {\mathcal U}_{11} - {\mathcal U}_{22} \ge \frac{1}{2} I_1 \omega_1^2 + \frac{1}{2} I_2 \omega_2^2.
\end{equation}

For the current analysis we note that if bodies 1 and 2 deform into a lower energy configuration then this can in fact increase the overall free energy available and potentially allow for larger mass ratios to escape.
The current analysis explores the energetic limits of these systems without necessarily proposing this as a true physical solution for their existence.

To model this we treat the parent body as a triaxial ellipsoid with semi-major axes $a_0 \ge b_0 \ge c_0$, which allows us to compute the self-potential of the body and the spin rate at which the body should undergo
disaggregation (see Scheeres, 2004, for these details). Then, for a given mass fraction split between the bodies, to maximize the free energy we must take
${\mathcal U}_{ii} \rightarrow - (3/5)(G M_i^2/R_i)$, which is the self-potential of a sphere of mass $M_i$ and radius $R_i$ and minimizes the self-potential over all possible shapes.
Similarly, to maximize the resulting spin rates we take the moments of inertia towards that of a spherical body
$I_i \rightarrow (2/5) M_i R_i^2$, which minimizes the maximum moment of inertia over all possible constant density ellipsoidal shapes.
Finally, for the constant density constraint and mass conservation we note that
$M_1 = (1-f)M, R_1 = (1-f)^{1/3}R, M_2 = f M$ and $R_2 = f^{1/3} R$, where $R = (a_0 b_0 c_0)^{1/3}$ is the geometric mean radius of the original body.

\begin{figure}
%\vspace{1cm}
\includegraphics[width=\textwidth]{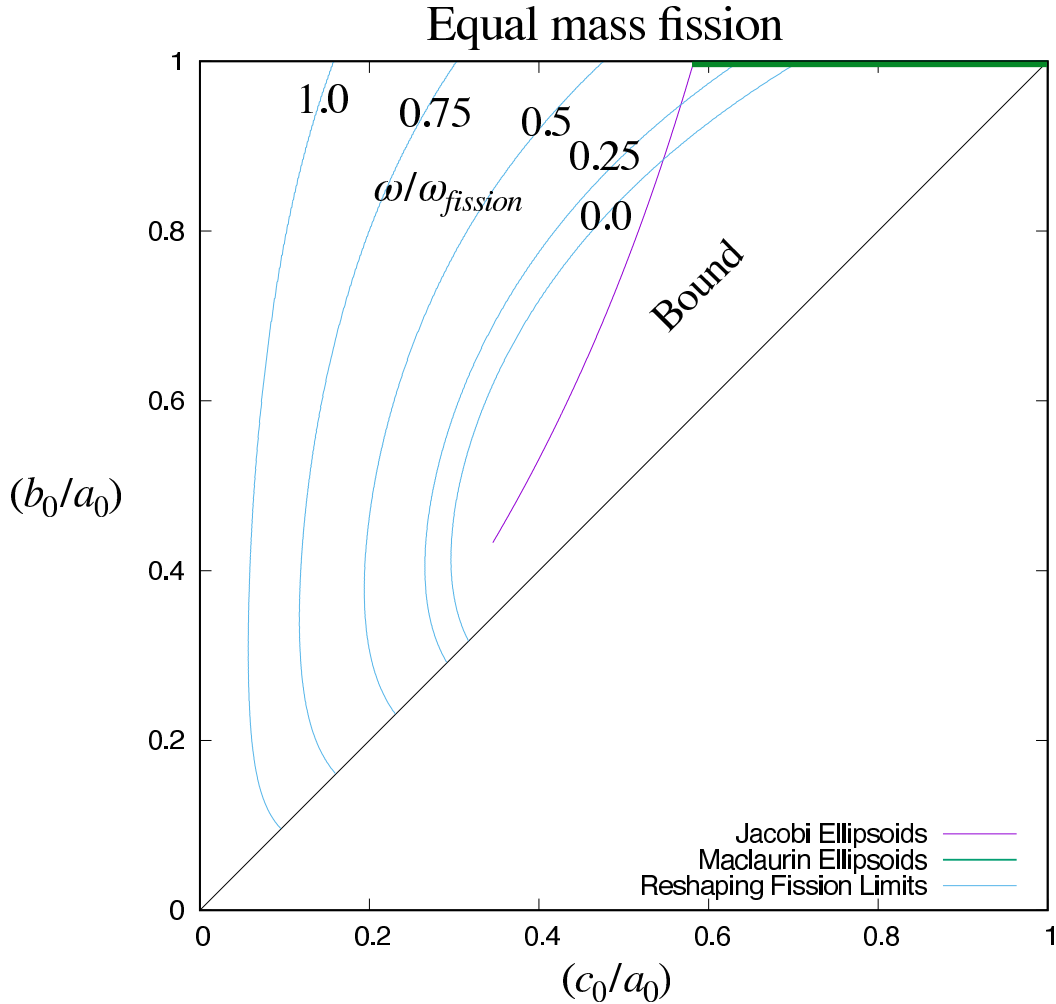}
\caption{\rm The parent body is assumed to be an ellipsoid with the semi-major axes $a_0 \ge b_0 \ge c_0$.
The region to the right of the 0.0 curve is bound and fissioned components cannot escape, while the curves to the left of this region allow escape with the indicated relative spin rates for each body.
}
\label{fissionlimitWfig}
\vspace{1cm}
\end{figure}

Then, for an assumed initial ellipsoidal shape we can calculate whether the bodies can escape and what their expected spin rates are if they do escape.
Figure~\ref{fissionlimitWfig} shows this calculation for $f=0.5$, meaning that the bodies split into equal masses.
This figure shows the possible final rotation rate (in terms of the surface disruption spin rate) as a function of the axis ratios of the initial ellipsoid.
It also assumes that the two bodies spin at the same rate after escape. Also plotted are the size ratios of the Maclaurin and Jacobi ellipsoids.
We note that Maclaurin and Jacobi ellipsoids near the bifurcation point between them can have enough energy to allow their fissioned components to escape if those bodies are allowed to reshape into a spherical shape,
however the spin rate of those bodies would be expected to be low. If a body is significantly distended, however, then the bodies can escape with a larger spin rate.
Thus, from this simple analysis we see that it may be energetically possible for our outlier asteroid pairs to be formed from a fission event --- albeit a very specific set of physical transformations would be required.

\begin{figure}
%\vspace{1cm}
\includegraphics[width=\textwidth]{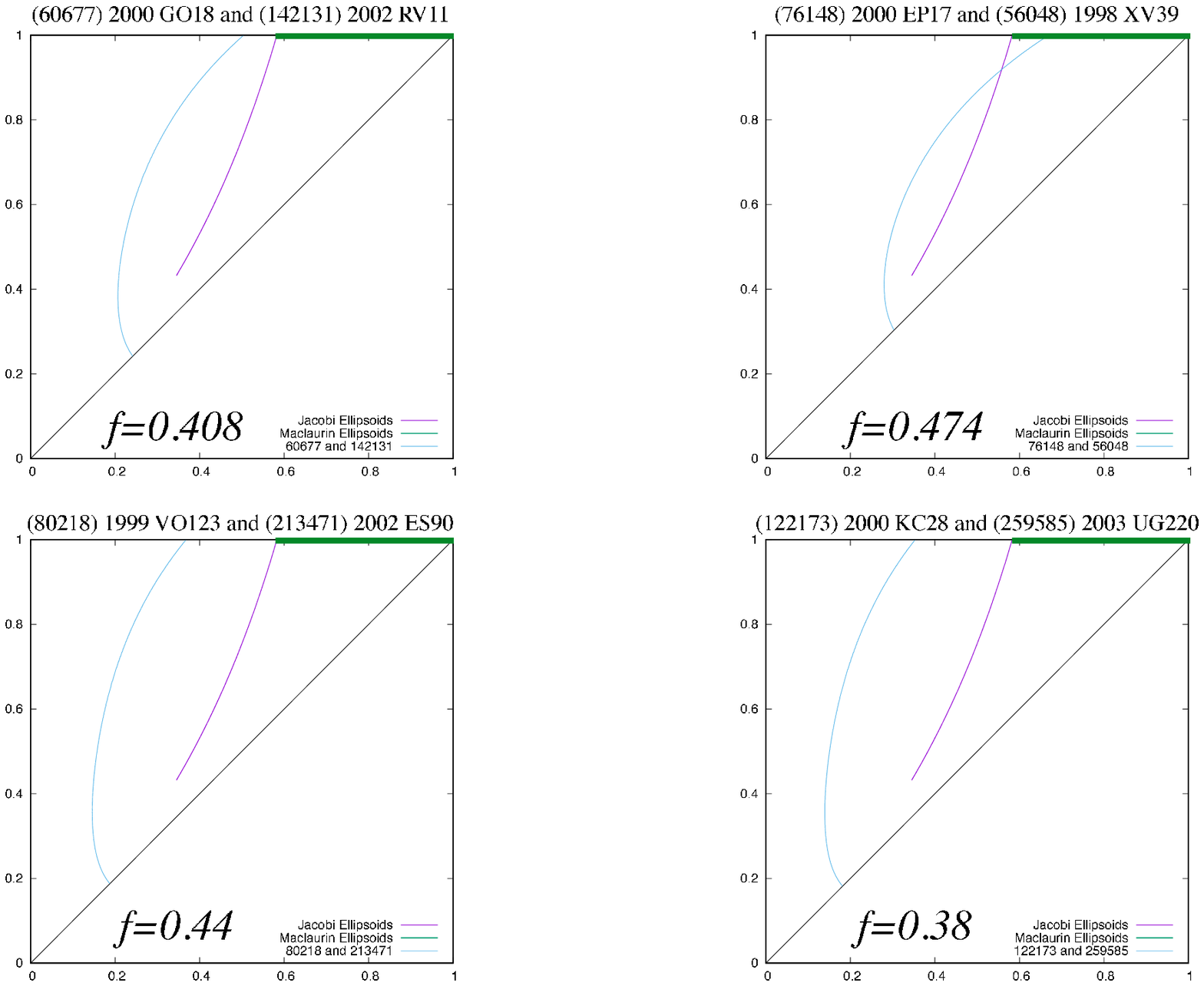}
\caption{\rm Diagrams for each of the specific outlier asteroid pairs. (See caption of Figure~\ref{fissionlimitWfig} for description of content).
The blue line represents the possible initial ellipsoid shape ratios that have sufficient energy to lead to the current asteroid pair.
}
\label{fissionlimitSfig}
\vspace{1cm}
\end{figure}

Given this result, we can compute similar energy curves for each of our candidate outlier asteroid pairs.
Here we assume the specific mass fraction for each pair, assume their respective spin rates (normalized by a spin period of 2 hours which we take as a proxy for these being
likely mostly S type asteroids), and then compute the line of progenitor ellipsoid shapes that would supply sufficient energy for the current configuration (assuming the individual bodies are spheres).
Figure~\ref{fissionlimitSfig} presents these results. For the different pairs Table~\ref{OutlierPairsParamTab} presents the different mass fractions and normalized spin rates.

We see that for the three more rapidly spinning pairs this would require a relatively flattened parent asteroid, requiring an oblate body to have $c_0/a_0$ from 0.4 to 0.6 across these three.
For the slowly rotating body, we see that a Maclaurin Spheroid could lead to this situation.
However, we note that the members of the 4 outlier asteroid pairs are actually not spherical as assumed in the calculations above, but they have certain equatorial elongations producing the non-zero observed lightcurve amplitudes
(and their polar flattenings are largely unconstrained).
Thus, if this scenario with pair component reshaping after fission is true, then the original parent body must be even more flattened than suggested by the above calculation.
Finally, we admit that we do not know a mechanism that could reshape the fissioned components into relatively unelongated bodies as we observe in the 4 outlier asteroid pairs.
We conclude that a formation process for the 4 outlier asteroid pairs remains unknown.

\clearpage

\bigskip
\subsection{Asteroid pairs with binary primaries}

An extremely interesting finding is that many of the fastest rotating primaries of asteroid pairs have also bound, orbiting secondaries.  We found 13 such cases, see Table~\ref{AstPairSatstable}.
In fact, of the 34 asteroid pairs with $P_1 < 3.4$~h in our sample, the 13 ones with binary primaries represent a fraction of 38\%.  And considering that the binary asteroid detection probability of the photometric
method is substantially less than 100\% (see Pravec et~al.~2012a), it is likely that a true fraction of binary systems among the fastest rotating primaries of asteroid pairs is actually at least 50\%.\footnote{A few more possible
binaries among asteroid pair primaries are marked with ``?'' in the column ``Sat.$_1$'' in Table~\ref{AstPairsDatatable}. They await confirmation with further observations.}
It may be comparable to the binary fraction among the fastest rotating near-Earth asteroids larger than 0.3~km that is $(66^{+10}_{-12})$\% (Pravec et~al.~2006).
The 13 asteroid pairs where the primary has a bound, orbiting secondary are marked with green crosses in Fig.~\ref{P1vsDeltaHfig}.

Other than the fast primary rotations, the binary systems among asteroid pair primaries share also other common features with many known near-Earth and small main belt binary asteroids that were reported, e.g.,
in Pravec et~al.~(2006, 2012a, 2016) and Pravec and Harris~(2007).
Specifically, the bound secondaries are relatively small with $D_{1,s}/D_{1,p} < 0.5$, their normalized total angular momentum content is close to critical with $\alpha_L = 0.9$ to 1.3, the primaries are nearly
spheroidal with $a_{1,p}/b_{1,p} \leq 1.2$, the secondaries have low to moderate equatorial elongations with $a_{1,s}/b_{1,s} \leq 1.5$, and the orbital periods of the bound secondaries are in the realm of
tens of hours.\footnote{An exception is the second, distant satellite of (3749) Balam that has an orbital period on an order of a few $10^3$~h.}
It is also notable that, with an exception of (3749) Balam and (21436) Chayoichi, the orbits and rotations of the bound secondaries appear relaxed with eccentricities close to 0 and synchronous spin states.
With the estimated ages of these pairs from 140 to about 1000~kyr, it may place constraints on relaxation timescales in such small a few-km diameter asteroid binaries.  Of the two exceptional asteroid pair primaries
with unrelaxed bound secondaries,
(21436) Chayoichi is quite young with the estimated age of the pair 21436-334916 about 30~kyr, thus possibly there was simply not enough time yet for tidal circularization of the secondary's orbit.
The case of (3749) Balam is very interesting as it has also a second, smaller, distant satellite, and the inner, close secondary is on a slightly eccentric orbit with $e = 0.03$--0.08 (3-$\sigma$ range), but it appears to be in a synchronous spin state\footnote{Note that the tidal circularization of the secondary
orbit is a slower process than tidal synchronization of secondary rotation.} (see Sect.~\ref{3749sect}).

Asteroid pairs having both a bound, orbiting and unbound, escaped secondary might be an outcome of the secondary fission process proposed by Jacobson and Scheeres (2011).  In Pravec et~al.~(2018), the secondary fission process was
proposed to be involved in formation of young asteroid clusters.  We suspect that the asteroid pairs with binary primaries could be ``failed clusters'' where only one of the two formed secondaries was ejected.
However, there is one significant common feature of the paired binary systems that needs to be explained: The bound secondaries
occur only around the fastest rotating asteroid pair primaries with $P_1 < 3.4$~h, but not around slightly slower rotating ones, see the concentration of pairs with binary primaries in the narrow horizontal band in Fig.~\ref{P1vsDeltaHfig}.
%There is no known reason for why the secondary fission process would not work for secondaries orbiting somewhat slower rotating primaries with periods $> 3.4$~h.
%If it was secondary fission what formed the orbiting secondaries, some mechanism must be involved to stabilize their orbits around the fastest rotating primaries, but not around somewhat slower rotating ones with periods $> 3.4$~h.
%In such case, the asteroid pairs with binary primaries could be considered as ``failed clusters''.

\begin{figure}
%\vspace{1cm}
\includegraphics[width=\textwidth]{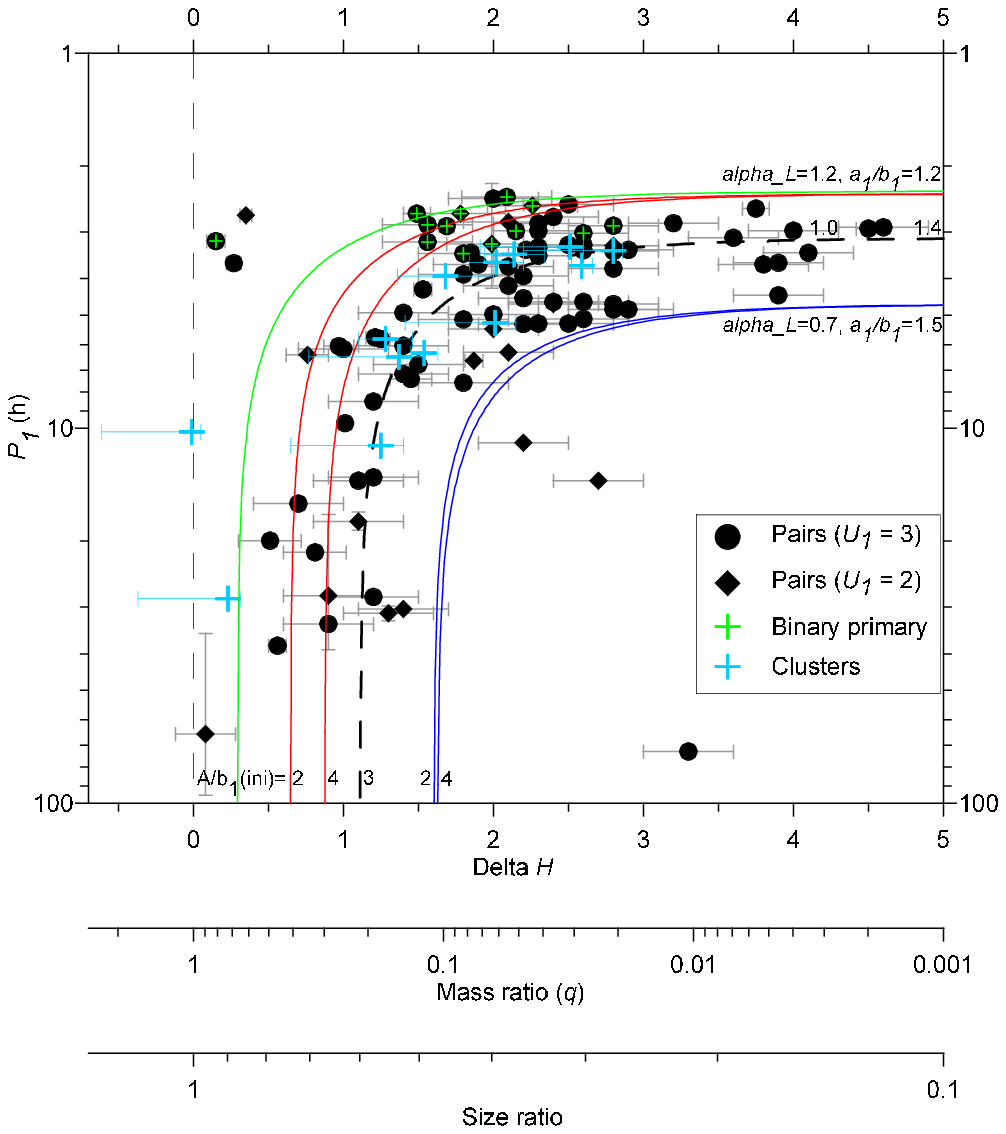}
\caption{\rm Primary rotation periods vs mass ratios of asteroid pairs and clusters, with the data for pairs with binary primaries corrected for hypothetical ejection of the orbiting secondaries, see text.
}
\label{P1vsDeltaHbincorrclustersfig}
\vspace{1cm}
\end{figure}

To look into the hypothesis that the asteroid pairs having also bound, orbiting secondaries are ``failed clusters'', we corrected the mass ratios and primary rotation periods of the 13 pairs with binary primaries for what
they would be if the bound orbiting secondary escaped and the system became a true asteroid cluster.  The resulting mass ratio was calculated as
\begin{equation}
q_{\rm corr} = \left( \frac{D_{2}}{D_{1,p}} \right)^3 + \left( \frac{D_{1,s}}{D_{1,p}} \right)^3,
\end{equation}
where $D_{2}/D_{1,p}$ and $D_{1,s}/D_{1,p}$ are from Table~\ref{AstPairSatstable}.\footnote{For (3749) Balam that has two orbiting secondaries, the corrected mass ratio was calculated with adding also the mass ratio of the second satellite.}
The resulting corrected primary rotation period was calculated from Eq.\,21 of Pravec et~al.~(2018) where for the initial spin rate we took the current (observed) spin rate of a given binary primary ($P_{1,p}$ from Table~\ref{AstPairSatstable}),
for $q$ we took $(D_{1,s}/D_{1,p})^3$, for the parameters $A_{\rm ini}/b_1$ and $a_1/b_1$ we took $a_{\rm orb}/D_{1,p}$ and $a_{1,p}/b_{1,p}$, all from Table~\ref{AstPairSatstable}, and for $c_1/b_1$ and $\rho$ we assumed the
values 1.2 and 2~g/cm$^3$ (see Pravec et~al.~2018, also the discussion in Supplementary Information of Pravec et~al.~2010).  The corrected data are plotted in Fig.~\ref{P1vsDeltaHbincorrclustersfig}, where for comparison we also plotted
the data for asteroid clusters from Pravec et~al.~(2018).  Comparing it with Fig.~\ref{P1vsDeltaHfig},
it is apparent that the points for the asteroid pairs with binary primaries shifted generally to the left (to higher mass ratios) by a substantial amount in most cases, which
is due to that the orbiting secondaries have mostly about comparable masses to the escaped ones (see the 4th and 5th column in Table~\ref{AstPairSatstable}), but they shifted only slightly to lower spin rates as only
a small fraction, on an order of a few percent, of the primary's rotational energy would need to be transferred to the orbiting secondary to put it to an escape parabolic trajectory.

From Fig.~\ref{P1vsDeltaHbincorrclustersfig}, we see that if the asteroid pairs with binary primaries became asteroid clusters, by ejecting the currently bound secondary with transfering a fraction of the primary's rotational
energy to the secondary's motion, they would have similar mass ratios as the existing asteroid clusters, but the primaries would still rotate substantially faster.
If the asteroid pairs with binary primaries are indeed ``failed clusters'', there must be involved a mechanism that stabilizes some secondary orbits around the fastest rotating primaries with $P_1 < 3.4$~h, but not around
somewhat slower rotating ones.\footnote{The general population of small asteroid binaries with primary diameters $\lesssim 10$~km, which are also suggested to be outcomes of rotational fission of critically spinning rubble pile parent asteroids,
also shows a tendency to very fast rotating primaries. In our current data set of small asteroid binaries, 71\% have primary periods $< 3.4$~h.  The 29\% of asteroid binaries with primary periods $> 3.4$~h
may be more evolved systems with the primary rotations slowed down, e.g., by tides or YORP; their ages are not known.  See, e.g., Pravec et~al.~(2006, 2012a, 2016), Pravec and Harris~(2007).}
We will look for such mechanisms in future studies.
One such mechanism was also proposed in Jacobson and Scheeres~(2011), where they hypothesized that a secondary that underwent secondary fission could have one of its components fall onto the primary asteroid,
which would increase its spin rate and potentially give it a more uniform shape. If this ``failed cluster'' and binary component's mass were included in the original fissioned system it would increase the mass ratio
further while decreasing the current primary spin period.  It is not clear how best to model such an interaction, and thus we leave this specific study for the future.

Another hypothesis for the asteroid pairs with binary primaries is that they could be formed by a cascade fission of the primary.  The scenario is following.
There was formed a satellite (orbiting secondary) of the primary in a spin fission event at an earlier time in the past, with the primary rotating sub-critically after the satellite formation.
Then the primary was spun up by YORP to the critical spin rate again and underwent another fission event.  The new secondary started chaotically orbiting the primary and it gravitationally interacted with both the primary and
the older secondary.  One of the two secondaries was then ejected from the system, becoming the unbound secondary (the smaller member of asteroid pair), and the other secondary's orbit around the primary was stabilized, so the system
became an asteroid pair with binary primary.  We will explore this hypothesis in the future.
We note that a cascade disruption process was also suggested for the asteroid cluster of (14627) Emilkowalski by Pravec et~al.~(2018) who found that two of the six secondaries of the cluster separated from the primary relatively recently,
about 320~kyr ago, while the other four secondaries separated at an earlier time, 1-4 Myr ago.

%\newpage
\bigskip
\section{Concluding remarks}

Our studied sample of asteroid pairs is predominated by differentiated asteroid types (mostly S types).  While the paucity of C and other primitive asteroid types in our sample is at least in part an observational bias due to the selection effect against
observations of small low-albedo asteroids in the main belt as discussed in Section~\ref{AlbedoColorTaxonSect}, with only one definitive C type detected and a few others suggested from their low albedos or neutral reflectance colors
(see Table~\ref{AstPairsTaxColorstable}), it will be needed to find and study more of them in the future.  Possible differences between observed properties of C/C-like and S/S-like asteroid pairs would be very interesting to find as they could
provide an information on how their different material properties affect the asteroid fission process.  However, due to the low apparent brightness of small low-albedo primitive types ---especially the small secondaries of asteroid pairs---,
their thorough studies will probably require larger telescopes (2+~m) than we had available for this study.

The two asteroid pairs for which we determined spin vectors for both pair components show the same sense of rotation for both components, which is consistent with the theory of their formation by rotational fission.  In the pair 6070--54827, the component spin vectors were not co-linear at the time of separation
of the two asteroids, but they were tilted by about $38^\circ$.  In the case of 2110--44612, we found that the angle between the initial spin vectors of the two asteroids was between 0 and $\sim 30^\circ$.  It will be needed
to obtain a sample of asteroid pairs with spin vector determinations for both pair members so that we can study what are typical angles between the initial spin vectors of the pair components, to get constraints for further
development of the asteroid pair formation theory.  Again, it will probably require relatively large telescopes (2+~m) as most asteroid pair secondaries are too small for thorough observations with smaller telescopes.

It is remarkable that all the paired asteroids for which we got sufficient observational coverage to check for possible deviations from single periodicity in their lightcurves showed just one-period rotational lightcurves, there
was present no apparent tumbling (i.e., non-principal axis rotation) in any of them.
We note that for $D = 2$~km and $P = 4.5$~h, which are about the median diameter and rotation period of paired asteroids in our sample, the damping time of a non-principal axis (NPA) rotation is estimated to be about $5 \times 10^5$~yr (Pravec et~al., 2014, and references therein).
So, if there was a NPA rotation set in some smaller, slower rotating, or younger paired asteroid in its formation or subsequent evolution (before the pair members separated), we would still see it now.  The fact that we see no tumbling
gives an important constraint to the theories of pair formation and evolution.  Note that with the photometric technique, we can resolve a NPA rotation with rotational axis misalignment angle of about 15\dg or larger (Henych and Pravec, 2013), so there
might be present a low amplitude tumbling in the paired asteroids that we could not detect.

A formation process for the 4 outlier high-mass ratio pairs remains unknown.  While it is energetically possible that they could be formed by rotational fission of a flattened parent asteroid with the components reshaped following fission
as we showed in Section~\ref{OutlierPairsSect}, a mechanism for how they could undergo such very specific physical transformation is unknown.  It will be needed to determine more of their properties, especially shapes, which may give
us more ground for studying how they actually formed.  Particularly interesting may also be to explain the existence of the satellite (bound secondary) of the primary of the high-mass ratio pair 80218--213471.

\bigskip
\bigskip
%\newpage
{\bf \large Acknowledgements}

The work at Ond\v{r}ejov Observatory and Charles University Prague and observations with the Danish 1.54-m telescope on the ESO La Silla station were supported by the Grant Agency of the Czech Republic, Grant 17-00774S.
Petr Fatka was supported by the Charles University, project GA~UK~No.~842218.
%Computational resources for the orbital integrations were provided by the CESNET LM2015042 and the CERIT Scientific Cloud LM2015085, provided under the programme ``Projects of Large Research, Development, and Innovations Infrastructures''.
Access to computing and storage facilities owned by parties and
projects contributing to the National Grid Infrastructure MetaCentrum
provided under the programme "Projects of Large Research, Development,
and Innovations Infrastructures" (CESNET LM2015042),
and the CERIT Scientific Cloud LM2015085, is greatly appreciated.
Operations at Sugarloaf Mountain Observatory and Blue Mountains Observatory were supported by a Gene Shoemaker NEO grant from the Planetary Society.
We thank to A.~Golubaev for his contribution to processing of the observations from Kharkiv Observatory.
The observations at Maidanak Observatory were supported by grants F2-FA-F026 and VA-FA-F-2-010 of the Ministry of Innovative Development of Uzbekistan.
Jose Luis Ortiz acknowledges a support by the Spanish project AYA2017-89637-R and Andalusian project P12-FQM1776.
The research is partly based on observations taken at the Sierra
Nevada Observatory, which is operated by the Instituto de Astrof\'isica
de Andaluc\'ia-CSIC, as well as on observations taken at Calar Alto
Observatory, which is jointly operated by the Max-Planck-Institut f\"ur
Astronomie Heidelberg and the Instituto de Astrof\'isica de
Andaluc\'ia-CSIC, and on observations taken at La Hita Observatory,
which is jointly operated by Astrohita and Instituto de Astrof\'isica de
Andaluc\'ia-CSIC.
The work at Tatransk\'a Lomnica was supported by the Slovak Grant Agency for Science VEGA, Grant No.~2/0023/18, and project ITMS No.~26220120029,
based on the supporting operational Research and development program financed from the European Regional Development Fund.
David Polishook is grateful to the AXA research fund for their generous postdoctoral fellowship.
Josef Hanu\v{s}' work was also supported by the Charles University Research program No.~UNCE/SCI/023.
The work at Abastumani was supported by the Shota Rustaveli National Science Foundation, Grant FR/379/6-300/14.
The work at Modra was supported by the Slovak Grant Agency for Science VEGA, Grant 1/0911/17.
We thank the AGORA association which administrates the 60-cm telescope at R\'eunion--Les Makes observatory,
under a financial agreement with Paris Observatory.
We thank A.~Peyrot and J.-P.~Teng for local support, A.~Klotz and J.~Berthier for robotizing, and P.~Thierry for mechanization of the telescope.
We thank the corps of loyal observers who recorded data at Whitin
Observatory: Kathryn Neugent, Molly Wasser, Amanda Zangari, Kirsten
Levandowski, and Emily Yax.  Observer K.~Neugent was supported by a
Sophomore Early Research Program grant from the Wellesley College Office
of the Dean of the College.  Service observers at Whitin Observatory were
supported in part by grants from the Massachusetts Space Grant Consortium.

\bigskip
{\bf References}

Breiter, S., Nesvorn\'y, D., Vokrouhlick\'y, D., 2005. Efficient Lie-Poisson integrator for secular spin dynamics of rigid bodies. Astron. J. 130, 1267--1277.

Bus, S. J., Binzel, R. P., 2002. Phase II of the Small Main-Belt Asteroid Spectroscopic Survey. A Feature-Based Taxonomy. Icarus 158, 146--177.

Carvano, J. M., Hasselmann, P. H., Lazzaro, D., Moth\'e-Diniz, T., 2010. SDSS-based taxonomic classification and orbital distribution of main belt asteroids. Astron. Astrophys. 510, A43 (12pp).

Colombo, G., 1966. Cassini's second and third laws. Astron. J. 71, 891--896.

Delb\'o, M., dell'Oro, A., Harris, A. W., Mottola, S., Mueller, M., 2007. Thermal inertia of near-Earth asteroids and implications for the magnitude of the Yarkovsky effect.  Icarus 190, 236--249.

Duddy, S. R., et al., 2012. Physical and dynamical characterisation of the unbound asteroid pair 7343--154634. Astron. Astrophys. 539, A36 (11pp).

Duddy, S. R., et al., 2013. Spectroscopic observations of unbound asteroid pairs using the WHT.  Mon. Not. Royal Astron. Soc. 429, 63--74.

Farnocchia, D., et al., 2013. Near Earth Asteroids with measurable Yarkovsky effect. Icarus 224, 1--13.

Hanu\v{s}, J., Delb\'o, M., \v{D}urech, J., Al\'{\i}-Lagoa, V., 2015. Thermophysical modeling of asteroids from WISE thermal infrared data -– Significance of the shape model and the pole orientation uncertainties.
Icarus 256, 101--116.

Hanu\v{s}, J., et al., 2018. Spin states of asteroids in the Eos collisional family. Icarus 299, 84--96.

Henrard, J., Murigande, C., 1987. Colombo's top. Celest. Mech. 40, 345--366.

Henych, T., Pravec, P., 2013. Asteroid rotation excitation by subcatastrophic impacts. Mon. Not. Royal Astron. Soc. 432, 1623--1631.

Ivezi\'c, \v{Z}., et al., 2001. Solar System objects observed in the Sloan Digital Sky Survey Commissioning Data. Astron. J. 122, 2749--2784.

Jacobson, S. A., Scheeres, D. J., 2011. Dynamics of rotationally fissioned asteroids: Source of observed small asteroid systems. Icarus 214, 161--178.

Jester, S., et al., 2005. The Sloan Digital Sky Survey view of the Palomar-Green Bright Quasar Survey. Astron. J. 130, 873--895.

Kaasalainen, M., Torrpa, J., Muinonen, K., 2001. Optimization methods for asteroid lightcurve inversion II. The complete inverse problem. Icarus 153, 37--51.

Kne\v{z}evi\'c, Z., Milani, A., 2003. Proper element catalogs and asteroid families. Astron. Astrophys. 403, 1165--1173.

Kne\v{z}evi\'c, Z., Lemaitre, A., Milani, A., 2002. The determination of asteroid proper elements. In Asteroids III, ed. W. F. Bottke et al. (Tucson: University of Arizona Press), 603--612.

Lagerros, J. S. V., 1996.  Thermal physics of asteroids. I. Effects of shape, heat conduction and beaming.  Astron. Astrophys. 310, 1011--1020.

Levison, H. F., Duncan, M. J., 1994. The long-term dynamical behavior of short-period comets. Icarus 108, 18--36.

Marchis, F., et al., 2008. (3749) Balam. IAU Circ. 8928.

Masiero, J. R., et al., 2011. Main belt asteroids with WISE/NEOWISE. I. Preliminary albedos and diameters. Astrophys. J. 741, 68--89.

Merline, W. J., et al., 2002. S/2002 (3749) 1. IAU Circ. 7827.

Milani, A., Groncchi, G. F., 2010. Theory of orbit determination. Cambridge University Press, Cambridge.

Moskovitz, N. A., 2012. Colors of dynamically associated asteroid pairs. Icarus 221, 63--71.

Nesvorn\'y, D., Vokrouhlick\'y, D., 2006. New candidates for recent asteroid breakups. Astron. J. 132, 1950--1958.

Polishook, D., 2014a. Spin axes and shape models of asteroid pairs: Fingerprints of YORP and a path to the density of rubble piles. Icarus 241, 79--96.

Polishook, D., 2014b. Spins, lightcurves, and binarity of eight asteroid pairs: 4905, 7745, 8306, 16815, 17288, 26416, 42946, and 74096. Minor Planet Bull. 41, 49--53.

Polishook, D., et al., 2014a. Observations of ``fresh'' and weathered surfaces on asteroid pairs and their implications on the rotational-fission mechanism. Icarus 233, 9--26.

Polishook, D., et al., 2014b. Rotationally resolved spectroscopy of asteroid pairs: No spectral variation suggests fission is followed by settling of dust. Icarus 243, 222--235.

Pravec, P., Harris, A. W., 2007. Binary asteroid population. 1. Angular momentum content. Icarus 190, 250--259.

Pravec, P., Vokrouhlick\'y, D., 2009. Significance analysis of asteroid pairs. Icarus 204, 580--588.

Pravec, P. et al., 2006. Photometric survey of binary near-Earth asteroids. Icarus 181, 63--93.

Pravec, P. et al., 2010. Formation of asteroid pairs by rotational fission. Nature 466, 1085--1088.

Pravec, P., et al., 2012a. Binary asteroid population. 2. Anisotropic distribution of orbit poles of small, inner main-belt binaries. Icarus 218, 125--143.

Pravec, P., et al., 2012b. Absolute magnitudes of asteroids and a revision of asteroid albedo estimates from WISE thermal observations. Icarus 221, 365--387.

Pravec, P., et al., 2013. (8306) Shoko. IAU Circ. 9268.

Pravec, P., et al., 2014. The tumbling spin state of (99942) Apophis. Icarus 233, 48--60.

Pravec, P., et al., 2015. (26416) 1999 XM84. CBET 4088.

Pravec, P., et al., 2016. Binary asteroid population. 3. Secondary rotations and elongations. Icarus 267, 267--295.

Pravec, P., et al., 2018. Asteroid clusters similar to asteroid pairs. Icarus 304, 110--126.

Rein, H., Liu, S.-F., 2012. REBOUND: an open-source multi-purpose N-body code for collisional dynamics. Astron. Astrophys. 537, A128 (10pp).

Rein, H., Tamayo, D., 2015. WHFAST: a fast and unbiased implementation of a symplectic Wisdom–Holman integrator for long-term gravitational simulations. Mon. Not. Royal Astron. Soc. 452, 376--388.

Ro{\. z}ek, A., Breiter, S., Jopek, T.J., 2011. Orbital similarity functions --– application to asteroid pairs. Mon. Not. R. Astron. Soc. 412, 987–-994.

Scheeres, D. J., 2002. Stability of Binary Asteroids.  Icarus 159, 271--283.

Scheeres, D. J., 2004. Bounds on rotation periods of disrupted binaries in the Full 2-Body Problem.  Celest. Mech. Dyn. Astron. 89, 127--140.

Scheeres, D. J., 2007. Rotational fission of contact binary asteroids. Icarus 189, 370--385.

Slivan, S. M., et al., 2008. Rotation rates in the Koronis family, complete to $H \approx 11.2$. Icarus 195, 226--276.

Vachier, F., Berthier, J., Marchis, F., 2012. Determination of binary asteroid orbits with a genetic-based algorithm. Astron. Astrophys. 543, A68 (9pp).

Vokrouhlick\'y, D., 1999. A complete linear model for the Yarkovsky thermal force on spherical asteroid fragments. Astron. Astrophys. 344, 362--366.

Vokrouhlick\'y, D., 2009. (3749) Balam: a very young multiple asteroid system. Astrophys. J. 706, L37--L40.

Vokrouhlick\'y, D., Nesvorn\'y, D., 2008. Pairs of asteroids probably of a common origin. Astron. J. 136, 280--290.

Vokrouhlick\'y, D., Nesvorn\'y, D., 2009. The common roots of asteroids (6070) Rheinland and (54827) 2001 NQ8. Astron. J. 137, 111--117.

Vokrouhlick\'y, D., Nesvorn\'y, D., Bottke, W. F., 2006. Secular spin dynamics of inner main-belt asteroids. Icarus 184, 1--28.

Vokrouhlick\'y, D., et al., 2011. Spin vector and shape of (6070) Rheinland and their implications. Astron. J. 142:159 (8pp).

Vokrouhlick\'y, D., et al., 2017. Detailed analysis of the asteroid pair (6070) Rheinland and (54827) 2001 NQ8. Astron. J. 153:270 (17pp).

Vra\v{s}til, J., Vokrouhlick\'y, D., 2015. Slivan states  in the Flora region? Astron. Astrophys. 579, A14.

Wolters, S. D., et al., 2014. Spectral similarity of unbound asteroid pairs. Mon. Not. Royal Astron. Soc. 439, 3085--3093.

Wright, E. L., et al., 2010. The Wide-field Infrared Survey Explorer (WISE): Mission description and initial on-orbit performance. Astron. J. 140, 1868--1881.

Zappal\`a, V., Cellino, A., Farinella, P., Kne\v{z}evi\'c, Z., 1990. Asteroid families. I. Identification by hierarchical clustering and reliability assessment. Astron. J. 100, 2030--2046.

\v{Z}i\v{z}ka, J., et al., 2016. Asteroids 87887-415992: The youngest known asteroid pair?  Astron. Astrophys. 595, A20 (11pp).

\newpage
{\appendix
\begin{center}
\large{APPENDIX}
\end{center}

\section{Thermophysical modeling of (1741) Giclas, (2110) Moore-Sitterly and (4905) Hiromi}
\label{AppendTPM}

Thermal infrared data from the NASA's Wide-field Infrared Survey Explorer (Wright et~al.~2010) are available for three of the paired asteroids for which we derived above their shape models.
Shape model together with the rotation state are necessary inputs for the thermophysical model (TPM) that we utilized to analyze the thermal infrared fluxes.
We made use of the TPM implementation by Delb\'o et~al.~(2007a) based on the previous development of Lagerros~(1996).
In our work, we followed the procedure of Hanu\v{s} et~al.~(2015, 2018).  The detailed description can be found in the aforementioned papers.

We applied the TPM to the thermal infrared data of asteroids (1741) Giclas, (2110)~Moore-Sitterly and (4905)~Hiromi.
For the first two, the best-fitting TPM solution agreed rather well with the thermal infrared fluxes as $\chi^2_\mathrm{red}\sim1-2$.
For both, we obtained thermal inertia values of $\sim$100 J\,m$^{-2}$\,s$^{-1/2}$\,K$^{-1}$ that are consistent with typical values for asteroids within similar size range (Hanu\v{s} et~al.~2018).
Moreover, our sizes are also consistent within the errorbars to the WISE radiometric sizes (Masiero et~al.~2011).
The TPM fit for (4905) Hiromi is poor ($\chi^2_\mathrm{red}\sim8$), therefore the thermophysical properties we provide for this asteroid should be taken with a grain of salt.
The derived thermophysical properties of these three asteroids are summarized in Table~\ref{TMPTab}.
The columns in the table are asteroid number and name, the number of WISE data points in filters W3 $N_\mathrm{W3}$ and W4 $N_\mathrm{W4}$, volume-equivalent diameter $D$,
WISE radiometric size $D_\mathrm{WISE}$, thermal inertia $\Gamma$ in J\,m$^{-2}$\,s$^{-1/2}$\,K$^{-1}$ units, geometric albedo $p_\mathrm{V}$,
surface roughness expressed as Hapke's mean surface slope $\overline \theta$, reduced chi-square of the best fit $\chi^2_\mathrm{red}$, absolute magnitude $H$ and slope $G$,
and the heliocentric distance $r_\mathrm{hel}$ when the thermal infrared observations were acquired.
We note that there are two possible pole solutions and thus shape models for
asteroids (1741)~Giclas and (2110)~Moore-Sitterly, so we applied the TPM independently with the two shape models as inputs.
In both cases, the TPM results are consistent within the two possible shape solutions, which does not allow us to prefer any of them.

\vspace{1cm}
\begin{landscape}
%\scriptsize{
\begin{longtable}{clcclccccccccc}
\caption{TPM results for three paired asteroids.} \\
\hline
\multicolumn{2}{c} {Asteroid} & \multicolumn{1}{c} {Pole} & \multicolumn{1}{c} {$N_\mathrm{W3}$} & \multicolumn{1}{c} {$N_\mathrm{W4}$} & \multicolumn{1}{c} {$D$} & \multicolumn{1}{c} {$D_\mathrm{WISE}$} & \multicolumn{1}{c} {$\Gamma$} & \multicolumn{1}{c} {$p_\mathrm{V}$} & \multicolumn{1}{c} {$\overline \theta$} & \multicolumn{1}{c} {$\chi^2_\mathrm{red}$} & \multicolumn{1}{c} {$H$} & \multicolumn{1}{c} {$G$} &  \multicolumn{1}{c} {$r_\mathrm{hel}$} \\
\multicolumn{2}{l} { } &  &  &  &  \multicolumn{1}{c} {(km)} & \multicolumn{1}{c} {(km)} & \multicolumn{1}{c} {(SI units)} &  &  & \multicolumn{1}{c} {} &  &  & \multicolumn{1}{c} {(au)} \\ \hline\hline
%\multicolumn{14}{l}{} \\
\hline
\endfirsthead
\multicolumn{14}{c}{Table \ref{TMPTab}: {\it cont.}}\\
\hline
\multicolumn{2}{c} {Asteroid} & \multicolumn{1}{c} {Pole} & \multicolumn{1}{c} {$N_\mathrm{W3}$} & \multicolumn{1}{c} {$N_\mathrm{W4}$} & \multicolumn{1}{c} {$D$} & \multicolumn{1}{c} {$D_\mathrm{WISE}$} & \multicolumn{1}{c} {$\Gamma$} & \multicolumn{1}{c} {$p_\mathrm{V}$} & \multicolumn{1}{c} {$\overline \theta$} & \multicolumn{1}{c} {$\chi^2_\mathrm{red}$} & \multicolumn{1}{c} {$H$} & \multicolumn{1}{c} {$G$} &  \multicolumn{1}{c} {$r_\mathrm{hel}$} \\
\multicolumn{2}{l} { } &  &  &  &  \multicolumn{1}{c} {(km)} & \multicolumn{1}{c} {(km)} & \multicolumn{1}{c} {(SI units)} &  &  & \multicolumn{1}{c} {} &  &  & \multicolumn{1}{c} {(au)} \\ \hline\hline
%multicolumn{14}{l}{} \\
\hline
\endhead
\hline
\endfoot
\hline
%\multicolumn{4}{l}{}
\endlastfoot
\label{TMPTab}
(1741)&	         Giclas &	 1 &	 17 &	 10 &	 12.8$^{+0.6}_{-0.3}$ &	12.5$\pm$0.2 &	 100$^{+10}_{-30}$ &	 0.220$^{+0.010}_{-0.021}$ &	 	38.8 &	 1.1 &	 11.62 &	 0.24 &	 3.1 \\
(1741)&	         Giclas &	 2 &	 17 &	 10 &	 12.4$^{+1.3}_{-0.3}$ &	12.5$\pm$0.2 &	 80$^{+30}_{-25}$ &	 0.231$^{+0.013}_{-0.043}$ &	 	38.8 &	 1.4 &	 11.62 &	 0.24 &	 3.1 \\
(2110)&	 Moore-Sitterly &	 1 &	 12 &	  8 &	 6.0$^{+0.5}_{-0.8}$ &	5.4$\pm$0.5 &	 110$^{+70}_{-65}$ &	 0.172$^{+0.059}_{-0.030}$ &	 	16.1 &	 2.0 &	 13.54 &	 0.24 &	 2.3 \\
(2110)&	 Moore-Sitterly &	 2 &	 12 &	  8 &	 6.4$^{+0.4}_{-0.9}$ &	5.4$\pm$0.5 &	 130$^{+70}_{-75}$ &	 0.152$^{+0.052}_{-0.020}$ &	 	16.1 &	 1.5 &	 13.54 &	 0.24 &	 2.3 \\
(4905)&	         Hiromi &	 1 &	 14 &	 13 &	 10.0$^{+0.9}_{-1.0}$ &	8.4$\pm$0.6  &	 45$^{+55}_{-45}$ &	 0.183$^{+0.039}_{-0.031}$ &	 	26.7 &	 8.2 &	 12.43 &	 0.24 &	 2.4 \\
\hline
\end{longtable}
%}
\end{landscape}

\newpage
\section{New asteroid clusters}
\label{appendA}

As a by-product of our search for asteroid pairs, we found also 3 new asteroid clusters.  We studied them with the methods described in Pravec et~al.~(2018) and we outline our results below.
Members of the pairs, their absolute magnitudes, distances from the primary in the space of mean elements and estimated ages are listed in Table~\ref{ClusterMembersTab}.

\bigskip
\subsection{Cluster of (5478) Wartburg}

We discovered this new cluster as a by-product of our search for asteroid pairs.
The distances in the space of mean elements of the two secondaries (479358) 2013~XN8 and 2008~SS185 from the primary (5478) Wartburg are
$d_{\rm mean} = 2.26$~m/s and 16.66~m/s, respectively.
With our backward orbital integrations we confirmed their relation, see Fig.~\ref{5478enchist}.
It suggests that the two secondaries separated from the primary about 300~kyr ago.
For (5478) and (479358), we measured their mean absolute magnitudes $H = 13.03 \pm 0.09$ and $17.72 \pm 0.06$, respectively, with the phase relation slope parameter $G = 0.27 \pm 0.10$ for (5478).
The color index of (479358) is $(V-R) = 0.508 \pm 0.020$.
We also measured their rotation periods $8.5522 \pm 0.0003$ and $6.1820 \pm 0.0003$~h with lightcurve amplitudes 0.49 and 1.08~mag, respectively.

As we identified the second secondary 2008~SS185 as belonging to this cluster only recently, we considered 5478--479358 as an asteroid pair before.  In Fig.~14 of Pravec et~al.~(2018), it was the rightmost point at
$\Delta H = 4.7$ and $P_1 = 8.5522$~h.  Now we know that it is not a pair, but a cluster, and the point shifts to the left to $\Delta H = 3.88 \pm 0.18$ ($q = 0.0047 \pm 0.0012$) in the plot.  However, it is possible that we do not
know all members of this cluster yet and that more will be discovered in the future.  We will analyze this cluster in detail in a future paper.

\bigskip
\subsection{Cluster of (10484) Hecht}

The two asteroids (10484) Hecht and (44645) 1999~RC118 were identified as a pair by Pravec and Vokrouhlick\'y~(2009), and they were further studied in Pravec et~al.~(2010, 2012b).
Recently we found that asteroid 2014~WV530 belongs to it and so the system is actually a cluster.
The distances in the space of mean elements of the two secondaries (44645) 1999~RC118 and 2014~WV530 from the primary (10484) Hecht are
$d_{\rm mean} = 2.35$~m/s and 7.90~m/s, respectively.
With our backward orbital integrations we confirmed their relation, see Fig.~\ref{10484enchist}.
It suggests that the two secondaries separated from the primary about 0.5~Myr ago.
We will study this cluster in detail in a future paper.

\bigskip
\subsection{Cluster of (157123) 2004 NW5}

We discovered this new cluster as a by-product of our search for asteroid pairs.
The distances in the space of mean elements of the two secondaries (385728) 2005~UG350 and 2002~QM97 from the primary (157123) 2004~NW5 are
$d_{\rm mean} = 19.66$~m/s and 2.96~m/s, respectively.
With our backward orbital integrations we found a moderate number of converging clones (Fig.~\ref{157123enchist}), which
suggests that the two secondaries might separate from the primary at different times, about 1800 and 150~kyr ago, respectively.
However, it is possible that the apparent anomalous time distribution of the clone encounters
is affected by that the cluster lies in a relatively chaotic dynamics zone of the main belt.
Moreover, we also consider the possibility that the largest known member of this pair (157123) may not be actually a primary of this cluster, but just the largest secondary,
while a real primary may be somewhat displaced from the three known members of the cluster and it still has to be found.
That means, it could be a case similar to the cluster of (6825)~Irvine where the three secondaries form a tighter concentration that is somewhat displaced from the primary (Pravec et~al.~2018).
For (157123), we measured its rotation period $3.5858 \pm 0.0005$~h, lightcurve amplitude 0.65~mag, the color index $(V-R) = 0.482 \pm 0.023$ and
the mean absolute magnitude $H = 16.93 \pm 0.07$, assuming $G = 0.24 \pm 0.11$ that is the mean $G$ value and range for S types, which is a likely classification for this asteroid.
We will study this cluster in detail in a future paper.

\newpage
\vspace{1cm}
\begin{longtable}{llrc}
\caption{Cluster members, absolute magnitudes, distances from the primary and estimated ages.} \\
\hline
Asteroid           & ~$H$   &  $d_{\rm mean}$   &  $T_{\rm sep}$   \\
                   &        &    (m/s)          &    (kyr)          \\
\hline
\endfirsthead
\multicolumn{4}{c}{Table \ref{ClusterMembersTab}: {\it cont.}}\\
\hline
Asteroid           & ~$H$   &  $d_{\rm mean}$   &  $T_{\rm sep}$   \\
                   &        &    (m/s)          &    (kyr)          \\
\hline
\endhead
\hline
\endfoot
\hline
%\multicolumn{4}{l}{}
%\multicolumn{4}{l}{Asteroids denoted with a star are one-opposition only.}
\endlastfoot
\label{ClusterMembersTab}
~~~(5478) Wartburg                   & 13.03 &   0.00  &         \\
~~~~~~~~~~~~2008 SS185               & 17.2  &  16.66  &    $295_{-123}^{+480}$      \\
(479358) 2013 XN8                    & 17.72 &   2.26  &    $270_{-128}^{+636}$      \\
\hline
~(10484) Hecht                       & 14.18 &   0.00  &         \\
(44645) 1999 RC118                   & 15.0  &   2.35  &    $395_{-133}^{+560}$      \\
~~~~~~~~~~~~2014 WV530               & 18.0  &   7.90  &    $590_{-287}^{+588}$      \\
\hline
(157123) 2004 NW5                    & 16.93 &   0.00  &         \\
(385728) 2005 UG350                  & 17.5  &  19.66  &   $1786_{-524}^{+645}$      \\
~~~~~~~~~~~~2002~QM97                & 18.6  &   2.96  &    $146_{-88}^{+380}$       \\
\hline
\end{longtable}

\clearpage

\begin{figure}
%\vspace{1cm}
\includegraphics[width=\textwidth]{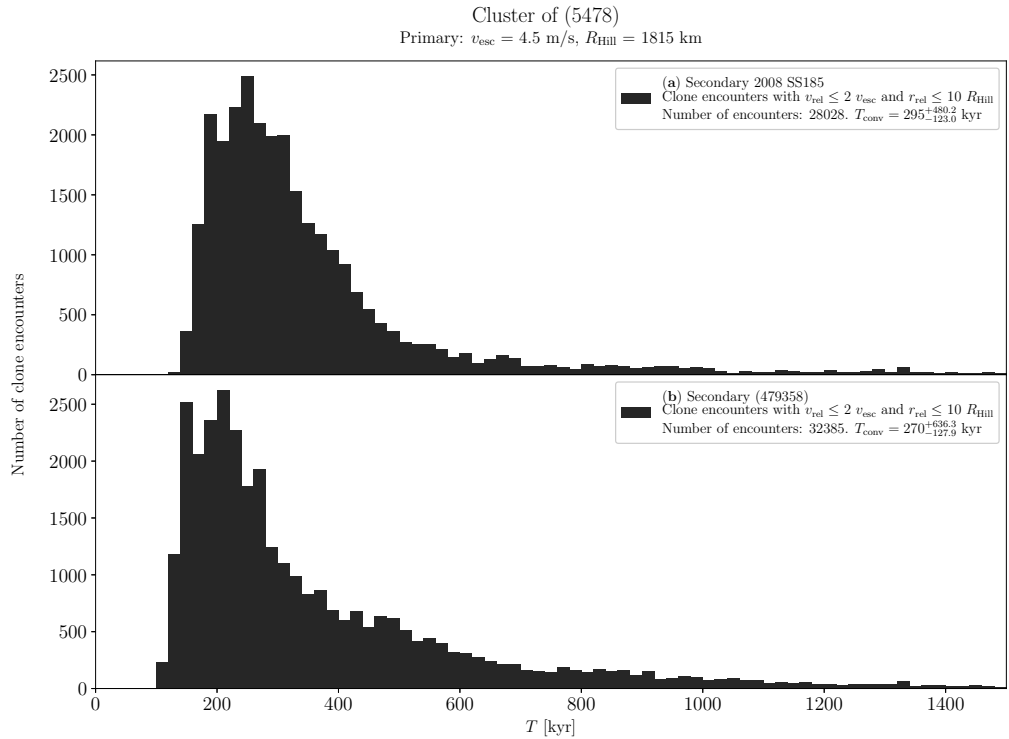}
\caption{\rm Distribution of past times of close and slow primary--secondary clone encounters for the two secondaries
2008~SS185 and (479358) 2013~XN8 of the cluster of (5478) Wartburg.
}
\label{5478enchist}
\vspace{1cm}
\end{figure}

\begin{figure}
%\vspace{1cm}
\includegraphics[width=\textwidth]{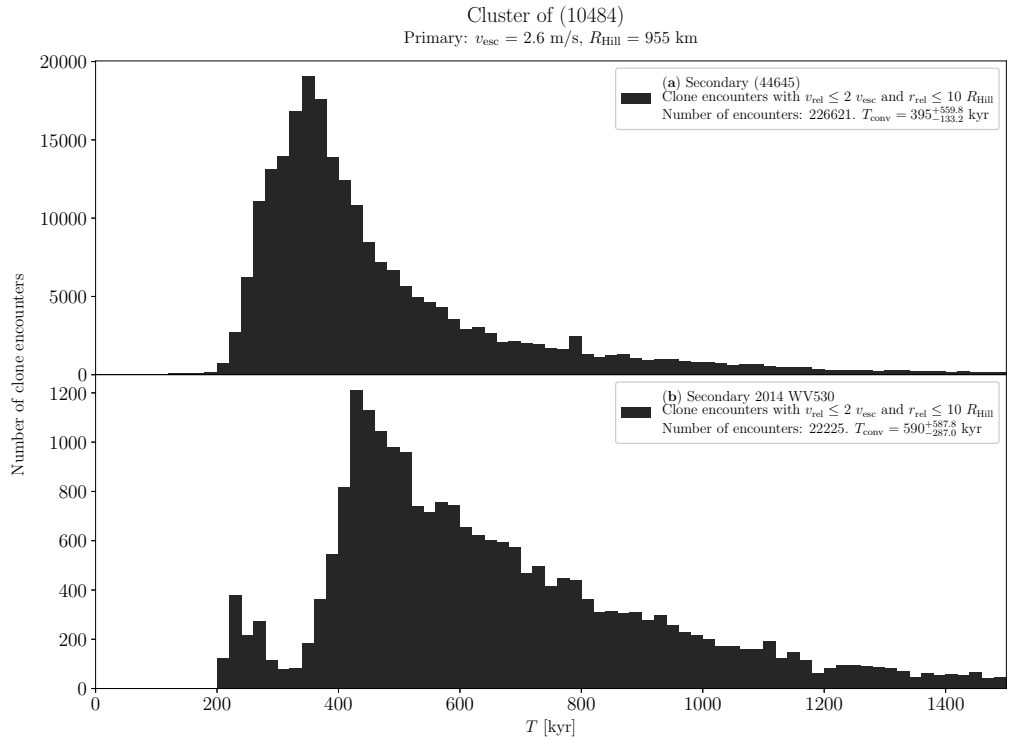}
\caption{\rm Distribution of past times of close and slow primary--secondary clone encounters for the two secondaries
(44645) 1999~RC118 and 2014~WV530 of the cluster of (10484) Hecht.
}
\label{10484enchist}
\vspace{1cm}
\end{figure}

\begin{figure}
%\vspace{1cm}
\includegraphics[width=\textwidth]{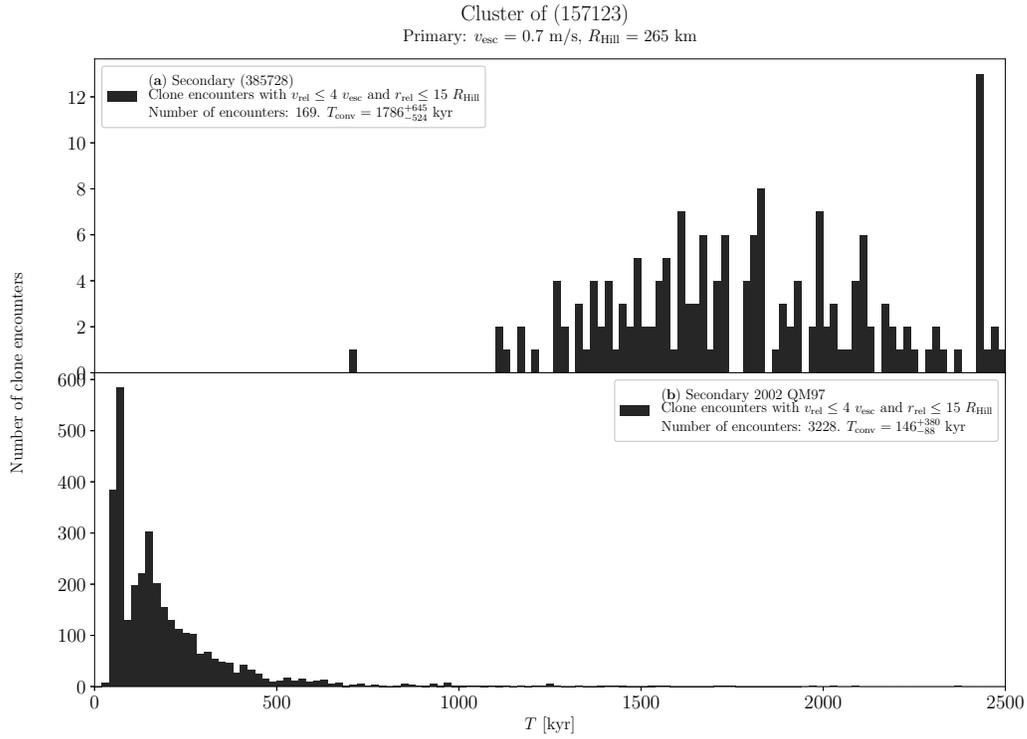}
\caption{\rm Distribution of past times of close and slow primary--secondary clone encounters for the two secondaries
(385728) 2005~UG350 and 2002~QM97 of the cluster of (157123) 2004 NW5.
}
\label{157123enchist}
\vspace{1cm}
\end{figure}

\clearpage

\section{Spurious pairs}

In our previous papers Pravec and Vokrouhlick\'y~(2009), Pravec et~al.~(2010, 2018), there were published two asteroid pairs that we considered to be real pairs at those times, but that we found spurious or needed further
confirmation from our more detailed analyses recently.  They are discussed here.

\bigskip
\subsection{Pair 1979--13732}

The candidate asteroid pair of (1979) Sakharov and (13732) Woodall was proposed by Pravec and Vokrouhlick\'y~(2009) from analysis of their osculating elements.  We revisited it, finding
that the distance of these two asteroids in the space of mean elements is $d_{\rm mean} = 23.70$~m/s and calculating that the probability that this pair is a random coincidence of two unrelated asteroids
from the background population in the space of mean elements is 7\%.
Our backward orbital integrations revealed no clones of the two asteroids that would approach mutually to within $15 R_{\rm Hill}$ at relative velocities $< 5 v_{\rm esc}$ in the past 1.5~Myr,
and we note that the secular angles $\Omega$ and $\varpi$ of their nominal orbits diverge as we go further to the past.
We consider this pair as spurious and we did not include it to the asteroid pair sample we study in this work.

\bigskip
\subsection{Pair 130778--490593}

We found a potential asteroid pair of (130778) 2000~SX369 and (490593) 2009~WL169 from their proximity in the space of mean elements.
The distance of these two asteroids in the space of mean elements is $d_{\rm mean} = 12.85$~m/s and the probability that this pair is a random coincidence of two unrelated asteroids
from the background population is 5\%.
Our backward orbital integrations showed a very low number of clone encounters even with the loosened limits for $r_{\rm{rel}}$ and $v_{\rm{rel}}$, specifically, a total of 72 clone encounters spread across a range of 1300~kyr.
%If this pair is real, it is older than $\sim200$ kyr
We observed (130778) from La Silla during 2015-11-03 to 2016-01-08 and found that it has an extremely long period of $320 \pm 2$~h with a lightcurve amplitude of 0.28~mag at solar phases 6\dg to $26^\circ$.
This is longer by one to two orders of magnitude than the rotation periods we found for other asteroid pairs in our sample.  (In Fig.~14 of Pravec et~al.~2018, it is the lowermost point at $\Delta H = 2.4$ and $P_1 = 320$~h.)
We consider this pair as questionable and its reality needs to be confirmed with further thorough studies.

}

%\addtocounter{table}{-1}

\newpage
%\addtolength{\topmargin}{1.4cm}
%\addtolength{\topmargin}{1cm}
\addtolength{\topmargin}{0.4cm}
\addtolength{\topmargin}{0.2cm}
\begin{landscape}
\tiny{
\begin{center}

\end{center}
}
\end{landscape}

\newpage
\addtolength{\topmargin}{0cm}
\begin{landscape}
\tiny{
\begin{center}
\begin{longtable}{lllccccl}
\caption{Asteroid pairs albedos, taxonomy and colors} \\
\hline
\multicolumn{1}{c}{Asteroid 1}  & \multicolumn{1}{c}{Asteroid 2}  & \multicolumn{1}{c}{$p_{V,1}$~~~~~} & Taxon.$_1$ & Taxon.$_2$ & $(V-R)_1$ & $(V-R)_2$ & Note \\
\hline
\endfirsthead
\multicolumn{8}{c}{Table~\ref{AstPairsTaxColorstable}: {\it cont.}}\\
\hline
\multicolumn{1}{c}{Asteroid 1}  & \multicolumn{1}{c}{Asteroid 2}  & \multicolumn{1}{c}{$p_{V,1}$~~~~~} & Taxon.$_1$ & Taxon.$_2$ & $(V-R)_1$ & $(V-R)_2$ & Note \\
\hline
\endhead
\hline
\endfoot
\hline \\
\multicolumn{8}{l}{\parbox{1.00\textwidth}{
1. Polishook et~al.~(2014a) found that (9068) is an X type.  Considering its position in the Hungaria asteroid group, it is probably an E type. \\
2. For (11286), X type is a likely classification from the SDSS measurements.  Its medium albedo suggests that it is actually an M type. \\
3. For both (13284) and (154828), S type is a likely classification from the SDSS measurements. \\
4. For (291188), we derived that it is an L or S type from the SDSS measurements. \\
5. For (16126), S type is a likely classification from the SDSS measurements. \\
6. Polishook et~al.~(2014a) found that (17288) is an Sw type.  We derived its color index from the SDSS measurements. \\
7. Polishook et~al.~(2014a) found that (19289) is a Q type. \\
8. For (38184), the SDSS colors give an L or S type. \\
9. Polishook et~al.~(2014a) found that (38707) is a V type. \\
10. For (322672), Q type appears a likely classification from the SDSS colors. \\
11. For (55913), X type is a likely classification from the SDSS measurements.  Considering that it belongs to the Hungaria family, it is likely an E type. \\
12. For (56700), S type is a likely classification from the SDSS measurements, though a L type cannot be entirely excluded. \\
13. For (63440), an X type was found both from the spectral measurements by Polishook et~al.~(2014a) and from our Lowell color measurements.  Considering its position in the Hungaria asteroid group, is is probably an E type. \\
14. Polishook et~al.~(2014) found that (74096) is an S type. \\
15. The reported geometric albedo is for (56048). \\
16. Polishook et~al.~(2014a) found that both (88604) and (60546) are S types. \\
17. For (92336), X type is a likely classification from the SDSS colors.  Considering its position in the Hungaria asteroid group, is is probably an E type. \\
%18. Taxon.$_1$, $(V-R)_1$ derived from the SDSS measurements. \\
%19. Taxon.$_2$, $(V-R)_2$ derived from the SDSS measurements. \\
18. Taxon.$_1$ and/or $(V-R)_1$ derived from the Lowell measurements. \\
19. Taxon.$_2$ and/or $(V-R)_2$ derived from the Lowell measurements. \\
}}
\endlastfoot
\label{AstPairsTaxColorstable}
~~(1741) Giclas           & (258640) 2002 ER36    & $0.225^{+0.019}_{-0.037}$ & S  &     &    $0.466 \pm 0.010$ &                    & Sect.~\ref{1741sect} \\
~~(2110) Moore-Sitterly   & ~(44612) 1999 RP27    & $0.162^{+0.069}_{-0.030}$ & S  &Sq/Q &    $0.45~ \pm 0.02~$ &      $0.444 \pm 0.010$ & Sect.~\ref{2110sect} \\
~~(2897) Ole Romer        & (182259) 2001 FZ185   & $0.24 \pm 0.05$ &    &     &                      &      $0.457 \pm 0.017$ & \\
~~(3749) Balam            & (312497) 2009 BR60    & $0.30 \pm 0.07$ & Sq &     &                      &                    & Sect.~\ref{3749sect} \\
~~(4765) Wasserburg       & (350716) 2001 XO105   &                 & X/E&     &    $0.40~ \pm 0.03~$ &                    & Sect.~\ref{4765sect} \\
~~(4905) Hiromi           &~~~(7813) Anderserikson& $0.183^{+0.039}_{-0.031}$ & Sw &  S  &    $0.49~ \pm 0.03~$ &      $0.46~ \pm 0.03~$ & Sect.~\ref{4905sect} \\
~~(5026) Martes           &~~~~~~~~~~~~2005 WW113 &                 & Ch &     &    $0.409 \pm 0.016$ &                    & Sect.~\ref{5026sect} \\
%~(5478) Wartburg         & (479358) 2013 XN8     &                 &    &     &                      &      $0.508 \pm 0.020$ & Sect.~\ref{5478sect} \\
~~(6070) Rheinland        & ~(54827) Kurpfalz     &                 & Sq &  Q  &                      &      $0.424 \pm 0.020$ & Sect.~\ref{6070sect} \\
\end{longtable}
\end{center}
}
\end{landscape}

\newpage
\addtolength{\topmargin}{0cm}
\begin{landscape}
\tiny{
\begin{center}
\begin{longtable}{llccccl}
\caption{Asteroid pair poles} \\
\hline
\multicolumn{1}{c}{Asteroid 1}  & \multicolumn{1}{c}{Asteroid 2}  & $L_1 (^\circ), B_1 (^\circ)$ & Unc.$_1$ ($3 \sigma$) & $L_2 (^\circ), B_2 (^\circ)$ &  Unc.$_2$ ($3 \sigma$) & Note \\
\hline
\endfirsthead
\multicolumn{7}{c}{Table~\ref{AstPairsPolestable}: {\it cont.}}\\
\hline
\multicolumn{1}{c}{Asteroid 1}  & \multicolumn{1}{c}{Asteroid 2}  & $L_1 (^\circ), B_1 (^\circ)$ & Unc.$_1$ ($3 \sigma$) & $L_2 (^\circ), B_2 (^\circ)$ &  Unc.$_2$ ($3 \sigma$) & Note \\
\hline
\endhead
\hline
\endfoot
\hline \\
\multicolumn{7}{l}{\parbox{1.00\textwidth}{
Note: For (2110), (44612), (4905), (56232) and (60744), we give mean radii or semiaxes of the uncertainty areas of the nominal poles, but their boundaries are actually irregular, see their plots in the Electronic Supplementary Information.
The pole uncertainties for (9783), (25021), (26416) and (44620) are long $\times$ short semiaxes of the pole uncertainty areas.
\\
}}
\endlastfoot
\label{AstPairsPolestable}
~~(1741) Giclas           & (258640) 2002 ER36    & $105, +30$ or $288, +24$ & $\pm 5, \pm 10$ or $\pm 5, \pm 10$ &               &      & Sect.~\ref{1741sect} \\
%~~(2110) Moore-Sitterly   & ~(44612) 1999 RP27    & $91, -75$ or $270, -77$  & 20\dg mean radius & $8, -73$ or $193, -69$ & $^{+52}_{-78},\,^{+23}_{-12}$ or $^{+37}_{-73},\,^{+29}_{-11}$  & Sect.~\ref{2110sect} \\
~~(2110) Moore-Sitterly   & ~(44612) 1999 RP27    & $91, -75$ or $270, -77$  & 20\dg mean radius & $8, -73$ or $193, -69$ & $42^\circ \times 18^\circ$  & Sect.~\ref{2110sect} \\
~~(3749) Balam            & (312497) 2009 BR60    & $49, -69$ or $250, -71$  & $\pm 10,\,^{+19}_{-6}$ or $\pm 25,\,^{+21}_{-9}$ &                          &      & Sect.~\ref{3749sect} \\
~~(4765) Wasserburg       & (350716) 2001 XO105   & $235, +8$                & $^{+5}_{-10}, \pm 10$ &                          &      & Sect.~\ref{4765sect} \\
~~(4905) Hiromi           &~~~(7813) Anderserikson& $185, -87$               & 12\dg mean radius &                          &      & Sect.~\ref{4905sect} \\
~~(5026) Martes           &~~~~~~~~~~~~2005 WW113 & $11, +62$ or $197, +47$  & $^{+19}_{-11},\,^{+18}_{-12}$ or $\pm 15, \pm 15$ &                          &      & Sect.~\ref{5026sect} \\
~~(6070) Rheinland        & ~(54827) Kurpfalz     & $124, -87$               & $10^\circ$ radius & $72, -49$ or $242, -46$ & 10\dg in $L_2$, 15\dg in $B_2$ & Sect.~\ref{6070sect} \\
~~(7343) Ockeghem         & (154634) 2003 XX38    & $39, +57$ or $231, +52$  & $^{+30}_{-10}, \pm 20$ or $^{+30}_{-10}, \pm 20$  &                         &                                & Sect.~\ref{7343sect} \\
\end{longtable}
\end{center}
}
\end{landscape}

\newpage
\addtolength{\topmargin}{0cm}
\begin{landscape}
%\begin{table}
%\scriptsize{
\tiny{
\begin{center}
\begin{longtable}{llccrcclllcclcccl}
\caption{Asteroid pair primaries satellites} \\
\hline
\multicolumn{1}{c}{Asteroid 1}  & \multicolumn{1}{c}{Asteroid 2} &  $D_{1,p}$   & $\frac{D_2}{D_{1,p}}$ & $\frac{D_{1,s}}{D_{1,p}}$ & $\frac{a_{\rm orb}}{D_{1,p}}$ & $e$~~ & ~$P_{1,p}$ &    ~$P_{\rm orb}$ &    ~$P_{1,s}$    & $\alpha_L$ & $A_{1,p}$  &   ~$A_{1,s}$  & \multicolumn{1}{c}{SolPh}  & $\frac{a_{1,p}}{b_{1,p}}$  &    $\frac{a_{1,s}}{b_{1,s}}$ & Note \\
                                &                                &  (km)    &                  &  &                 &     & ~(h)   &      ~~(h)        &     ~(h)        &            & (mag)  &     (mag) & \multicolumn{1}{c}{( \dg)}  &            &             & \\
\hline
\endfirsthead
\multicolumn{17}{c}{Table~\ref{AstPairSatstable}: {\it cont.}}\\
\hline
\multicolumn{1}{c}{Asteroid 1}  & \multicolumn{1}{c}{Asteroid 2} &  $D_{1,p}$   & $\frac{D_2}{D_{1,p}}$ & $\frac{D_{1,s}}{D_{1,p}}$ & $\frac{a_{\rm orb}}{D_{1,p}}$ & $e$~~ & ~$P_{1,p}$ &    ~$P_{\rm orb}$ &    ~$P_{1,s}$    & $\alpha_L$ & $A_{1,p}$  &   ~$A_{1,s}$  & \multicolumn{1}{c}{SolPh}  & $\frac{a_{1,p}}{b_{1,p}}$  &    $\frac{a_{1,s}}{b_{1,s}}$ & Note \\
                                &                                &  (km)    &                  &  &                 &     & ~(h)   &      ~~(h)        &     ~(h)        &            & (mag)  &     (mag) & \multicolumn{1}{c}{( \dg)}  &            &             & \\
\hline
\endhead
\hline
\endfoot
%\hline \\
\multicolumn{17}{l}{\parbox{1.00\textwidth}{
~\\
%Note: The values in the fifth column are 3-$\sigma$ upper limits on eccentricity.\\
}}
\endlastfoot
\label{AstPairSatstable}
~~(3749) Balam            & (312497) 2009 BR60    &    4.1   &  0.16  &   0.46   &  3.1  &0.03--0.08&    2.8049167 &  33.38    &   33.39   & (1.31) &   0.11  &  0.04   &   8  &  1.11   &  1.20 & Sect.~\ref{3749sect} \\
~~(3749) Balam            & (312497) 2009 BR60    &    4.1   &  0.16  &   0.24   & 55~~~ & 0.3--0.8 &    2.8049167 &  2600~~~  &           & (1.29) &   0.11  &         &   8  &  1.11   &       & Sect.~\ref{3749sect} \\
~~(6369) 1983 UC          & (510132) 2010 UY57    &    3.3   &  0.17  &   0.37   &  3.4  &    0~~   &    2.39712   &  39.80    &   39.80   &  1.30  &   0.06  &  0.06   &   5  &  1.06   &  1.53 & Sect.~\ref{6369sect} \\
~~(8306) Shoko            &~~~~~~~~~~~~2011 SR158 &    2.4   &  0.27  &$\ge 0.40$&  3.3  &    0~~   &    3.35015   &  36.20    &   36.20   & (1.19) &   0.13  &  0.05   &   6  &  1.14   &  1.28 & Sect.~\ref{8306sect} \\
~~(9783) Tensho-kan       & (348018) 2003 SF334   &    5.1   &  0.24  &   0.24   &  2.8  &    0~~   &    3.0108    &  29.5663  &           &  0.95  &   0.19  &         &   3  &  1.19   &       & Sect.~\ref{9783sect} \\
~(10123) Fide\"oja        & (117306) 2004 VF21    &    3.2   &  0.37  &   0.36   &  4.3  &    0~~   &    2.8662    &  56.46    &           &  1.16  &   0.08  &         &   4  &  1.08   &       & Sect.~\ref{10123sect} \\
~(21436) Chaoyichi        & (334916) 2003 YK39    &    1.9   &  0.29  &   0.36   &  5.5  &0.16--0.22&    2.8655    &  81.19    &           &  1.21  &   0.10  &         &  13  &  1.09   &       & Sect.~\ref{21436sect} \\
~(25021) Nischaykumar     & (453818) 2011 SJ109   &    2.0   &  0.28  &   0.28   &  2.4  &    0~~   &    2.5344    &  23.4954  &   23.50   &  1.09  &   0.07  &  0.04   &  18  &  1.05   &  1.46 & Sect.~\ref{25021sect} \\
\end{longtable}
\end{center}
}
%\vspace{11cm}
%\end{table}
\end{landscape}

\vspace{1cm}
\begin{longtable}{ccccc}
\caption{The outlier high-mass ratio asteroid pairs with their fundamental properties for the computation in Fig.~\ref{fissionlimitSfig}.} \\
\hline
Asteroid Pair      & $f$    & Normalized $\omega_1$ & Normalized $\omega_2$ & Section \\
\hline
\endfirsthead
\multicolumn{5}{c}{Table \ref{OutlierPairsParamTab}: {\it cont.}}\\
\hline
Asteroid Pair      & $f$    & Normalized $\omega_1$ & Normalized $\omega_2$ & Section \\
\hline
\endhead
\hline
\endfoot
\hline
%\multicolumn{4}{l}{}
%\multicolumn{4}{l}{Asteroids denoted with a star are one-opposition only.}
\endlastfoot
\label{OutlierPairsParamTab}
60677--142131  & 0.41 & 0.55 & 0.43 & \ref{60677sect} \\
76148--56048   & 0.47 & 0.03 & 0.29 & \ref{56048sect} \\
80218--213471  & 0.44 & 0.64 & 0.71 & \ref{80218sect} \\
122173--259585 & 0.38 & 0.74 & 0.71 & \ref{122173sect} \\
\hline
\end{longtable}

\end{document}